\documentclass[a4paper,fleqn,usenatbib]{mnras}

\usepackage{mathptmx}

\usepackage{amssymb}
\usepackage{graphicx}

\title[Obscured AGNs in Major Mergers and Interactions]{Incidence of {\it WISE}-Selected Obscured AGNs in Major Mergers and Interactions from the SDSS}

\author[M. E. Weston et al.]{
Madalyn E. Weston,$^{1}$\thanks{E-mail: mew9bc@mail.umkc.edu}
Daniel H. McIntosh,$^{1}$
Mark Brodwin$^{1}$
\newauthor
Justin Mann$^{1,2}$
Andrew Cooper$^{1,3}$
Adam McConnell$^{1}$
Jennifer L. Nielsen$^{1,2}$
\\
$^{1}$Department of Physics \& Astronomy, University of Missouri-Kansas City, Kansas City, MO 64110, USA.\\
$^{2}$ Department of Physics \& Astronomy, University of Kansas, Lawrence, KS 66045, USA.\\
$^{3}$ Department of Physics \& Astronomy, University of North Carolina at Chapel Hill, Chapel Hill, NC 27599, USA.\\
}

\date{Accepted XXX. Received YYY; in original form ZZZ}

\pubyear{2016}

\begin{document}
\label{firstpage}
\pagerange{\pageref{firstpage}--\pageref{lastpage}}
\maketitle


\begin{abstract}
We use the {\it Wide-field Infrared Survey Explorer} ({\it WISE}) and the Sloan Digital Sky Survey (SDSS) to confirm a connection between dust-obscured active galactic nuclei (AGNs) and galaxy merging. Using a new, volume-limited ($z\leq0.08$) catalog of visually-selected major mergers and galaxy-galaxy interactions from the SDSS, with stellar masses above $2\times10 ^ {10} ~ {\rm M}_{\odot}$, we find that major mergers (interactions) are 5--17 (3--5) times more likely to have red $[3.4]-[4.6]$ colors associated with dust-obscured or `dusty' AGNs, compared to non-merging galaxies with similar masses. Using published fiber spectral diagnostics, we map the $[3.4]-[4.6]$ versus $[4.6]-[12]$ colors of different emission-line galaxies and find one-quarter of Seyferts have colors indicative of a dusty AGN. We find that AGNs are five times more likely to be obscured when hosted by a merging galaxy, half of AGNs hosted by a merger are dusty, and we find no enhanced frequency of optical AGNs in merging over non-merging galaxies. We conclude that undetected AGNs missed at shorter wavelengths are at the heart of the ongoing AGN-merger connection debate. The vast majority of mergers hosting dusty AGNs are star-forming and located at the centers of ${\rm M}_{\rm halo}<10 ^ {13} ~ {\rm M}_{\odot}$ groups. Assuming plausibly short duration dusty-AGN phases, we speculate that a large fraction of gas-rich mergers experience a brief obscured AGN phase, in agreement with the strong connection between central star formation and black hole growth seen in merger simulations.
\end{abstract}

\begin{keywords}
merger -- interaction -- AGN -- WISE -- infrared.
\end{keywords}


\section{Introduction}
\label{sec:intro} 

A connection between major galaxy mergers and active galactic nuclei (AGNs) remains a debate in galaxy evolution research. In simulations, major encounters between similar mass (typically $\leq$4:1 mass ratio), gas-rich galaxies are predicted to drive gas to the centers of interacting and merging systems triggering new star formation (SF) and fueling an AGN \citep{Volonteri+03,Hopkins+08, DiMatteo+05,Springel+05, DeBuhr+11}. Many observational studies support the correlation between AGNs and merging systems \citep{Karouzos+10, Treister+12, Cotini+13, Ellison+13b, Nazaryan+14, Satyapal+14, Rosario+15}. However, other researchers, particularly those working at shorter wavelengths, do not find such a connection \citep{Cisternas+11, Kocevski+12, Fan+14, Scott+14, Villforth+14}. One possible reason for this debate is dust obscuration caused by SF in the center of some merging systems \citep{Goulding+09}, indicating that longer wavelengths, such as the infrared, may be able to provide a better understanding of the AGN-merger connection. In this study, we perform a simple test of whether or not highly disturbed galaxies and major pairs with indicators of tidal activity have an excess obscured-AGN frequency.

In the larger context of galaxy evolution, there are many compelling reasons to expect a connection between major gas-rich mergers and supermassive black hole (SMBH) growth. The hierarchical assembly of massive structures is a key feature of $\Lambda$CDM cosmology \citep{White+78}. The growth of large-scale structures is predicted to drive galaxy mergers \citep{Kauffmann+93, Baugh+96, Cole+00}, which in turn have long been tied to the formation of galactic bulges and spheroidal galaxies \citep{Lake+86, Shier+98, Rothberg+06}. The masses of galaxy spheroids are strongly correlated with the masses of their SMBHs \citep{Magorrian+98}. Moreover, massive, bulge-dominated galaxies are predominantly passive and old \citep{Kauffmann+03}, requiring one or more processes to shut down star production and maintain SF quenching. An oft-cited theoretical quenching process is AGN feedback \citep{Schawinski+07, DiMatteo+08, Kaviraj+11}, which is the release of energy from black hole accretion, either through gas outflows \citep{Granato+14}, often associated with gas-rich mergers \citep{DiMatteo+05, Springel+05, Hopkins+06}, or heating of the interstellar medium \citep{Hopkins+10a}. Such feedback is especially important if gas-rich mergers trigger new strong SF activity as simulations predict \citep{Kauffmann+00, DiMatteo+05, Springel+05, Hopkins+08, Cox+08}.

All of the above has been neatly folded into the modern merger hypothesis and nicely summarized by \citet{Hopkins+08}. Briefly, \citet{Hopkins+08} produced the following gas-rich major-merger model, in which quasar activity from z = 0 -- 6 was accurately reproduced through the merger process. First, two equal stellar mass, gas-rich galaxies pass one another, causing morphological disturbances, such as tidal tails, and causing a small rise in SF ({\it interacting} phase). As the galaxies are gravitationally pulled back into one another, the gravitational torques cause gas inflows into the nucleus, triggering both SF and black hole growth \citep{Barnes+91,Mihos+96}, with the rates of both limited by the amount of gas in the merger ({\it ongoing merger} phase, \citealt{Kennicutt+98}). As the gas supply is used by SF and black hole accretion in this coalescence, it will shroud the center of the merger with large obscuring columns of dust. When the gas supply is terminated $10^7 - 10^8$ years later \citep{Hopkins+08}, the system will enter a brief unobscured quasar phase \citep{Storchi-Bergmann+01, Schawinski+09}, lasting 10 - 100 Myr \citep{Martinez-Sansigre+09}. This will be followed by a halt in the black hole accretion, leaving behind a decaying galaxy. Over time, the merger remnant will evolve into a spheroid or, if the gas content in the surrounding halo is still high enough for SF, a spheroidal disk \citep{Barnes+02, Springel-Hernquist+05, Hopkins+09}.

The connection to AGN activity predicted by simulations is supported by many studies that find high incidences of tidal features associated with AGNs \citep{Treister+12, Cotini+13, Ellison+13b, Satyapal+14, Rosario+15}, including \citet{Karouzos+10}, who found that nearly 30\% of active galaxies showed signs of an interaction or merger in the optical and infrared. \citet{Nazaryan+14} performed an optical study of 180 Markarian galaxy pairs and found that ongoing mergers are 7 times more likely to contain an AGN than interacting pairs. In addition, AGN activity increased with decreasing pair separation. \citet{Treister+12} found that the AGN-merger connection has a luminosity dependence, such that mergers overall made up only 10\% of their AGN sample, but increased to $\sim$ 70 - 80\% for the most luminous AGNs ($L_{\rm bol} > 46$ erg/s). \citet{Satyapal+14} used a Wide-Field Survey Explorer ({\it WISE}) color-color cut to find that 9\% of post-mergers and 1\% of close projected pairs at redshift z $<$ 0.2 are obscured AGNs. Their control population in the same redshift range was found to be only 0.5\% obscured AGNs, suggesting that optically obscured AGNs are more prevalent in merging systems. This connection is further supported by the work of \citet{Kocevski+15}, who used X-ray spectral analysis to select Compton-thick AGNs with z $<$ 1.5 and found that obscured AGNs are three times more likely to display merger or interaction signatures compared to unobscured AGNs. In this work, we aim to replicate the \citet{Satyapal+14} study to lend support to the dusty AGN-merger connection, as well as add valuable information about the properties of dusty AGN-mergers.

While simulations and many observational studies find a link between merging systems and AGNs, the topic is still a debate. Many studies, especially those in the X-ray regime, do not find a correlation between mergers and AGN activity \citep{Cisternas+11, Kocevski+12, Bohm+13, Fan+14, Scott+14, Villforth+14}. \citet{Villforth+14} studied the morphologies of 60 X-ray AGNs at 0.5 $<$ z $<$ 0.8 and found that, compared to a simulated control, AGN host galaxies showed no increase in asymmetries or disturbances. Their sample showed a maximum of 6\% of AGNs were related to a major merger. In stark contrast to the simulation-predicted AGN-merger connection, \citet{Scott+14} used optical data to analyze the Seyfert fraction as a function of pair separation in an interacting sample. They found a decrease in Seyferts as pair separation dropped, indicating a drop in AGN activity as interactions begin to merge. The majority of connection-lacking studies in the X-ray and optical regimes all suffer from one important implication of the simulation: they can miss AGNs due to dust obscuration caused by SF in the nucleus.

The relationship between SF and interacting systems is a well-documented phenomenon \citep{Kennicutt+87, Barton+00, Lambas+03, Alonso+04, Ellison+08, Ellison+11}. The tidal forces in the interaction funnel cold gas to the center of the merging system, which collapses into stellar nurseries. SF has been found to increase with decreasing pair separations in interacting pairs with $d_{\rm sep}\leq150$\,kpc \citep{Patton+13}. Dust is produced in stellar nurseries and distributed through the interstellar medium via two mechanisms: AGB stars and supernovae \citep{Clemens+13}. As these methods would imply, the SF rate in galaxies correlates to cold dust mass; the more SF, the more dust in the galaxy. A large portion of a galaxy's bolometric luminosity can be absorbed by this dust and re-emitted in the infrared \citep{Kennicutt+98, Treister+10}.

The addition of dust from SF in a merging system can make AGNs hard to detect. As the simulations predict \citep{Springel+05, Hopkins+06, Hopkins+08, DeBuhr+11}, SF occurring in the center of a merging galaxy may also produce large obscuring columns, which limit the detection of the AGN. Because of this obscuration, shorter wavelength regimes can be ineffective at isolating obscured AGNs \citep{Treister+10, Goulding+11}, which would be the AGN type most likely to occur in the major merger model \citep{Hopkins+08}. Optical spectroscopic surveys can miss as much as half of the AGN population due to obscuration, particularly that caused by dust from SF \citep{Goulding+09}. Below 10 keV, the X-ray regime can also be affected by dust obscuration \citep{Treister+10, Kocevski+15}. \citet{Koss+10} use the {\it SWIFT} BAT AGN survey (hard X-rays; 14--195 keV) and do find an AGN-merger connection, verifying that soft X-rays might not be adequate for obscured AGN selection in mergers. The likelihood of dust obscuration from increased SF in merging and interacting systems leads to the use of the infrared wavelength regime to isolate AGNs, as countless other studies have done \citep{Stern+05, Goulding+09, Jarrett+11, Stern+12, Assef+13, Yan+13}. While the infrared regime has drawbacks of its own (see \S \ref{sec:55discussion_IRlimitations}), it will allow us to isolate a subpopulation of {\it dusty} AGNs \citep{Cardamone+08}. Following work by \citet{Satyapal+14}, we search for obscured AGNs in a complete sample of visually selected major mergers and interactions from the SDSS, using near- and mid-infrared data from {\it WISE} \citep{Wright+10}. {\it WISE} provides all-sky coverage at wavelengths between 3 and 22 microns with sensitivities better than IRAS and DIRBE, making it the best database to look for obscured AGNs in SDSS galaxies.

Using the infrared in this study allows us to quantify the incidence of {\it dusty} AGNs in merging and interacting systems. Using a sample of interacting pairs and ongoing mergers visually selected from a large parent catalog of $\sim$ 65,000 Sloan Digital Sky Survey (SDSS) galaxies, combined with mid-infrared colors from the {\it WISE} All-Sky survey, we isolate a population of dusty AGNs and compare their relative frequency to that of a control sample. We find strong evidence in support of the AGN-merger connection. We also analyze the properties of the infrared-selected AGNs to identify any unique trends. In addition, we discuss the AGN-merger connection for both optical and infrared selection to emphasize the dusty AGN-merger connection. In \S \ref{sec:2sample}, we describe our sample of merging and interacting galaxies. In \S \ref{sec:3WISEanalysis}, we show our analysis and highlight the results of different {\it WISE} color-color methods to select AGNs. We give a summary of {\it WISE} AGN properties for our interacting and merging sample in \S \ref{sec:4nature}. We provide an overall summary of our results and how they compare to literature in \S \ref{sec:5discussion} and \S \ref{sec:6summary}. Throughout this paper, we assume a $\Lambda$CDM cosmological model with ${\Omega}_{m}$ = 0.3, ${\Omega}_{\Lambda}$ = 0.7, and a Hubble constant of $H_{0} = 70$\,km ${\mathrm{s}^{-1}}\mathrm{Mpc^{-1}}$.


\section{Sample \& Data}
\label{sec:2sample} 

If major mergers and galaxy-galaxy interactions produce high levels of central obscuration, it may be difficult to detect the presence of an AGN at {\it shorter} wavelengths. As described in the Introduction, \citet{Satyapal+14} tested this scenario using data from {\it WISE} \citep{Wright+10} applied to a sample of spectroscopic galaxy-galaxy pairs and Galaxy Zoo post-mergers selected from the SDSS. To further test this idea, we likewise use {\it WISE} to identify obscured AGNs and focus on their connection to galaxies with visible signs of {\it ongoing} merging activity selected from a complete catalog of over 60,000 high-mass, low-redshift galaxies in the SDSS. By limiting our selection to $z\leq 0.08$ galaxies for which we can robustly detect tidal features, our samples of major mergers and interactions are complementary to the Galaxy Zoo merging sample of \citet{Darg+10a}. Moreover, our study is complementary to the analysis of \citet{Satyapal+14} in that we focus on active mergers as opposed to post-mergers, and close pairs identified by tidal distortions rather than by velocity differences. Here, we describe our sample selection methodology and provide details about the SDSS and {\it WISE} data we employ in our analysis. We note that all {\it WISE} magnitudes and colors are in the Vega system, while SDSS photometry is in AB mags.


\subsection{Major Mergers \& Interactions from the SDSS}
\label{sec:21SDSSsample} 

We analyze two separate subpopulations: (1) {\it major mergers} consisting of individual isolated galaxies in the last throes of merging, with very disturbed morphologies and sometimes double nuclei, often referred to as ``train wrecks''; and (2) {\it major interactions} between two galaxies with a mass ratio $\leq$4:1 and signs of tidal disturbances commonly associated with merging. The critical difference between these two subpopulations is whether or not the system clearly involves two distinct galaxies ({\it interacting}), as opposed to a single object or a poorly resolved pair ({\it actively merging or coalescing}).

Despite its inherent subjectivity, the visual identification of such dynamical encounters remains the most robust method, especially for systems involving gas and disks that produce long tidal tails, connecting bridges of material between a pair, loops, warps and other strong asymmetries. Indeed, comparisons of classification methods show that visual classification remains the gold standard for validating quantitative methods for identifying galaxy interactions and mergers \citep[e.g.,][]{Conselice+03a,Lotz+04}. Owing to the variety, strength and time dependence of tidal distortions and asymmetries associated with major galaxy interactions, we elect the coarse merger and interaction classifications described above to minimize subjectivity and perform a simple test of whether or not highly disturbed and/or major pairs with indications of tidal activity have an excess obscured AGN frequency. We note that other studies, such as \citet{Veilleux+02}, have attempted more detailed classification systems to identify a larger number (five in their case) of distinct interaction and merger stages in a much smaller sample of rare, low-redshift ultraluminous infrared galaxies (ULIRGs). To maximize the statistical signal of obscured AGN frequency among our subpopulations we decide against subdividing our samples further.

To identify a statistical sample of major mergers and galaxy-galaxy interactions, we visually inspect images of {\it all} 63,454 SDSS DR4 \citep{Adelman-McCarthy+06} galaxies from the stellar mass-limited and volume-limited catalog of \citet{McIntosh+14}. This parent sample contains galaxies with stellar masses ${\rm M}_{\rm star} > 2 \times 10 ^ {10} ~ {\rm M}_{\odot}$ and redshifts $0.01 < z \leq 0.08$ taken from the New York University Value-Added Galaxy Catalog (NYU-VAGC, \citealt{Blanton+05}) reprocessing of the SDSS DR4 Main galaxy sample \citep{Strauss+02} selection of spectroscopic targets with ${r\leq17.77}$ mag. Stellar masses for this catalog were calculated using stellar mass-to-light ratios from \citet{Bell+03} applied to SDSS Petrosian magnitudes and $(g-r)$ colors, as described in detail in \citet{McIntosh+14}. In what follows, we describe our visual identification scheme, discuss spectroscopic redshift completeness of galaxy-galaxy pairs, and summarize our final selections of major mergers and interactions.

\subsubsection{Visual Identification Scheme}
\label{sec:scheme}

We use the SDSS Image List Tool\footnote{\texttt http://cas.sdss.org/astro/en/tools/chart/list.asp} to visually examine a square (143\,kpc on a side)\footnote{Corresponds to 100\,kpc if $H_{0} = 100$\,km ${\mathrm{s}^{-1}}\mathrm{Mpc^{-1}}$.} image centered on each galaxy. This window allows the identification of isolated mergers and possible major interacting companions with $r$-band magnitude differences of $|{\Delta r}| \leq 1.5$ (a proxy for stellar mass ratio $\leq$4:1 under the assumption of a constant stellar mass-to-light ratio), and at large projected separations comparable to other pair studies; e.g., \citet{Satyapal+14} used $d_{\rm sep} \leq 80$\,kpc. Although more time-consuming, this method ensures that we identify all plausible major interaction candidates, including those that would be missed using relative velocity criteria owing to the large spectroscopic incompleteness of SDSS pairs (see \S~\ref{sec:zCompl}). \citet{McIntosh+14} found that the identification of faint tidal features in the SDSS $gri$-combined color images at fixed sensitivity scaling is fairly robust for the masses and redshifts we probe. Each galaxy was independently inspected by D.H.M. plus three students (A.C., A.M., J.M.) and assigned one of the following three types:
\begin{itemize}
	\item {\it Interacting} $=$ galaxy involved in a clear major interaction with a companion based on either (a) both galaxies in a pair show obvious tidal signatures such as tails, bridges, strong asymmetries like warped disks in spiral-spiral (Sp-Sp) interactions, or off-center isophotes in early-type (E-E) interactions; or (b) in the case of `mixed' (E-Sp) interactions, the disk galaxy exhibits strong tidal distortions that can be attributed to its near proximity to a dense E companion. This type is similar to the \citet{Veilleux+02} stage III classification.
	\item {\it Possibly interacting} $=$ galaxy involved in a plausible interaction with a major companion based on more subtle features that depend on pair distinction as follows: (Sp-Sp) either both galaxies show weak signs of interaction, or one galaxy exhibits clear tidal signatures but its major companion appears undisturbed; (E-Sp) a weakly asymmetric disk and an undisturbed early-type galaxy; and (E-E) the nuclei of each galaxy overlaps the outer stellar envelope of its companion. This type is similar to either stage I or II of the \citet{Veilleux+02} system.
	\item {\it Merger} $=$ an individual object with strongly-disturbed global morphology consistent with the coalescing, yet still {\it unsettled}, final phase of merging two galaxies. Common features include large-scale asymmetries extending from the core outwards, double nuclei, multiple asymmetric dust lanes, and multiple tidal tails or loops. This type is closely aligned with the \citet{Veilleux+02} stage IV classification or in cases of double nuclei the close binary subclass of stage III.
\end{itemize}

\begin{figure*}
	\center{\includegraphics[scale=0.85, angle=0]{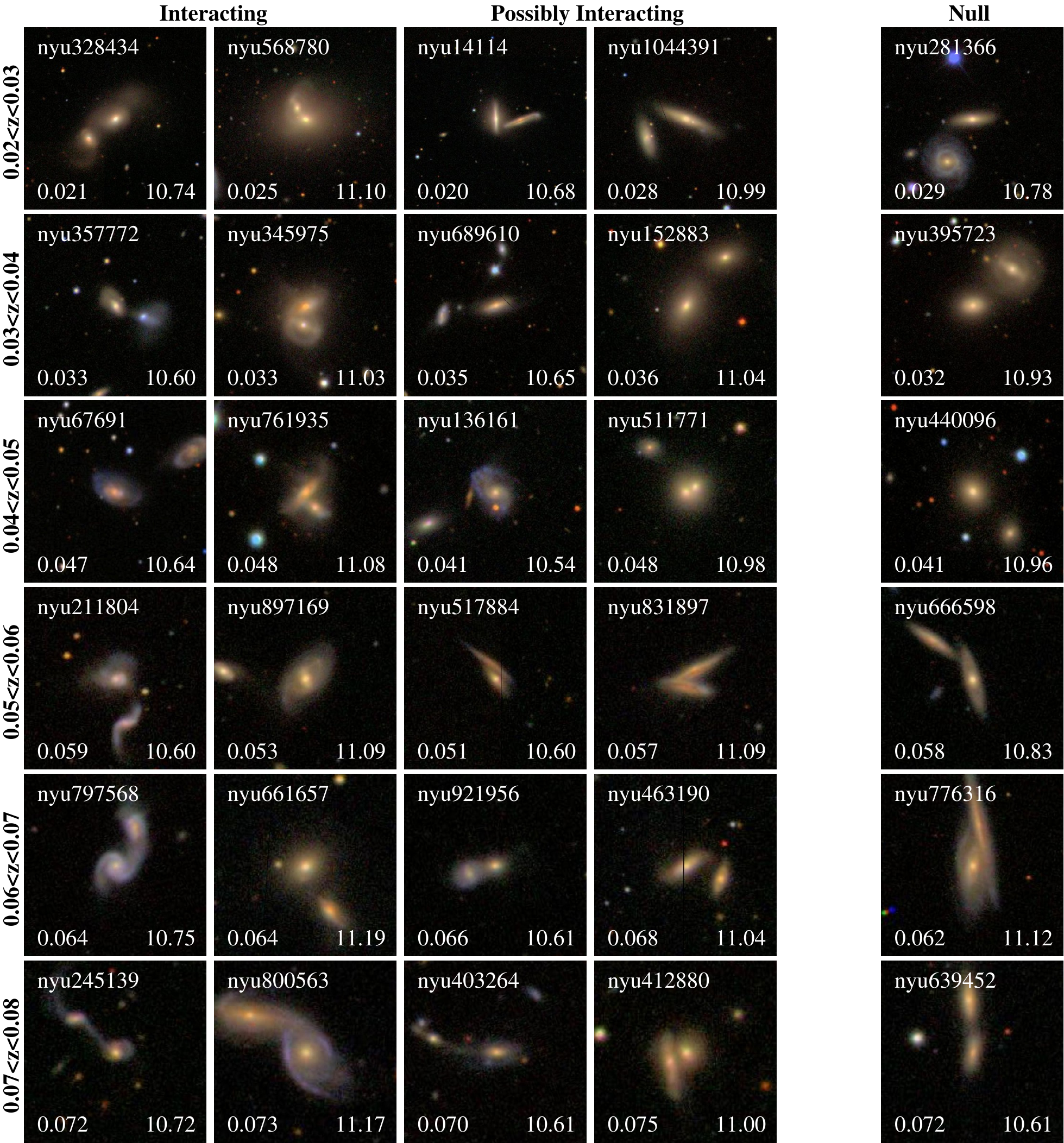}}
	\caption[]{Examples of visual classifications of major galaxy-galaxy pairs from left to right: clear interaction (first two columns), possible interactions (middle two columns), and null or non-interacting (far right column). Each column contains images from six $\Delta z=0.01$ redshift bins between $0.02<z\leq 0.08$, selected from the subset of pairs with spectroscopic redshift information for the primary (${\rm M}_{\rm star,1}>{\rm M}_{\rm star,2}$) galaxy centered in each image. The examples are selected near the 25\% and 75\%-tile of the mass distribution for each redshift bin; nulls are selected randomly. All images are fixed physical size (85\,kpc $\times$ 85\,kpc) cutouts of $gri$-combined color images with fixed sensitivity scaling downloaded from the SDSS Image List Tool. Each image includes the primary galaxy identification number \citep[from the DR4 NYU-VAGC,][]{Blanton+05}, SDSS spectroscopic redshift (lower left), and $(g-r)$-derived stellar mass (lower right, $\log_{10}({\rm M}_{\sun})$ units).
	\label{fig:iExamples}}
\end{figure*}

The visual classifications netted a total of 3533 galaxies (5.6\% of sample) with merger (468) or either major interaction (3065) signatures identified by at least one classifier. An expert (D.H.M.) re-examined this sample, blind to the initial set of classifications, and reclassified 22\% of interaction candidates as non-interacting; i.e., belonging to a null pair. Any galaxies with a minor ($|{\Delta r}|>1.5$) nearby companion that may be responsible for observed tidal disturbances were excluded from the interacting subset and included in the possibly interacting selection. We distinguish pairs of interacting galaxies with small projected separations from mergers with double nuclei by requiring that both galaxies in a close pair be readily resolved as distinct object detections by the SDSS pipeline. The pipeline is known to have systematic magnitude errors for pairs with very close separations \citep[$<3$ arcsec,][]{Masjedi+06}; therefore, pairs with smaller separations were moved from the merger to the clear interaction sample. After matching all galaxies to their major companion, we achieve a sample of 1908 unique pairs: 400 clear interactions based on high classifier agreement (minimum of three out of four; one must be D.H.M.), 1075 possible interactions with lower agreement, and 433 null pairs. In Figure \ref{fig:iExamples}, we present examples of the three types of visually classified pairs. We describe our refined sample of mergers in \S~\ref{sec:RefinedSamps}.

\subsubsection{Redshift Completeness of Small-Separation Major Pairs and Interaction Identification Validation}
\label{sec:zCompl}

\begin{figure}
	\center{\includegraphics[scale=0.5, angle=0]{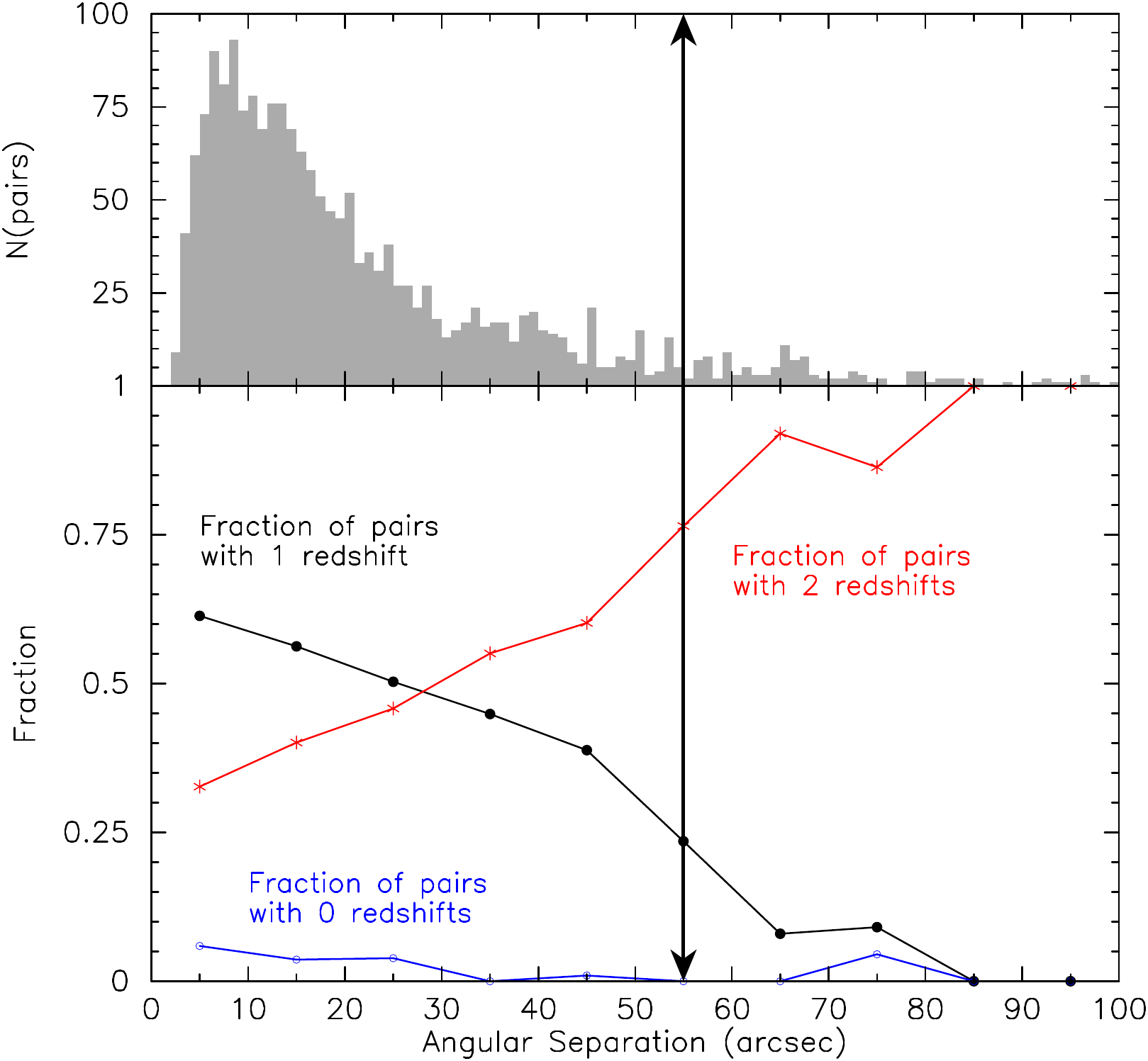}}
	\caption[]{{\it Top:} Distribution of angular separations for the subset of 1908 visually-classified pairs. {\it Bottom:} The fraction of pairs per 10-arcsecond bin that have no spectroscopic data (blue), spectroscopic data for only one galaxy (black), and spectroscopic data for each galaxy (red) are plotted as a function of angular separation. The 55 arcsecond minimum separation (vertical bold arrows) for placement of SDSS spectroscopic fibers on a single plate \citep{Blanton+03b} produces increased redshift incompleteness in pairs with decreasing angular separation.
	\label{fig:SpecIncompleteness}}
\end{figure}

We find a high degree of spectroscopic incompleteness in the SDSS when selecting close pairs of galaxies \citep[see also,][]{McIntosh+08}. The SDSS spectroscopy has 92\% overall completeness for Main sample targets that is independent of galaxy luminosity. The primary source of incompleteness results from the $55\arcsec$ minimum separation for fiber placement (i.e., ``fiber collisions'') in the mechanical spectrograph \citep{Blanton+03b}. The fiber collisions are known to cause slight biases in regions of high galaxy number density \citep{Hogg+04}, such as in massive groups and clusters. The issue is more severe for galaxies in close pairs. The vast majority (91\%) of the 1908 major pairs that we study are comprised of two galaxies meeting Main target criteria of $r\leq 17.77$\,mag, yet, only two-thirds have a spectrum. We explore the dependence of SDSS spectroscopic completeness with projected separation for the full set of visually classified pairs in Figure \ref{fig:SpecIncompleteness}. We divide the pairs into subsets with two, one and zero SDSS redshifts. We find that the two-redshift completeness for pairs decreases steadily from $\sim90\%$ at angular separations of $\theta=80\arcsec$ to below 40\% for $\theta <15\arcsec$. Overall, only 646 (33.9\%) of the 1908 major pairs have SDSS redshifts for both galaxies, thus, a simple selection of close spectroscopic pairs for identifying interactions from the SDSS would be significantly incomplete.

To improve spectroscopic completeness of the interaction samples and to better test the validity of our visual identification of these systems, we include 47 additional spectroscopic redshifts compiled in \citet{Yang+07} from the \citet[2dF,][]{Colless+01}, \citet[PSCz,][]{Saunders+00}, and \citet[RC3,][]{deVaucouleurs+91}. Moreover, a comprehensive search of the NASA/IPAC Extragalactic Database (NED) netted 201 additional spectroscopic redshifts. This improves the number of pairs with two redshifts to 871 (45.6\%), while 969 pairs have only one and 68 have no redshift information.

Since our visual classifications are blind to redshift constraints, we test the reliability of our identification of interacting galaxies using the updated subset of 871 pairs with two redshifts (i.e., spectroscopically complete). In Table~\ref{Delv_table}, we give the percentages for the three pair types that have $\Delta v\leq 500$\,km\,s$^{-1}$ and $\Delta v > 1000$\,km\,s$^{-1}$. We see that clear and possible interactions have a small tail of large velocity outliers that are clearly not physical pairs. These results demonstrate that clear visual interaction cues correspond to close physical pairs (in redshift space) with high confidence, while null interaction signatures often correspond to clearly unphysical pairs with large radial velocity separations. We speculate that the null pairs with small $\Delta v$ may be flybys or may be beginning to interact, but have not yet experienced the initial first close pass which produces clear tidal distortions.

\begin{table}
	\caption{The percentages of different visual pair types (Col. 1) that have small (Col. 2) and large (Col. 3) velocity separations. Values are calculated for the subset of 871 major pairs with dual spectroscopic redshift information.}
	\label{Delv_table}
	\begin{tabular}{lcc}
		\hline
		Pair Type & $\Delta v\leq 500$\,km\,s$^{-1}$ & $\Delta v > 1000$\,km\,s$^{-1}$\\
		(1) & (2) & (3) \\
		\hline 
		\hline 
		Interaction & 92.6\% & 2.3\%\\
		Possible Interaction & 78.6\% & 15.5\%\\
		Null & 56.7\% & 32.6\%\\
		\hline
	\end{tabular}
\end{table}

\subsubsection{Refined Samples}
\label{sec:RefinedSamps}

\begin{figure*}
	\center{\includegraphics[scale=0.95, angle=0]{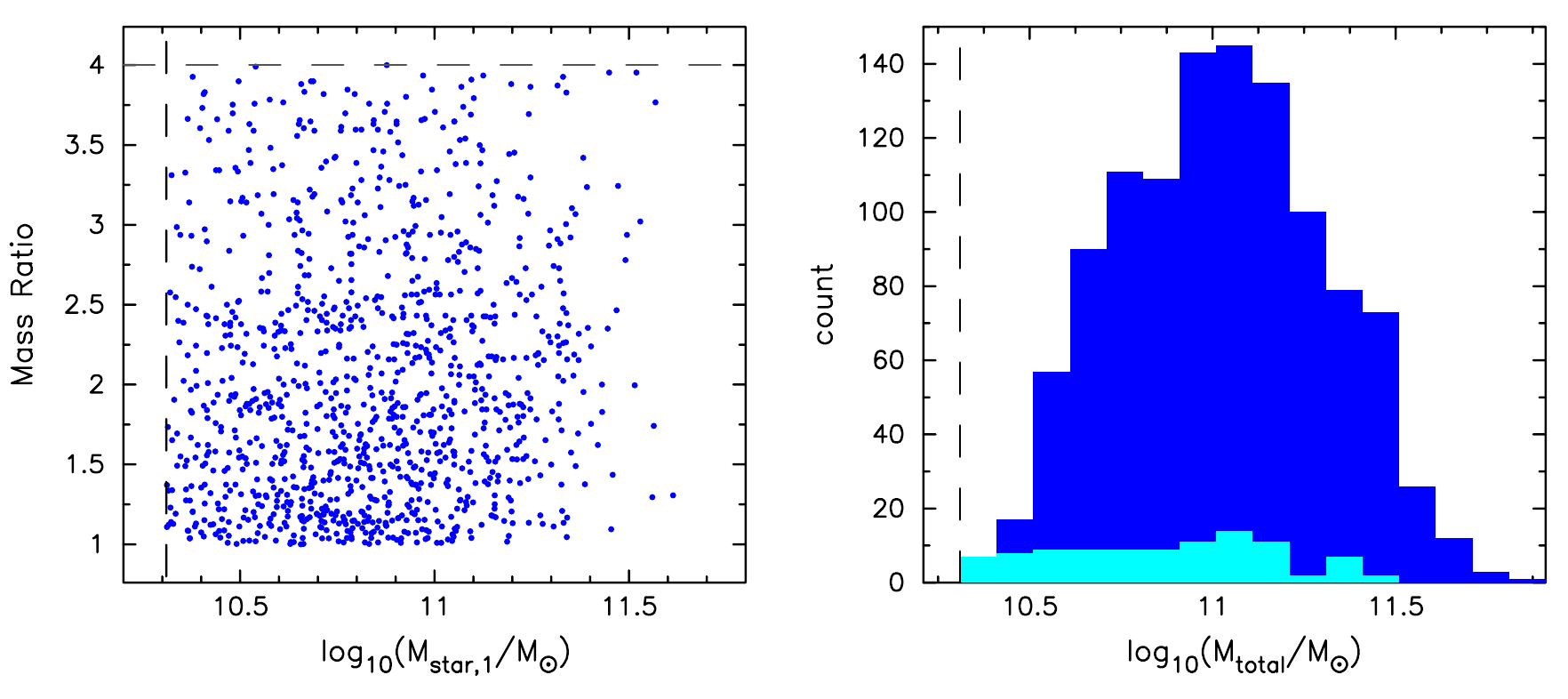}}
	\caption{Left: For interactions, the stellar mass ratio of the system versus the stellar mass of the primary (most massive) galaxy. The vertical dashed line at ${\rm M}_{\rm star,1} > 2 \times 10 ^ {10} ~ {\rm M}_{\odot}$ represents the stellar mass cutoff for this sample. The horizontal dashed line at a stellar mass ratio of four represents the stellar mass ratio cutoff for a major interaction. Right: Stellar mass distribution for the full sample. Ongoing mergers are shown in light blue. The total stellar mass of the interacting pairs (${\rm M}_{\rm star,1}+{\rm M}_{\rm star,2}$) is shown in dark blue. The vertical dashed line at ${\rm M}_{\rm total} > 2 \times 10 ^ {10} ~ {\rm M}_{\odot}$ represents the stellar mass cutoff for this sample.}
	\label{Mratio_Mstar_plot}
\end{figure*}

We construct a refined sample of interactions (clear and possible) from the 1908 major pairs described above. We remove null pairs, dual  spectroscopic pairs with $\Delta v > 1000$\,km\,s$^{-1}$, and a handful of pairs with $d_{\rm sep}>100$\,kpc (97\% of the interaction sample has $d_{\rm sep}<71.4$\,kpc.). For interactions that lack spectroscopic redshifts for one galaxy, we assume both galaxies are at the same physical distance, adopt the known redshift and recompute ${\rm M}_{\rm star}$ as described \citet{McIntosh+14}. We further refine our interaction sample by eliminating pairs with actual stellar mass ratios ${\rm M}_{\rm star,1}/{\rm M}_{\rm star,2}>4$, where ${\rm M}_{\rm star,1}$ is the stellar mass of the primary and most massive galaxy in the pair. For the small subset of interactions with no spectroscopic redshift information for either galaxy, we use the \citet{Yang+07} estimates based on the nearest neighbor redshift. For completeness, this subset is included in the refined interaction sample; hereafter, we omit these systems from our analyses. In Figure~\ref{Mratio_Mstar_plot} (left panel), we show that the visually selected interactions sample the full range of major mass ratios ($1\leq {\rm M}_{\rm star,1}/{\rm M}_{\rm star,2}\leq 4$) at all primary galaxy masses that we probe. We tabulate coordinates, masses and spectroscopic redshift information for the full refined sample of interactions in Table~\ref{tab:i}. In what follows, we analyze a sample of 330 clear and 795 possible interactions.

\begin{figure*}
	\center{\includegraphics[scale=0.8, angle=0]{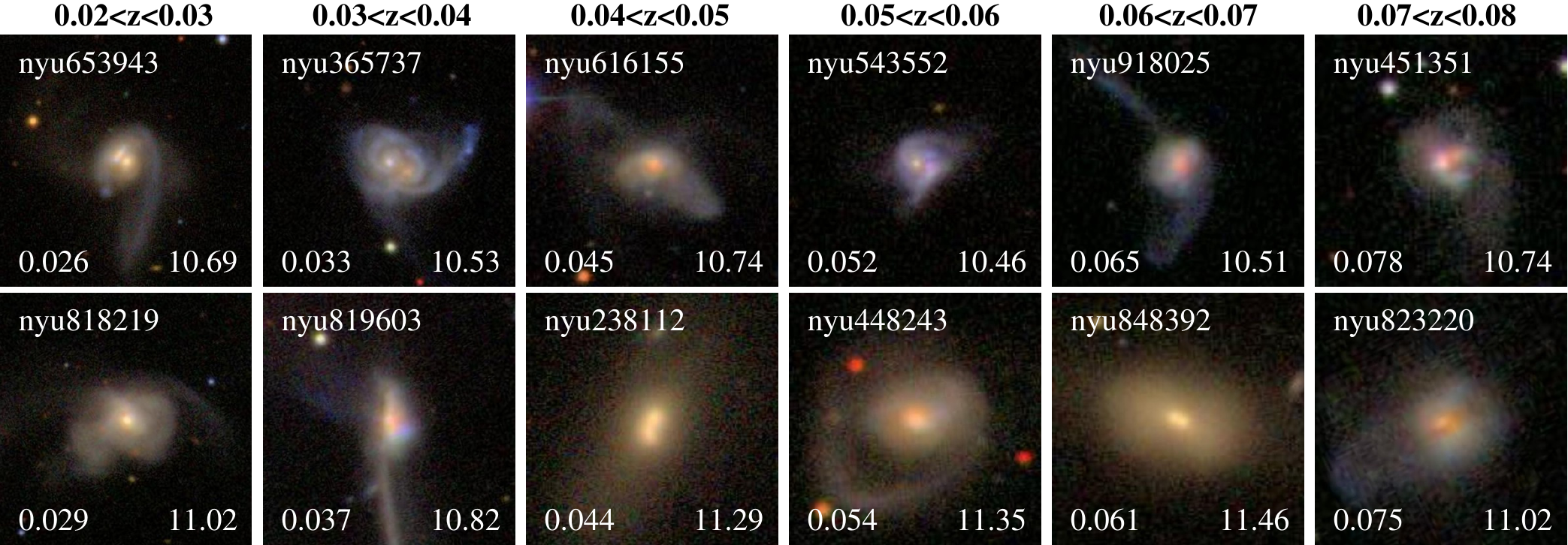}}
	\caption[]{Examples of major mergers spanning six $\Delta z=0.01$ redshift bins between $0.02<z \leq 0.08$. Examples are selected from the subset with spectroscopic redshifts near the 25\% and 75\%-tile of the mass distribution for each redshift bin. All images are fixed physical size (60\,kpc $\times$ 60\,kpc) cutouts of $gri$-combined color images with fixed sensitivity scaling downloaded from the SDSS Image List Tool. Each image includes the DR4 NYU-VAGC identification number, spectroscopic redshift (lower left), and $(g-r)$-derived stellar mass (lower right, $\log_{10}({\rm M}_{\sun})$ units).
	\label{fig:mergers}}
\end{figure*}

To construct a refined sample of mergers, we note that visual merger classifications have an additional source of subjective uncertainty. Unlike the galaxy-galaxy interaction identifications which are limited to clear {\it major} pairs, it becomes increasingly difficult to distinguish between different types of mergers the later one observes a merger remnant after the time of coalescence. The strong tidal distortions found in simulations of major Sp-Sp collisions rapidly disappear once the merger coalesces into a single spheroidal stellar mass distribution and dynamically relaxes. For example, simulations by \citet{Feldmann+08} demonstrate that both major and minor mergers between a variety of progenitor types produce thin (cold) loop and tidal tail features typically associated with gas-rich major interactions. In short, it is impossible using purely visual cues to make clear distinctions between a {\it post-merger} galaxy resulting from a major encounter or from a recent minor accretion onto a pre-existing early-type galaxy. Therefore, we exclude peculiar and disturbed galaxies with a well-defined spheroidal core and define our refined sample of 100 clear mergers that exhibit the strong globally-disturbed morphologies of coalescence. The right panel of Figure~\ref{Mratio_Mstar_plot} compares the stellar mass distribution for the mergers against the total (${\rm M}_{\rm star,1}+{\rm M}_{\rm star,2}$) mass of the major interactions. Mergers have a slightly smaller mean mass than the full sample of interactions, which includes a tail of massive E-E systems. We tabulate coordinates, masses and spectroscopic redshift information for the full refined sample of mergers in Table~\ref{tab:m}, and we display examples in Figure~\ref{fig:mergers} that sample the stellar masses and redshifts we probe.

We compare our combined merging and interacting sample to catalogs produced using Galaxy Zoo. \citet{Darg+10a} created a catalog of 3003 visually-selected binary mergers (akin to our interactions sample) with $z\leq 0.1$ from the SDSS DR6. A second refined catalog of 97 post-mergers from the Galaxy Zoo efforts was produced by \citet{Ellison+13b}. Taking into account differences in sample selections, we find 773 Darg et al. pairs that overlap with our parent sample of $z\leq 0.08$ galaxies from SDSS DR4, with at least one object in pairs having ${\rm M}_{\rm star} > 2 \times 10 ^ {10} ~ {\rm M}_{\odot}$. Our {\it major} interaction classifications agree with 58\% of these Darg et al. binary mergers. We visually examine all objects for which we disagree and confirm that three-quarters are null pairs as described in \S \ref{sec:scheme}.  We note that the remaining quarter have strong tidal features, but 85\% of these are clearly {\it minor} interactions/mergers, 10\% could be major mergers with mass ratios slightly in excess of 4:1, and a handful (5 pairs) appear to be bonafide major interactions with faulty mass ratios. Likewise, for the subset of 26 Ellison et al. post-mergers that fall within our parent sample selection we find nearly 60\% agreement and confirm through visual inspection that the remainder for which we disagree with their classification are examples of inconclusive major merger identifications described in the preceding paragraph; i.e., either plausible post {\it minor} mergers or examples of the peculiar and disturbed galaxies we excluded from our refined sample of {\it major} mergers.

\begin{table*}
	\caption{Basic data for the primary and companion galaxies in visually-identified major interactions. Columns: (1,7) DR4 NYU-VAGC identification numbers from \citet{Blanton+05} or SDSS objID; (2,3,8,9) epoch J2000.0 celestial coordinates; (4,10) stellar masses; (5,11) spectroscopic redshifts compiled by \citet{Yang+07}; and (6) visual interaction type, either I (clear) or I? (possible). In addition to SDSS redshifts, we include spectroscopic redshifts from \citet[2dF,][]{Colless+01}, \citet[PSCz,][]{Saunders+00}, and \citet[RC3,][]{deVaucouleurs+91}. For the subset of galaxies without spectroscopic data from these sources, a thorough search of NED was performed and published spectroscopic redshifts are given with their associated reference code in parenthesis as follows: (a)\citet{Karachentsev+85}, (b)\citet{Peterson+86}, (c)\citet{Owen+88}, (d)\citet{Sharples+88}, (e)\citet{Willick+90}, (f)\citet{Batuski+91}, (g)\citet{Beers+91}, (h)\citet{Fairall+92}, (i)\citet{Hickson+92}, (j)\citet{Davoust+95}, (k)\citet{Lu+95}, (l)\citet{Fisher+95}, (m)\citet{Marzke+96}, (n)\citet{Shectman+96}, (o)\citet{Keel+96}, (p)\citet{Small+97}, (q)\citet{Wu+98}, (r)\citet{Theureau+98}, (s)\citet{daCosta+98}, (t)\citet{Hill+98}, (u)\citet{Barton+98}, (v)\citet{Zabludoff+98}, (w)\citet{Slinglend+98}, (x)\citet{Grogin+98}, (y)\citet{Huchra+99}, (z)\citet{Wegner+99}, (aa)\citet{Falco+99}, (ab)\citet{Wegner+01}, (ac)\citet{Kopylova+01}, (ad)\citet{Xu+01}, (ae)\citet{Koranyi+02}, (af)\citet{Rines+02}, (ag)\citet{Miller+02}, (ah)\citet{Cappi+03}, (ai)\citet{Miller+03}, (aj)\citet{Davoust+04}, (ak)\citet{Francis+04}, (al)\citet{Gronwall+04}, (am)\citet{Rines+04}, (an)\citet{Mahdavi+04}, (ao)\citet{Springob+05}, (ap)\citet{Woods+06}, (aq)\citet{Papovich+06}, (ar)\citet{Yagi+06}, (as)\citet{Jones+09}, (at)\citet{Cava+09}, (au)\citet{Domingue+09}, (av)\citet{Kirshner+83}. Stellar masses were recomputed for these objects. This table is published in its entirety in the electronic edition of the journal. A portion is shown here for guidance regarding its form and content.}
	\label{tab:i}
	\begin{tabular}{lccccclcccc}
		\hline
		\multicolumn{5}{c}{Primary Galaxy} & & \multicolumn{5}{c}{Companion Galaxy} \\
		\cline{1-5} \cline{7-11} \\
		ID$_1$ & RA$_1$ & Dec$_1$ & $\log{\rm M}_{{\star},1}$ & $z_1$ & Type & ID$_2$ & RA$_2$ & Dec$_2$ & $\log{\rm M}_{{\star},2}$ & $z_2$ \\
		(1) & (2) & (3) & (4) & (5) & (6) & (7) & (8) & (9) & (10) & (11)\\
		\hline 
		\hline 
		nyu265583 & 00:05:11.71 & +15:38:46.2 & 11.12 & 0.054 & I & nyu265582 & 00:05:11.42 & +15:38:22.4 & 10.77 & na\\
		nyu248469 & 00:05:58.00 & $-$09:21:17.8 & 10.88 & 0.065 & I & nyu248468 & 00:05:57.46 & $-$09:21:28.3 & 10.83 & na\\
		nyu98607 & 01:24:08.10 & +14:02:04.1 & 11.29 & 0.056 & I? & nyu98608 & 01:24:08.86 & +14:02:05.4 & 10.84 & 0.054(u)\\
		\hline
	\end{tabular}
\end{table*}

\begin{table*}
	\caption{Basic data for visually-identified major mergers. Columns: (1) DR4 NYU-VAGC identification number; (2,3) epoch J2000.0 celestial coordinates; (4) stellar mass; (5) spectroscopic redshift; and (6) the redshift source (see Table~\ref{tab:i} for details). For the handful of mergers without spectroscopic data from these sources, a thorough search of NED was performed and published spectroscopic redshifts are given with their associated reference. Stellar masses were recomputed for these objects. This table is published in its entirety in the electronic edition of the journal. A portion (for the mergers in Fig.~\ref{fig:mergers}) is shown here for guidance regarding its form and content.}
	\label{tab:m}
	\begin{tabular}{lccccc}
		\hline
		ID & RA & Dec & $\log({\rm M}_{\star}/{\rm M}_{\sun})$ & $z$ & Redshift source\\
		(1) & (2) & (3) & (4) & (5) & (6)\\
		\hline 
		\hline 
		nyu653943 & 12:22:57.73 & +10:32:54.0 & 10.69 & 0.026 & SDSS\\
		nyu818219 & 14:45:45.12 & +51:34:51.2 & 11.02 & 0.030 & SDSS\\
		nyu365737 & 09:17:02.03 & +50:02:44.8 & 10.53 & 0.033 & PSCz\\
		nyu819603 & 13:44:42.15 & +55:53:13.8 & 10.82 & 0.037 & PSCz\\
		nyu616155 & 12:30:40.31 & +51:08:14.8 & 10.74 & 0.045 & SDSS\\
		nyu238112 & 23:56:26.52 & $-$11:05:09.5 & 11.29 & 0.044 & SDSS\\
		nyu543552 & 12:17:02.91 & +56:08:29.3 & 10.46 & 0.052 & SDSS\\
		nyu448243 & 22:12:03.50 & $-$07:34:44.2 & 11.35 & 0.054 & SDSS\\
		nyu918025 & 11:36:38.57 & +42:12:35.0 & 10.51 & 0.065 & SDSS\\
		nyu848392 & 13:21:18.70 & +12:12:46.7 & 11.46 & 0.061 & SDSS\\
		nyu451351 & 20:38:48.55 & $-$00:13:34.9 & 10.74 & 0.078 & SDSS\\
		nyu823220 & 14:24:59.77 & +54:31:06.5 & 11.02 & 0.075 & SDSS\\
		\hline
	\end{tabular}
\end{table*}

\subsection{SDSS Spectroscopic Emission Types}
\label{sec:22SDSSspec} 

The SDSS fiber spectroscopy provides a reliable method of isolating different galaxy types through emission-line analysis. We collect all available SDSS DR7 spectroscopy for our sample and use the MPA-JHU emission-line analysis \citep{Kauffmann+03, Brinchmann+04} to determine spectroscopic types where available. We use these spectroscopic types throughout our analysis.

As discussed in \S \ref{sec:zCompl}, the SDSS Main galaxy sample \citep{Strauss+02} has an overall completeness of 92\% for spectroscopy due to plate tiling to minimize fiber collisions \citep{Blanton+03b}. For our refined samples (see \S \ref{sec:RefinedSamps}), the DR4 spectroscopic completeness is 76\% for ongoing mergers and 91\% for interacting pairs. Of the interacting pairs, only 235 (21\%) have SDSS spectroscopy for both galaxies. These galaxy pairs have a median separation of 23 arcseconds. Sixty-six percent of interacting pairs have SDSS spectroscopy for only one of the two galaxies (median pair separation of 16 arcseconds). The drop in completeness coincides with the drop in median pair separation, confirming the predictions made by \citet{McIntosh+08} (see \S \ref{sec:zCompl}). The remaining 106 interacting pairs have no SDSS spectroscopic information for either galaxy.

\begin{figure}
	\center{\includegraphics[scale=0.68, angle=0]{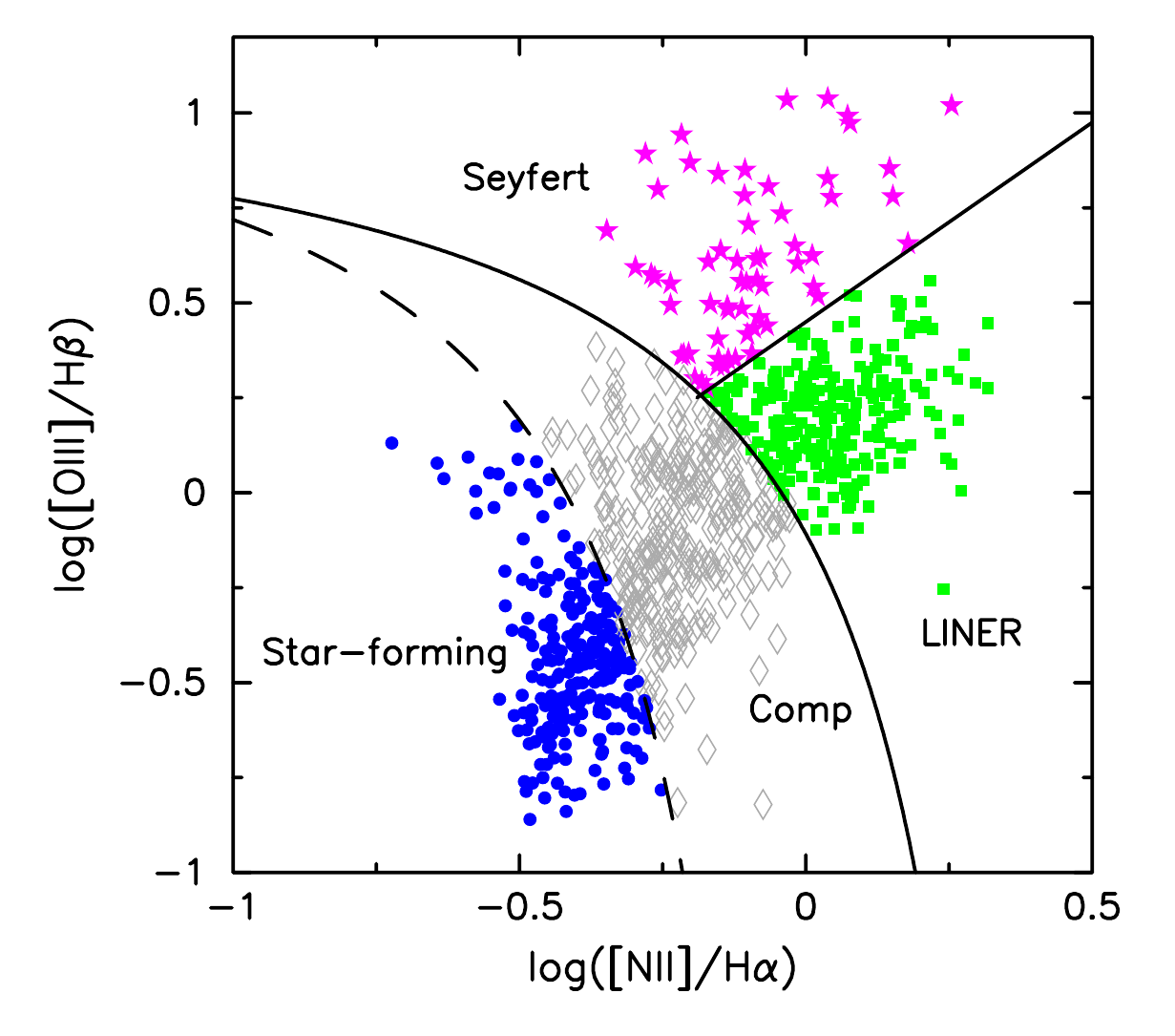}}
	\caption{BPT Diagram \citep{Baldwin+81} for the interacting and merging subset with spectroscopic data and emission detected in all four lines. Star-forming galaxies are shown as blue circles, Composite galaxies as grey diamonds, LINER as green squares, and Seyfert galaxies as pink stars. The dashed curve between star-forming and Composite galaxies is from \citet{Kauffmann+03}. The solid curve is from \citet{Kewley+01} and shows the theoretically derived division between Composite galaxies (some SF) and plausible AGNs. The additional line between the Seyfert and LINER populations is from \citet{Schawinski+07}.}
	\label{fig:BPT}
\end{figure}

For the subset of our sample with spectra, spectroscopic emission types were determined through MPA-JHU emission line flux measurements from the SDSS fiber spectra \citep{Kauffmann+03, Brinchmann+04}. We use the spectroscopic emission types described in detail in \citet{McIntosh+14}. Briefly, as shown in Figure~\ref{fig:BPT}, HII (pure star-forming), Composite (combination of SF and plausible AGN), and plausible AGN galaxies are determined using the criteria of \citet{Kauffmann+03} for galaxies with emission detected  at $\mathrm{S/N} \geq 3$ in the four lines of the standard BPT \citep{Baldwin+81} diagram. LINERs are separated from Seyferts using criteria from \citet{Schawinski+07}. Spectroscopically quiescent (hereafter quiescent) galaxies are defined as galaxies that lack detectable emission in all 4 BPT lines; i.e., these systems are non-star-forming and inactive. A small subset of spectroscopic galaxies do not meet the criteria for BPT analysis nor quiescent selection. These objects have either low-$\mathrm{S/N}$ in all four BPT lines or strong ($\mathrm{S/N} > 3$) emission in only {[}OII{]} and/or H{$\alpha$}. Following \citet{Yan+06}, we identify 'weak-LINER' galaxies with high {[}OII{]}/H${\alpha}$ equivalent-width ratios and strong emission in both lines, and weakly star-forming objects with low {[}OII{]}/H${\alpha}$ ratios or no detectable {[}OII{]} line flux. We find that these galaxies make up 13\% and 12\% of the LINER and star-forming populations, respectively. Throughout our analysis, we consider only Seyferts to be clear AGNs since recent work shows that many LINERs are not ionized exclusively by a central source \citep{Annibali+10,Kehrig+12,Yan_Blanton+12}.

\subsection{{\it WISE} Photometry}
\label{sec:23WISEphot} 

To identify dust-obscured AGNs, we use infrared colors from {\it WISE} observations. {\it WISE} is a 40-cm infrared space telescope funded by NASA and launched in December of 2009 \citep{Wright+10}. The {\it WISE} All-Sky Survey covered more than 99\% of the sky with sensitivities more than 100 times better than the InfraRed Astronomical Satellite (IRAS) in the 12 micron band. We use data from the {\it WISE} All-Sky Source Catalog \footnote{{\it WISE} data and documentation are available at http://irsa.ipac.caltech.edu/Missions/wise.html}. This catalog contains positional and photometric information for over 563 million objects detected in the {\it WISE} images. We matched the sample from \S \ref{sec:RefinedSamps} to the {\it WISE} All-Sky catalog to obtain profile-fit magnitudes in four infrared channels: 3.4 ${\mu}$m (W1), 4.6 ${\mu}$m (W2), 12 ${\mu}$m (W3), and 22 ${\mu}$m (W4) \citep{Wright+10}. These bands have minimum 5${\sigma}$ point source sensitivities of 0.008, 0.11, 0.8, and 6 mJy, respectively \citep{Jarrett+11}. The angular resolution of the first three bands (W1, W2, W3) is $\sim$ 6 arcsecond, while the 22 micron band (W4) has an angular resolution of $\sim$ 12 arcsecond. All {\it WISE} magnitudes are in the Vega system.

\begin{figure}
	\center{\includegraphics[scale=0.7, angle=0]{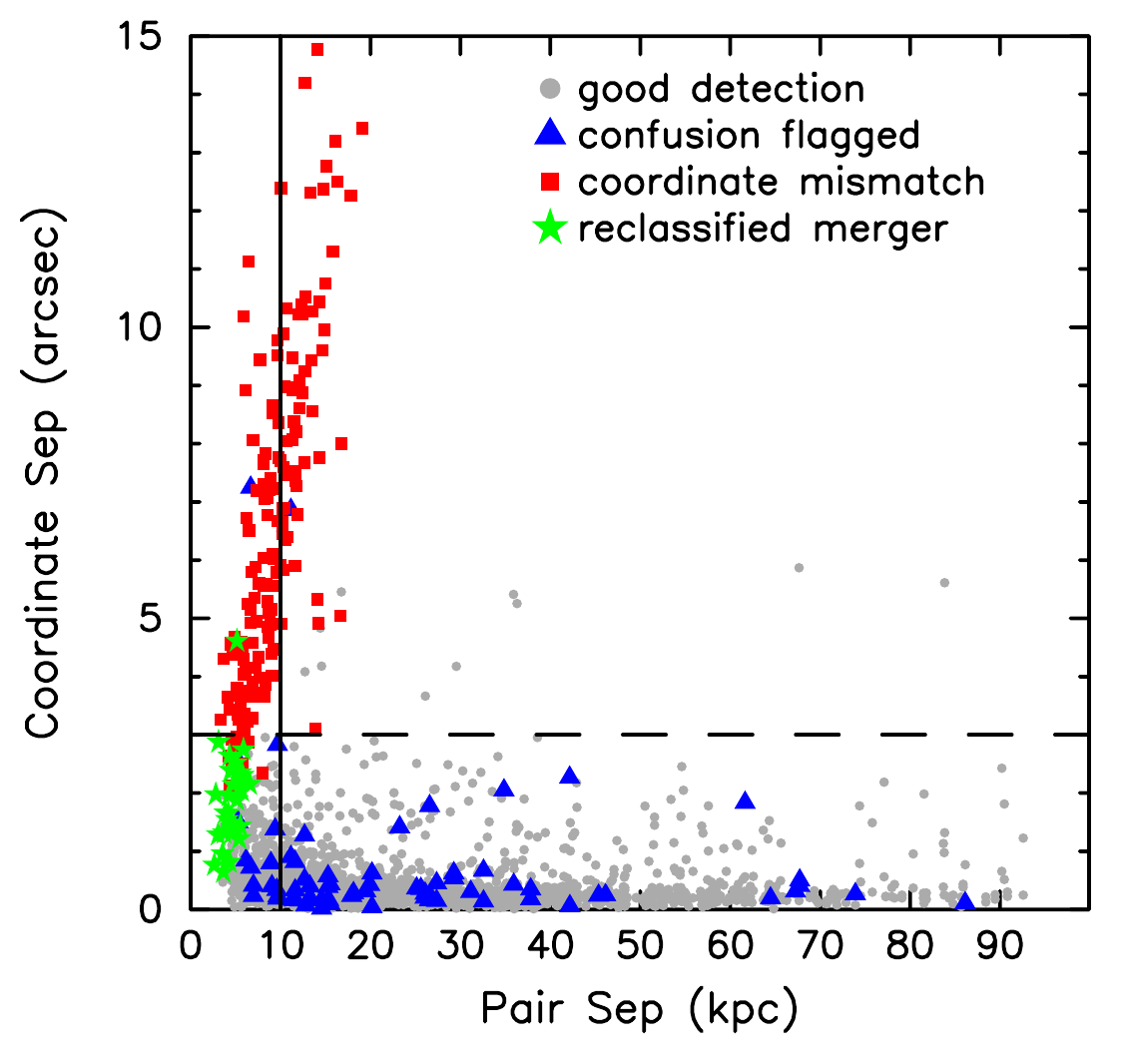}}
	\caption{Separation between the SDSS and the closest {\it WISE} detection coordinates for all pair galaxies versus separation between the primary and companion galaxies for interacting and possibly interacting galaxies. Grey circles represent galaxies with good {\it WISE} detections. Blue triangles show galaxies removed from the sample for contamination flagging. Galaxies removed from the sample for coordinate mismatches are shown as red squares (see Figure~\ref{WISE_bad_coords} and text for details). Pairs with both nuclei contained in a single 6 arcsecond {\it WISE} detection radius are reclassified as mergers and are shown in the figure as green stars (see examples in Figure~\ref{IntMergers}). Galaxies that fell below 10 kpc in pair separation (solid vertical line) or above 3 arcsecond in coordinate matching (dashed horizontal line) were checked in SDSS for correct location and flagged or removed from the sample as necessary.}
	\label{Separation_plot}
\end{figure}

\begin{figure*}
	\center{\includegraphics[scale=0.7, angle=0]{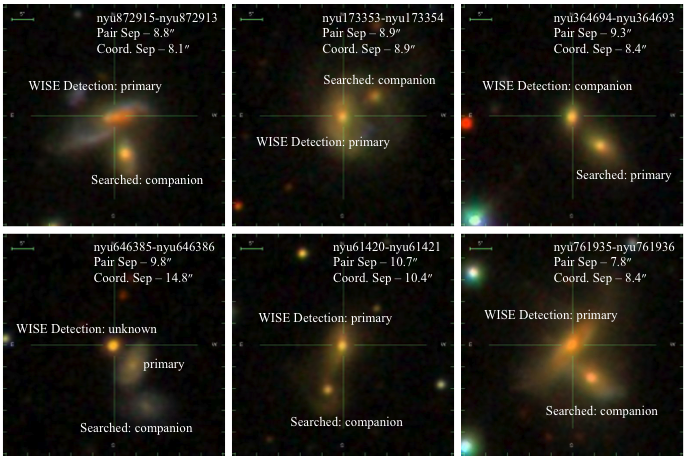}}
	\caption{Examples of interaction galaxies removed from the sample because of a coordinate mismatch between SDSS and {\it WISE}. Pair identification numbers (from the DR4 NYU-VAGC; \citealt{Blanton+05}), angular separations, and coordinate separation between the SDSS and {\it WISE} are given in the upper-right corner of each image. All images are $51 \times 51$ arcsecond cutouts of $gri$-combined color images, centered on the {\it WISE} detection, downloaded from the SDSS Image List Tool. \textquotedblleft{}Searched\textquotedblright{} objects are the galaxies we attempt to acquire {\it WISE} data for and \textquotedblleft{}{\it WISE} Detection\textquotedblright{} objects are the galaxy or object the {\it WISE} catalog matches as the closest object to the searched coordinates. In the case of nyu646385-nyu646386, we also label the primary galaxy.}
	\label{WISE_bad_coords}
\end{figure*}

\begin{figure*}
	\center{\includegraphics[scale=0.7, angle=0]{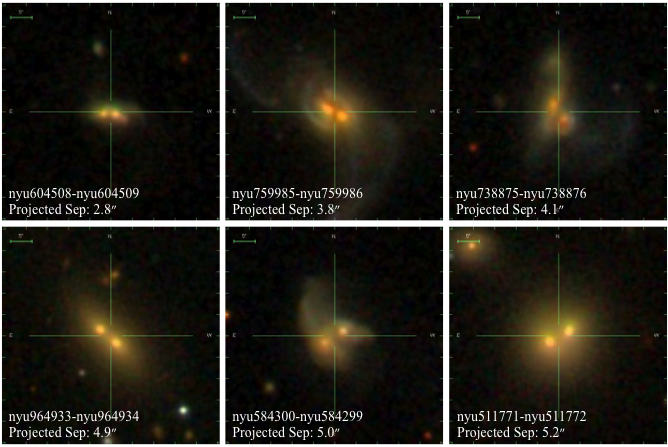}}
	\caption{Example of galaxy-galaxy interactions reclassified as mergers for having a separation smaller than the angular resolution of {\it WISE}, or with a {\it WISE} detection centered between the two galaxies. All images are $51 \times 51$ arcsecond cutouts of $gri$-combined color images, centered on the {\it WISE} detection, downloaded from the SDSS Image List Tool. Pair identification numbers (from the DR4 NYU-VAGC; \citealt{Blanton+05}) and angular separations are given in the lower-left corner of each image.}
	\label{IntMergers}
\end{figure*}

We search the {\it WISE} All-Sky Source catalog for galaxy coordinate matches using a 15 arcsecond search radius centered on the SDSS coordinates. Many SDSS-{\it WISE} studies use a 3 arcsecond coordinate match \citep{Donoso+12, Shao+13, Yan+13}; \citet{Donoso+12} find that the overall SDSS-{\it WISE} false detection rate drops to 0.05\% below the 3 arcsecond tolerance. However, we find that close projected pairs in our interaction and possible interaction subsets require closer scrutiny. Because of the {\it WISE} 6 arcsecond resolution, it is necessary to check the coordinate matches of each interacting galaxy with a pair separation below that limit. We determine galaxy matches based on closest match to the SDSS coordinates provided by the master catalog. We check all galaxies with a pair separation of less than 10 kpc, marked by a vertical solid black line in Figure~\ref{Separation_plot}, as well as all galaxies with a separation between the SDSS and {\it WISE} coordinates greater than 3 arcsecond, marked by the horizontal dashed black line. We eliminate one interacting pair and 157 individual galaxies from the sample for having no {\it WISE} data, usually owing to a nearby object, like the companion galaxy, being detected instead, as shown by red squares in Figure~\ref{Separation_plot}. Examples of galaxies removed from the sample for this reason are shown in Figure~\ref{WISE_bad_coords}. In most of these cases, the fainter companion, such as an early-type galaxy, is undetected by {\it WISE}. In addition to checking for accurate galaxy identifications, we cut twenty interacting pairs and thirty individual galaxies flagged as a spurious artifacts (cc\_flags) in the {\it WISE} catalog, as done in \citet{Stern+12}. We note that only the pair galaxy undetected in {\it WISE} is removed from the sample; this leaves 200 interacting pairs with only one {\it WISE} detection. We reclassify 32 galaxy interactions as ongoing mergers, meaning that both galaxy nuclei fall within the 6 arcsecond radius of a single {\it WISE} detection (examples shown in Figure~\ref{IntMergers}). As shown in Figure~\ref{IntMergers}, many of these systems meet the requirements of an ongoing merger, described in \S \ref{sec:21SDSSsample}, and may have been misclassified. Rather than eliminating these systems from the sample, we add them to the ongoing merger sample and note that because the {\it WISE} detection is split between the two nuclei (and doesn't necessarily contain the full nucleus of each) it may be missing some of the flux from the system. We also remove three complete galaxy pairs, ten primary detections, and nine companion detections for double-counting. Our sample contains a subset of 'triple pairs' (interaction systems containing three or more interacting galaxies). While designated as individual pair systems in our parent catalog, we limit our {\it WISE} data to only account for each galaxy once, whether it is listed in a second pair or not.

After these cuts, the final interaction sample with {\it WISE} data consists of 307 interacting pairs and 762 possibly interacting pairs. Of these, 224 interactions and 645 possible interactions have {\it WISE} detections for both galaxies, 83 interactions have only one detection, and 117 possible interactions have only one detection. This gives a final sample of 1069 interacting pairs and 130 ongoing mergers (including the reclassified interactions). The {\it WISE} All-Sky Source catalog requires a signal-to-noise ratio of five in at least one band for any source included. We find that for our sample, all galaxies have a SNR $\geq$ 4 in both the 3.4 and 4.6 micron bands, which are the critical bands for detecting obscured AGNs \citep{Jarrett+11, Assef+13, Donoso+14}. These cuts leave a final merger sample completeness of $\sim$ 98\%. Our full interaction sample (interactions and possible interactions) completeness is $\sim$ 89\%, a significantly lower completeness than expected. The closeness of pairs in an interaction affects the completeness, as discussed above and seen in Figures~\ref{Separation_plot} and ~\ref{IntMergers}.


\section{{\it WISE} Color Analysis}
\label{sec:3WISEanalysis} 

To test for a connection between dusty AGNs and merging or interacting galaxies, it is necessary to use a wavelength regime unaffected by dust attenuation in the center of these galaxies. Energy from an AGN emitted emitted at X-ray wavelengths can be absorbed by gas and dust in the galaxy and reemitted in the mid-IR \citep{Sanders+89, Pier+92}, making mid-IR AGN selection a valid candidate for isolating dusty AGNs. Analysis of the {\it WISE} {[}3.4{]}$-${[}4.6{]} vs. {[}4.6{]}$-${[}12{]} color-color space has been shown to be an effective method of distinguishing galaxy and AGN activity among extended sources and separating high-redshift QSOs from stars among point sources (e.g., \citealt{Wright+10}; \citealt{Yan+13}). In particular, this method is useful for isolating AGNs that escape optical detection due to dust obscuration in the nucleus \citep{Jarrett+11, Mateos+12, Assef+13, Gurkan+14}.


\subsection{{\it WISE} Colors for Spectroscopic Types}
\label{sec:31WISEcolors} 

The {\it WISE} infrared color-color space has been theoretically mapped by galaxy classification based on simulated spectral energy distributions \citep{Wright+10}. Here, we quantify the {\it WISE} color-color space for different galaxy spectroscopic types. We use a control sample of 42,642 `normal' (non-merging and non-interacting) galaxies from the SDSS parent sample described in \S~\ref{sec:2sample} that have a {\it WISE} detection within a 3 arcsecond coordinate match, a SNR $\geq$ 4 in both the 3.4 and 4.6 micron bands, and no spurious detection in the cc\_flag, and adequate spectral information for spectroscopic type classification (see \S \ref{sec:22SDSSspec}).

\begin{figure*}
	\center{\includegraphics[scale=0.85, angle=0]{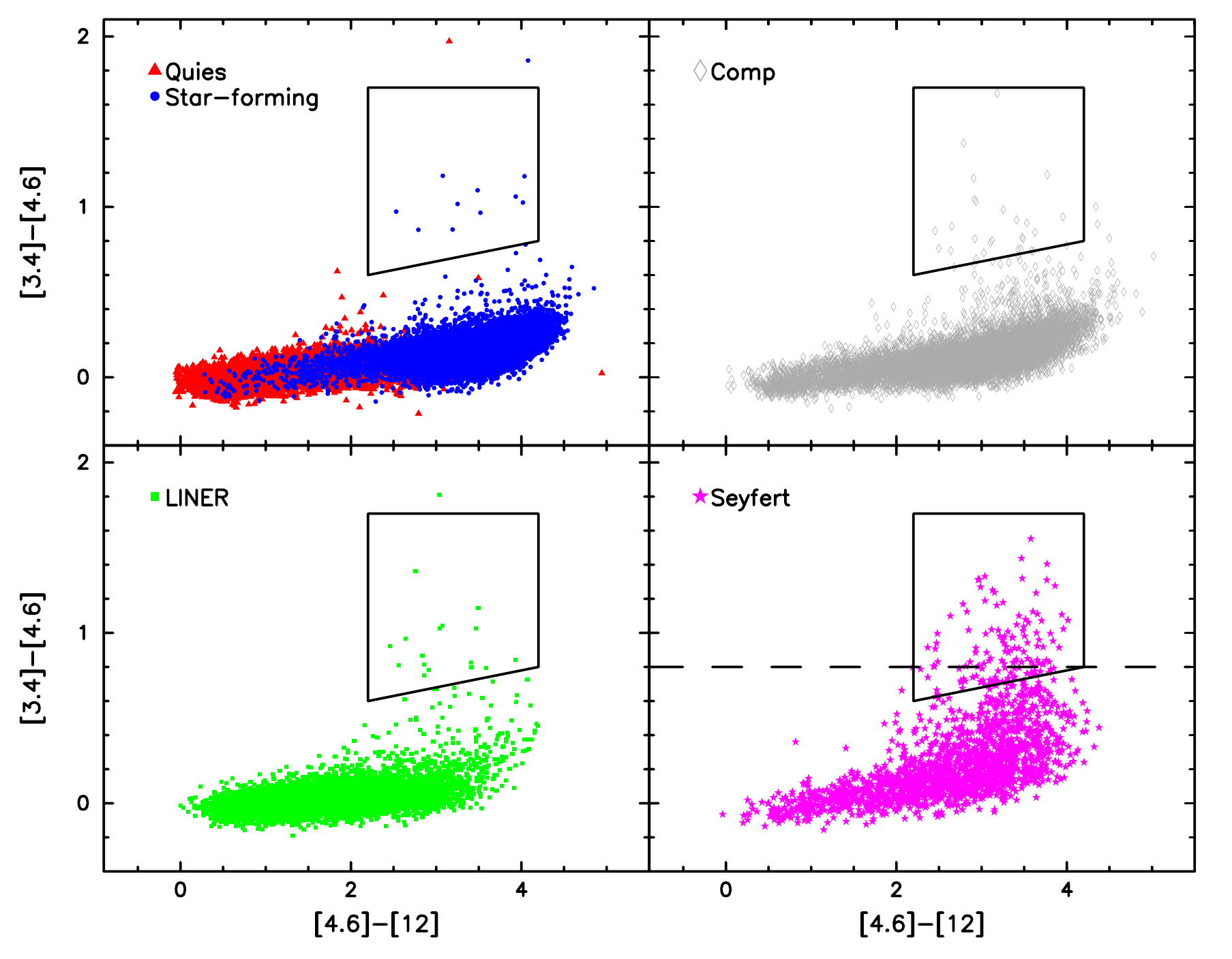}}
	\caption{{\it WISE} {[}3.4{]}$-${[}4.6{]} vs. {[}4.6{]}$-${[}12{]} color-color plot for different galaxy spectroscopic types. In the upper-left, quiescent (red triangles) and star-forming (blue circles) galaxies are plotted together to show the bimodality of the infrared color-color space. In the upper-right and lower-left, composite (grey diamonds) and LINER (green squares) galaxies are shown to populate the full color-color continuum. In the bottom-right, Seyfert galaxies (pink stars) rise above the continuum into hot dusty color-color space. The box is defined as \textquotedblleft{}{\it WISE} AGN\textquotedblright{} by \citet{Jarrett+11} and is populated only by heavily obscured AGN. The dashed line in the Seyfert panel at {[}3.4{]}$-${[}4.6{]} $>$ 0.8 is used by \citet{Yan+13} to distinguish Type-2 AGN.}
	\label{control_2x2}
\end{figure*}

As shown in Figure~\ref{control_2x2}, when the {\it WISE} {[}3.4{]}$-${[}4.6{]} vs. {[}4.6{]}$-${[}12{]} colors are plotted for different spectroscopic types, the majority of galaxies follow a roughly horizontal color continuum in the {[}4.6{]}$-${[}12{]} color from red and dead to star-forming, with little change in the {[}3.4{]}$-${[}4.6{]} color \citep{Wright+10}. Additionally, at redder {[}4.6{]}$-${[}12{]} colors galaxies with increasing dust obscuration extend vertically into very red {[}3.4{]}$-${[}4.6{]} colors \citep{Wright+10, Jarrett+11}. Conceptually, the {[}4.6{]}$-${[}12{]} color is a measure of dust heating by SF. We can expect quiescent, non-star-forming galaxies, to have low {[}4.6{]}$-${[}12{]} colors because they lack sufficient SF to heat the dust. The {[}3.4{]}$-${[}4.6{]} color is representative of dust heating by an AGN \citep{Jarrett+11}. Thus galaxies with a {[}3.4{]}$-${[}4.6{]} color close to zero have no AGN component, while galaxies rising above the continuum are more likely to host a dust-enshrouded AGN. This is shown in the Seyfert population of Figure~\ref{control_2x2}.

In Figure~\ref{control_2x2}, we show the {\it WISE} color-color space for the five spectroscopic types: quiescent and star-forming, composite, LINER, and Seyfert. As shown in the upper-left plot, the bluer color-color space is occupied mostly by quiescent galaxies (shown as red triangles) and star-forming galaxies (shown as blue circles) are on the opposite, redder side of {[}4.6{]}$-${[}12{]} color space. The majority of quiescent and star-forming galaxies are separated by {[}4.6{]}$-${[}12{]} $\sim$ 2.25, as can be seen by their $80{}^{th}$ percentile ranges in Table~\ref{WISE_living_spaces}. Composite galaxies (grey diamonds), which are presumed to be a combination of SF and AGN, span nearly the full range of the continuum. LINER galaxies (green squares) span the continuum, though few are found redder than {[}4.6{]}$-${[}12{]} = 3.5. Seyfert emission-line galaxies (shown as pink stars) span much of the continuum, yet a significant number of Seyfert galaxies fall above the general continuum at redder (hotter) {[}3.4{]}$-${[}4.6{]} colors. Above {[}3.4{]}$-${[}4.6{]} $>$ 0.5, the population is dominated by Seyfert galaxies. This population of AGNs at redder {[}3.4{]}$-${[}4.6{]} colors agrees well with the qualitative predictions of \citet{Wright+10}.  Seyfert galaxies falling above the continuum are likely to be dust-obscured, Type-2 Seyferts.

\begin{table*}
	\caption{{\it WISE} Color-Color locations of spectroscopic types based on the control sample of 42,642 normal high-mass SDSS galaxies described in \S~\ref{sec:31WISEcolors}. Columns (1) and (2): the emission type and total number of that emission type in the control sample. Columns (3) and (4): the {[}3.4{]}$-${[}4.6{]} color median value and $80{}^{th}$ percentile range. Columns (5) and (6): the {[}4.6{]}$-${[}12{]} color median value and $80{}^{th}$ percentile range.}
	\label{WISE_living_spaces}
	\begin{tabular}{lccccc}
		\hline
		Type & N & \multicolumn{2}{c}{{[}3.4{]}$-${[}4.6{]}} & \multicolumn{2}{c}{{[}4.6{]}$-${[}12{]}}\\
		\cline{3-4} \cline{5-6} \\
		& & Median & 80\%-tile range & Median & 08\%-tile range\\
		(1) & (2) & (3) & (4) & (5) & (6)\\
		\hline
		\hline
		Quies & 9413 & 0.02 & -0.06 -- 0.10 & 1.03 & 0.48 -- 2.08\\
		Composite & 9704 & 0.11 & -0.01 -- 0.26 & 2.90 & 1.51 -- 3.73\\
		Star-forming & 14323 & 0.16 & 0.06 -- 0.26 & 3.50 & 2.73 -- 4.02\\
		LINER & 7354 & 0.03 & -0.06 -- 0.12 & 1.72 & 0.82 -- 2.76\\
		Seyfert & 1848 & 0.18 & 0.00 -- 0.61 & 2.80 & 1.35 -- 3.62\\
		\hline
	\end{tabular}
\end{table*}


\subsubsection{{\it WISE} AGN Selection}
\label{sec:311controlWISEAGN} 

Several studies have shown that {\it WISE} color-color analysis is an effective method of isolating dusty AGNs from the general galaxy population \citep{Jarrett+11, Yan+13, Satyapal+14}. In the Seyfert panel of Figure~\ref{control_2x2}, we highlight two AGN selection methods. First, the dashed line represents {[}3.4{]}$-${[}4.6{]} $>$ 0.8, a cut used by \citet{Stern+12}, \citet{Assef+13}, and \citet{Yan+13} to distinguish Type-2 AGNs from the general population. This straight cut also requires a {\it WISE} AGN to have {[}4.6{]} $>$15.2 to eliminate possible contamination by high-redshift elliptical and Sbc galaxies \citep{Yan+13}. We find that 57 -- 74\% of galaxies falling above this line are classified as Seyferts. Throughout this study, the errors on all numbers, fractions, and percentages are 95\% binomial confidence intervals, unless otherwise stated. Second, the outlined box is defined using the criteria of \citet[][hereafter J11]{Jarrett+11} to identify {\it WISE} AGNs:
\begin{equation}
	2.2 < [4.6] - [12] < 4.2
\end{equation}
\vspace{-0.5cm}
\begin{equation}
	0.1 \times ([4.6] - [12]) + 0.38 < [3.4] - [4.6] < 1.7 .
\end{equation}
\noindent As discussed in J11, this box is chosen to enclose Seyfert galaxies and QSOs, while still lying above the star-forming galaxies at the blue end of the spectrum; recent SF can also cause a dust reddening effect. We find that 66 -- 79\% of galaxies falling in this box are classified as Seyfert galaxies. In Table~\ref{control_WAGN_table}, we show the results for each of these selection methods, broken down by spectroscopic types. We find similar results with both methods. We also include the {\it WISE} AGN selection method by \citet{Satyapal+14}; a straight color cut of {[}3.4{]}$-${[}4.6{]} $>$ 0.5. We find that 57 -- 66\% of galaxies selected using this method are emission-selected Seyfert galaxies, making it a less robust, but still competitive, selection method.

\begin{table*}
	\caption{Dusty AGN fractions from analysis of {[}3.4{]}$-${[}4.6{]} vs. {[}4.6{]}$-${[}12{]} color-color plotting for a control sample of 42,642 normal high-mass SDSS galaxies described in \S~\ref{sec:31WISEcolors} and split by spectroscopic type. Column (1): the galaxy emission type. Column (2): the percent of that emission type contained in the {\it WISE} AGN box defined by J11. Column (3): the percent of that emission type contained in the Extended {\it WISE} AGN box defined in \S \ref{sec:312extendedWISEAGN}. Column (4): the percent of that emission type above the {[}3.4{]}$-${[}4.6{]} $>$ 0.8 cut used by \citet{Yan+13}. Column (5): the percent of that emission type above the {[}3.4{]}$-${[}4.6{]} $>$ 0.5 cut used by \citet{Satyapal+14}.}
	 \label{control_WAGN_table}
	\begin{tabular}{lcccc}
		\hline
		Type & \multicolumn{4}{c}{Dusty AGN Criteria}\\
		\cline{1-5} \\
		& {\it WISE} AGN & Ext. {\it WISE} AGN & {[}3.4{]}$-${[}4.6{]}$>$ 0.8 & {[}3.4{]}$-${[}4.6{]}$>$ 0.5\\
		& Jarrett et al. (2011) & This work & Yan et al. (2013) & Satyapal et al. (2014)\\
		(1) & (2) & (3) & (4) & (5)\\
		\hline
		\hline
		Quies & $0.00{}_{-0.00}^{+0.04}$\% & $0.02{}_{-0.01}^{+0.06}$\% & $0.01{}_{-0.01}^{+0.05}$\% & $0.03{}_{-0.02}^{+0.06}$\%\\
		Composite & $0.26{}_{-0.09}^{+0.12}$\% & $1.30{}_{-0.21}^{+0.24}$\% & $0.21{}_{-0.08}^{+0.11}$\% & $0.97{}_{-0.18}^{+0.21}$\%\\
		Star-forming & $0.07{}_{-0.03}^{+0.06}$\% & $0.30{}_{-0.07}^{+0.10}$\% & $0.08{}_{-0.04}^{+0.06}$\% & $0.24{}_{-0.06}^{+0.10}$\%\\
		LINER & $0.22{}_{-0.09}^{+0.13}$\% & $0.71{}_{-0.17}^{+0.22}$\% & $0.19{}_{-0.08}^{+0.13}$\% & $0.48{}_{-0.14}^{+0.18}$\%\\
		Seyfert & $7.41{}_{-1.10}^{+1.29}$\% & $21.05{}_{-1.80}^{+1.92}$\% & $4.87{}_{-0.89}^{+1.08}$\% & $14.72{}_{-1.54}^{+1.69}$\%\\
		\hline
	\end{tabular}
\end{table*}

For this analysis, we choose the selection method defined by J11. As shown in Figure~\ref{control_2x2} and discussed above, this selection is dominated by emission-line Seyfert galaxies, confirming the use of the J11 box as a reliable method for isolating dusty AGNs. Indeed, the Seyfert population is the only spectroscopic type to have significant representation in the {\it WISE} AGN box. The majority of Seyferts fall outside this cut, and are presumably unobscured, Type 1 Seyferts. In addition, Composite galaxies, which are defined to be a combination of AGN and SF, make up around 13\% of the J11 box. The remainder of the sources consist of LINER and HII galaxies (non-AGNs), which each make up around 14\% of the population in the J11 box. Because their {\it WISE} colors show clear dust signatures, we suggest that their optical classification may be unreliable and they may be mislabeled AGNs. In addition to providing the color-color spaces of different emission types in {\it WISE} color space and testing the validity of multiple {\it WISE} AGN cuts, we also use our control population to examine the likelihood of a typical non-merging, non-interacting galaxy being classified as a {\it WISE} AGN. As shown in Table~\ref{sample_WAGN_table}, the overall likelihood of a normal galaxy being a {\it WISE} AGN is 0.38 -- 0.51\%. This result agrees well with the control analysis done by \citet{Satyapal+14}.


\subsubsection{Extended {\it WISE} AGN Selection}
\label{sec:312extendedWISEAGN} 

\begin{figure}
	\center{\includegraphics[scale=0.8, angle=0]{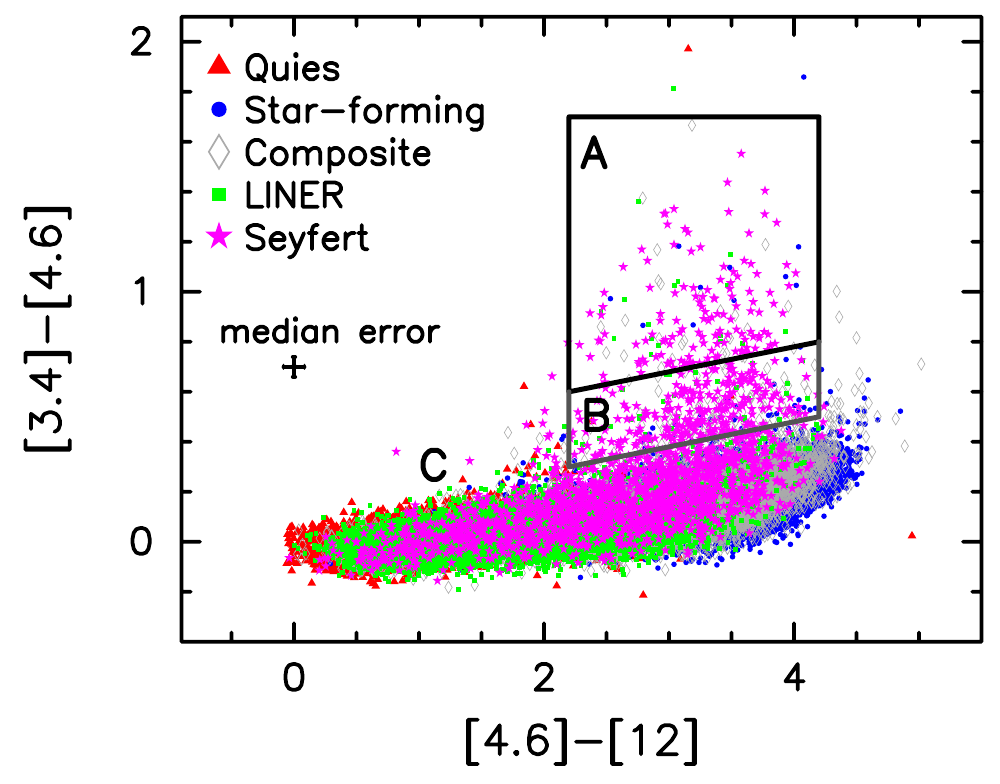}}
	\caption{{\it WISE} {[}3.4{]}$-${[}4.6{]} vs. {[}4.6{]}$-${[}12{]} plot for 42,642 control galaxies. Emission types are denoted as in Figure~\ref{control_2x2}. The black box is defined as \textquotedblleft{}{\it WISE} AGN\textquotedblright{} by J11. The lower box represents the boundary of the Extended {\it WISE} AGN box (see text for details). The median color errors for the full control sample are given under the key. Regions for analysis of {\it WISE} AGN are noted as follows: (A) {\it WISE} AGN selected by J11, (B) galaxies falling in the extended region, and (C) all galaxies not contained in Regions A or B.}
	\label{control_WISE_colors}
\end{figure}

We also use the control population to define a lower, more liberal cutoff for the {\it WISE} AGN box used by J11. This extension is used to include a population of galaxies that are presumably dustier than the normal continuum but still lie below the J11 box, similar to the lower cut of {[}3.4{]}$-${[}4.6{]} $>$ 0.5 used by \citet{Satyapal+14} and the Spitzer {[}3.6{]}$-${[}4.5{]} $>$ 0.5 cut used by \citet{Ashby+09}. In Figure~\ref{control_WISE_colors}, we show the {\it WISE} colors for all spectroscopic types from the control population. While the majority of Seyfert galaxies do follow the common trend, a noticeable fraction of Seyfert galaxies rise above the continuum, but are not red enough at {[}3.4{]}$-${[}4.6{]} to fall in Region A, as a {\it WISE} AGN. We define this new region, Region B, with Equation (2) from \S \ref{sec:311controlWISEAGN}, and Equation (3) three-tenths of a magnitude below. We find that 55 -- 64\% of galaxies that fall in Region B are emission-line Seyfert galaxies. The remaining color space, containing the main locus of {\it WISE} galaxies is labeled Region C. We combine Regions A and B into the Extended {\it WISE} AGN box. Of the galaxies contained in the Extended {\it WISE} AGN box, 58 -- 68\% are spectroscopically classified as Seyferts. The remaining population in the Extended {\it WISE} AGN box is made up mostly of Composite galaxies ($\sim$ 21\%) and LINERs ($\sim$ 8\%), both of which are considered possible or partial AGN systems. This high number of known AGNs and possible AGNs justifies this Extended {\it WISE} AGN cut as a plausible means of isolating dusty AGNs through {\it WISE} color-color analysis. In addition, we note that $\sim$ 99\% of galaxies fall outside the Extended {\it WISE} AGN box. We do note that some galaxies are red enough in {[}3.4{]}$-${[}4.6{]} to be associated with Extended {\it WISE} AGNs, but lie outside the {[}4.6{]}$-${[}12{]} window. As explained in J11, this small {[}4.6{]}$-${[}12{]} window is chosen to leave out extreme starburst galaxies and dusty spiral galaxies. With this Extended {\it WISE} AGN cut, the likelihood of a normal galaxy being a dusty AGN is 1.33 -- 1.55\% (see Table~\ref{sample_WAGN_table}). Results for this Extended cut for different spectroscopic types are summarized in Table~\ref{control_WAGN_table}.

\subsection{Incidence of Dusty AGNs in Mergers \& Interactions}
\label{sec:32mergersWISEAGN} 

\begin{figure*}
	\center{\includegraphics[scale=0.95, angle=0]{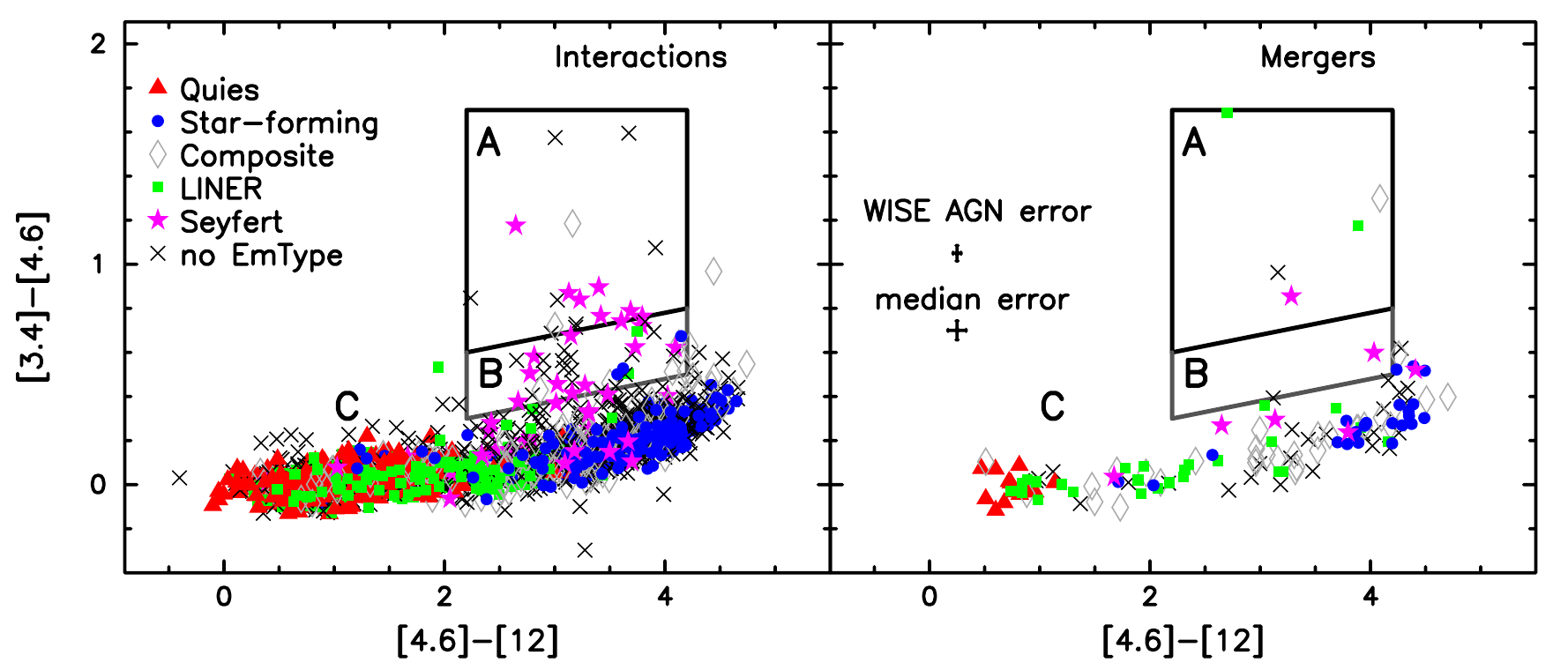}}
	\caption{{\it WISE} {[}3.4{]}$-${[}4.6{]} vs. {[}4.6{]}$-${[}12{]} plot of interacting (left) and merging (right) galaxies. Spectroscopic types are denoted as in Figure~\ref{control_2x2} and shown in the left panel. Black x\textquoteright{}s represent mergers and interactions with no spectroscopic type. The black box is defined as \textquotedblleft{}{\it WISE} AGN\textquotedblright{} by J11. The lower box represents the boundary of the Extended {\it WISE} AGN box. The median color errors for the {\it WISE} AGN population and the full merging and interacting sample are given in the right panel.}
	\label{sample_WISE_colors}
\end{figure*}

To compare the incidence of {\it WISE} AGNs among mergers and interactions to the frequency among non-merging, non-interacting (control sample) galaxies, we analyze the {\it WISE} {[}3.4{]}$-${[}4.6{]} and {[}4.6{]}$-${[}12{]} colors of our sample of visually identified mergers and interactions in Figure~\ref{sample_WISE_colors}. We include identifications without SDSS spectroscopic classification, shown as black x's. The majority of mergers and interactions follow the same general continuum of {\it WISE} colors as the control sample in Figures~\ref{control_2x2} and ~\ref{control_WISE_colors}, with galaxies of each emission type populating similar areas in the continuum as the control population. Additionally, like in the control sample, a small fraction of galaxies fall above the continuum and a handful are found in the J11 {\it WISE} AGN box. Similar to the control sample, a larger fraction of Seyfert galaxies fall above the continuum than any other spectroscopic type. Of the Seyfert sample of 61 galaxies, 8 -- 26\% are defined as {\it WISE} AGNs from this analysis. This analysis also identifies 14 new {\it WISE} AGNs, 5 of which were previously classified as LINERs (2) or Composite galaxies (3), and 9 that previously had no spectroscopic classification.

Statistics for the analysis of the merging and interacting sample are shown in Table~\ref{sample_WAGN_table}. Of the sample, $\sim$ 2--9\% of the ongoing mergers are classified as {\it WISE} AGNs. This is twice as high as the interacting population ($\sim$ 1--5\%) and three times higher than the possibly interacting population ($\sim$ 1--2\%). This result agrees well with \citet{Ellison+13b}, who found that AGN occurrence in post-merger galaxies is 3.75 times higher than in close pairs. When compared to our control sample, the ongoing merger population is 5 -- 17 times more likely to be a dusty AGN than a normal galaxy (interaction sample 2.6 -- 4.9 times more likely). When our Extended cut is used, the merger sample is 2 -- 7 times more likely to be a dusty AGN than a normal galaxy (interaction sample 2.9 -- 4.2 times more likely).

\begin{table*}
	\caption{Dusty AGN fractions from analysis of {[}3.4{]}$-${[}4.6{]} vs. {[}4.6{]}$-${[}12{]} colors for the merging and interacting sample with stellar masses ${\rm M}_{\rm star} > 2 \times 10^{10}~{\rm M}_{\odot}$ and redshifts $0.01 < z \leq 0.08$. Columns (1) and (2): the galaxy morphology type and number of sample galaxies of that type. Column (3): the percent of that morphology type contained in the {\it WISE} AGN box defined by J11. Column (4): the percent of that morphology type contained in the Extended {\it WISE} AGN box defined in \S \ref{sec:312extendedWISEAGN}. Column (5): the percent of that morphology type above the {[}3.4{]}$-${[}4.6{]} $>$ 0.8 cut used by \citet{Yan+13}. Column (6): the percent of that morphology type above the {[}3.4{]}$-${[}4.6{]} $>$ 0.5 cut used by \citet{Satyapal+14}.}
	\label{sample_WAGN_table}
	\begin{tabular}{lccccc}
		\hline
		Type & N & \multicolumn{4}{c}{Dusty AGN Criteria}\\
		\cline{2-6} \\
		& & {\it WISE} AGN & Ext. {\it WISE} AGN & {[}3.4{]}$-${[}4.6{]} $>$ 0.8 & {[}3.4{]}$-${[}4.6{]} $>$ 0.5\\
		& & Jarrett et al. (2011) & This work & Yan et al. (2013) & Satyapal et al. (2014)\\
		(1) & (2) & (3) & (4) & (5) & (6)\\
		\hline 
		\hline
		mergers & 130 & $3.9_{-2.2}^{+4.8}$\% & $5.4_{-2.8}^{+5.3}$\% & $3.9_{-2.2}^{+4.8}$\% & $9.2_{-3.9}^{+6.2}$\%\\
		int. pair & 307 & $2.3_{-1.2}^{+2.3}$\% & $6.8_{-2.3}^{+3.4}$\% & $2.0_{-1.1}^{+2.2}$\% & $6.5_{-2.2}^{+3.3}$\%\\
		poss. int. pair & 762 & $1.3_{-0.6}^{+1.1}$\% & $4.3_{-1.2}^{+1.7}$\% & $0.7_{-0.4}^{+0.8}$\% & $3.9_{-1.2}^{+1.7}$\%\\
		control & 42,642 & $0.44{}_{-0.06}^{+0.07}$\% & $1.43{}_{-0.10}^{+0.12}$\% & $0.32{}_{-0.05}^{+0.06}$\% & $1.03{}_{-0.09}^{+0.10}$\%\\
		\hline
	\end{tabular}
\end{table*}


\section{Nature of {\it WISE} AGNs in Mergers \& Interactions}
\label{sec:4nature} 

In the previous section, we demonstrated a strong statistical connection between dusty AGNs and galaxy merging, rather than a connection between merging and all AGNs. Yet, the physical nature of {\it WISE} AGNs has been sparsely studied, particularly in relation to mergers and interactions. To test whether merging and interacting {\it WISE} AGNs are unique, we compare key galaxy and pair property distributions for this special subset of galaxies to those of merging and interacting systems that are not {\it WISE} AGNs. For these comparisons, we use three subsamples of the merging and interacting population throughout this section and illustrated by regions shown in Figure~\ref{sample_WISE_colors} as follows: (i) {\it WISE} AGNs defined by the J11 criteria (Region A), (ii) our more inclusive Extended {\it WISE} AGN selection that includes additional presumably dusty systems (Regions A+B), and (iii) a control sample consisting of all merging and interacting galaxies not classified as Extended {\it WISE} AGNs, (Region C). 

The vast majority of galaxies in Region C lack any sign of warm dust heated by an AGN according to our {\it WISE} color analysis but do include a subset of emission-line Seyfert AGNs. These Seyferts are unobscured, and therefore are not an important population for our dusty AGN-merger connection. We test for differences using galaxy and pair properties (total stellar mass, stellar mass ratio, and pair separation), environmental properties (dark matter halo  mass and central or satellite rank), and an SDSS color-color based distinction between star-forming and quiescent galaxies. Additionally, for the subset of merging and interacting galaxies that are {\it WISE} and/or emission-line defined AGNs, we compare a measure of their accretion power. We note that we treat a pair of interacting galaxies as a single entity throughout this section.


\subsection{Galaxy \& Pair Properties}
\label{sec:41props} 

The likelihood of finding an AGN in a merging or interacting system has been found to depend on a variety of properties, including stellar mass ratios \citep{Ellison+11, Capelo+15}, pair separation \citep{Ellison+11, Satyapal+14}, and galaxy stellar mass \citep{Sabater+13}. For our sample of high-mass, low-redshift merging and interacting galaxies, we analyze the occurrence of {\it WISE} AGNs as a function of total stellar mass, stellar mass ratio, and pair separation.

\subsubsection{Total Stellar Mass}

\begin{figure*}
	\center{\includegraphics[scale=0.95, angle=0]{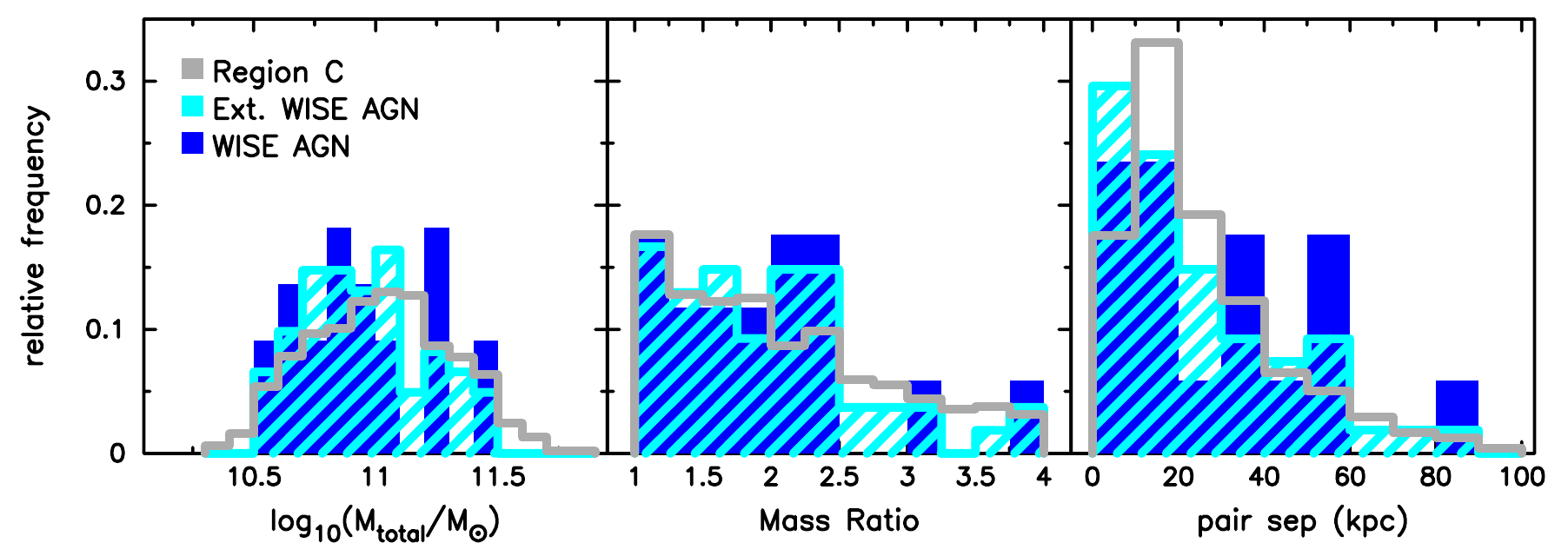}}
	\caption{Relative frequency distributions of interacting and merging galaxies for total stellar mass (for interactions, ${\rm M}_{\rm star,1} + {\rm M}_{\rm star,2}$) in units of $\log_{10}\left(M/M_{\odot}\right)$ (left), stellar mass ratio (middle), and projected pair separation in kpc (right). {\it WISE} AGNs are given in dark blue and Extended {\it WISE} AGNs are given in light blue hatching. The subsample of merging and interacting galaxies not classified as Extended {\it WISE} AGNs is given as a grey outline. For stellar mass ratio and pair separation, only systems classified as interacting pairs are considered.}
	\label{Props_mass_mratio_sep}
\end{figure*}

A correlation between AGNs and their host galaxy stellar mass has long been known. \citet{Kauffmann+03} found that $>80$\% of emission-line galaxies with $M_{star}>10^{11}$ $M_{\odot}$ meet the AGN classification in the BPT diagram; most of these are LINERs. The occurrence of radio-loud AGNs rises with increasing host galaxy stellar mass \citep{Best+05}, while \citet{Sabater+13} found this to be true for AGNs of any type. For our sample of mergers and interactions, the relative stellar mass distributions (combined stellar mass of primary and companion, ${\rm M}_{\rm star,1} + {\rm M}_{\rm star,2}$ in the case of interactions) are given in the left panel of Figure~\ref{Props_mass_mratio_sep} for the three subpopulations described above: {\it WISE} AGNs (Region A) in dark blue, Extended {\it WISE} AGNs (Regions A+B) in hatched light blue, and Region C as a grey outline. Binned in this way, the dusty AGNs appear to have a higher occurrence at ${\rm M}_{\rm total}\lesssim 10^{11}\,{\rm M}_\odot$ compared to Region C. To quantify this comparison, we make use of the two-sample Kolmogorov-Smirnov (K-S) test \citep{Press+92}, which returns a probability ($p$-value) of two separate distributions being statistically consistent with being drawn from the same parent distribution. Thus, for each property we compute separate $p$-values for the {\it WISE} and Extended {\it WISE} AGNs distributions each against the Region C control sample. A small $p$-value in effect ``proves'' to a $1-p$ significance level that two distributions are different (i.e. rejects the null hypothesis that they are the same). For example, $p=0.05$ means that two distributions are statistically different at the 95\% level. We tabulate the percent difference significance levels from the K-S tests in Table~\ref{KS_test_values}. The K-S tests show that the stellar mass distributions of the {\it WISE} AGN and Extended {\it WISE} AGN subsets are not statistically different from other mergers and interactions. Therefore, whether or not a merging or interacting galaxy is a dusty AGN does not appear to depend on stellar mass.

\begin{table*}
	\caption{Merging and Interacting Dusty AGNs versus Non-Dusty AGNs Property Comparison. K-S test p-values for the merging and interacting subsamples of {\it WISE} AGNs and Extended {\it WISE} AGNs. Column (1): the galaxy or pair property distribution tested. Columns (2) and (3): the number of galaxies in the property distribution for Region C and the median value of the Region C distribution for the property in units of  $\log_{10}\left(M/M_{\odot}\right)$ for stellar and halo mass, and kpc for pair separation. Column (4): the number of {\it WISE} AGN galaxies in the property distribution. Column (5): the median value of the {\it WISE} AGN distribution for the property in units of  $\log_{10}\left(M/M_{\odot}\right)$ for stellar and halo mass, and kpc for pair separation. Column (6): two-sided K-S probability (i.e. $1-p$), expressed as a percentage, that the {\it WISE} AGN and Region C distributions of a given property were not drawn from the same parent sample. Column (7): the number of Extended {\it WISE} AGN galaxies in the property distribution. Column (8): the median value of the Extended {\it WISE} AGN distribution for the property in units of  $\log_{10}\left(M/M_{\odot}\right)$ for stellar and halo mass, and kpc for pair separation. Column (9): two-sided K-S probability, expressed as a percentage, that the Extended {\it WISE} AGN and Region C distributions of a given property were not drawn from the same parent sample.}
	\label{KS_test_values}
	\begin{tabular}{c|cc|ccc|ccc}
		\hline
		Galaxy Property & \multicolumn{2}{c}{Region C} & \multicolumn{3}{c}{{\it WISE} AGN} & \multicolumn{3}{c}{Ext. {\it WISE} AGN}\\
		\cline{2-3} \cline{4-6} \cline{7-9} \\
		& Number & Median & Number & Median & (\%)diff & Number & Median & (\%)diff\\
		(1) & (2) & (3) & (4) & (5) & (6) & (7) & (8) & (9)\\
		\hline 
		\hline
		Stellar Mass & 1138 & 11.02 & 22 & 10.96 & 44.28 & 61 & 10.96 & 91.35\\
		Mass Ratio & 1015 & 1.90 & 17 & 1.90 & 37.20 & 54 & 1.88 & 65.15\\
		Pair Sep. & 1015 & 19.82 & 17 & 25.62 & 86.36 & 54 & 19.82 & 95.55\\
		Halo Mass & 1137 & 12.79 & 22 & 12.42 & 85.85 & 60 & 12.57 & 98.34\\
		Halo Mass (CEN) & 888 & 12.63 & 18 & 12.39 & 90.69 & 48 & 12.42 & 98.03\\
		\hline
	\end{tabular}
\end{table*}

\subsubsection{Stellar Mass Ratio}

Simulations show that the stellar mass ratio between two interacting galaxies has a strong effect on black hole activity \citep{Capelo+15} and SF \citep{Cox+08}. Using hydrodynamic simulations, \citet{Capelo+15} found that a lower stellar mass ratio (closer to equal stellar mass) can produce a higher black hole accretion rate. Similarly, \citet{Cox+08} found in simulations that both the strength of a merger-driven starburst and the duration of the starburst can vary based on the stellar mass ratio of the merger progenitors. In the middle panel of Figure~\ref{Props_mass_mratio_sep}, we show the relative stellar mass ratio (${\rm M}_{\rm star,1} / {\rm M}_{\rm star,2}$) distributions for different interaction subsamples. For each dusty AGN subset, we find higher frequencies at lower mass ratios in agreement with theoretical predictions and with a study of SDSS pairs \citep{Ellison+11}. Yet, we also find that Region C interactions follow a similar trend. This may be the result of a selection effect from the visual identification of interactions based primarily on the presences of tidal features, which are typically stronger in lower mass ratio systems. The K-S test results show that stellar mass ratio distributions of the interaction subsets are statically the same; thus, stellar mass ratio does not appear to play a significant role in the occurrence of dusty AGNs among visually interacting galaxies.

\subsubsection{Pair Separation}
\label{sec:pairsep}

Observational studies have shown that the AGN occurrence in galaxy pairs depends on the projected separation of the two galaxies. For example, \citet{Ellison+11} found that the AGN fraction in galaxy pairs with $d_{\rm sep} < 10$\,kpc is increased by a factor of $\sim$ 2.5 over non-pair galaxies, while \citet{Satyapal+14} found that the infrared AGN fraction in close pairs increases with decreasing pair separation. In the right-hand panel of Figure~\ref{Props_mass_mratio_sep}, we show the relative pair separation distributions for different interaction subsamples. For each subset, including Region C, the frequency peaks at pair separations $d_{\rm sep} < 20$\,kpc, with the Extended {\it WISE} AGN subpopulation peaking at $<10$\,kpc. The K-S test result for the Extended population shows a difference from the Region C control at a 2-sigma significance, suggesting that pair separation may play an important role in the occurrence of dusty AGNs in our sample of interacting galaxies. Our results imply that dusty AGNs prefer hosts in pairs with smaller separations compared to other interactions, in agreement with \citet{ Satyapal+14} and in accord with previous studies of AGNs in pairs \citep{Ellison+11, Ellison+13b}. Physically, smaller pair separations should drive stronger torques that will bring more gas to the nucleus of each galaxy and, ultimately, the post-merger remnant \citep{Mihos+96}. Given that the visual identification of interactions is more likely to pick pairs with small projected separations owing to the increased likelihood of stronger tidal features at pericentric passage, the 2-sigma difference from Region C interactions is noteworthy. As the {\it WISE} AGNs (Region A) are a subset of the Extended subpopulation (Region A$+$B), we speculate that a larger sample of {\it WISE} AGNs may also show a significant difference from Region C interactions. We test this idea by performing a number of K-S tests on pairs of distinct, but overlapping Gaussian distributions with sample size $N=22$ and find that only half of the time does this test return a difference more significant than 2-sigma.


\subsection{Environmental Properties}
\label{sec:42enviro} 

The environment of a galaxy can impact its properties. For example, depending on the mass of a galaxy's dark matter halo, virial shock heating can heat the halo gas \citep{Keres+05, Dekel+06}, preventing it from collapsing to the center of the galaxy and accreting onto the black hole. We use the SDSS DR4 galaxy group catalog described in detail in \citet{Yang+07} to examine whether or not the occurrence of dusty AGNs in galaxy mergers and interactions depends on group halo mass or central/satellite designation. 

The group catalog consists of environmental parameters for over 80,000 groups, complete for bright galaxies (${r\leq18}$ mag) in the NYU-VAGC at redshifts 0.01 $\le$ z $\le$ 0.20. For this study, we use the following properties: (i) the galaxy rank within its halo based on highest stellar mass, with most massive ranked as central (CEN) and all less massive galaxies ranked as satellite (SAT); and (ii) an estimate of the dark matter halo mass ${\rm M}_{\rm halo}$ of the host group, which is computed by ranking groups in terms of characteristic stellar mass and applying halo occupation statistics in the assumed $\Lambda$CDM cosmological model. We note that using either stellar masses or luminosities to define the group properties does not affect our results. For galaxy groups with $z \leq 0.08$, the \citet{Yang+07} catalog is complete for halos with ${\rm M}_{\rm halo} > 10 ^ {11.78} ~ {\rm M}_{\odot}$. 

\subsubsection{Halo Mass}

\begin{figure*}
	\center{\includegraphics[scale=0.95, angle=0]{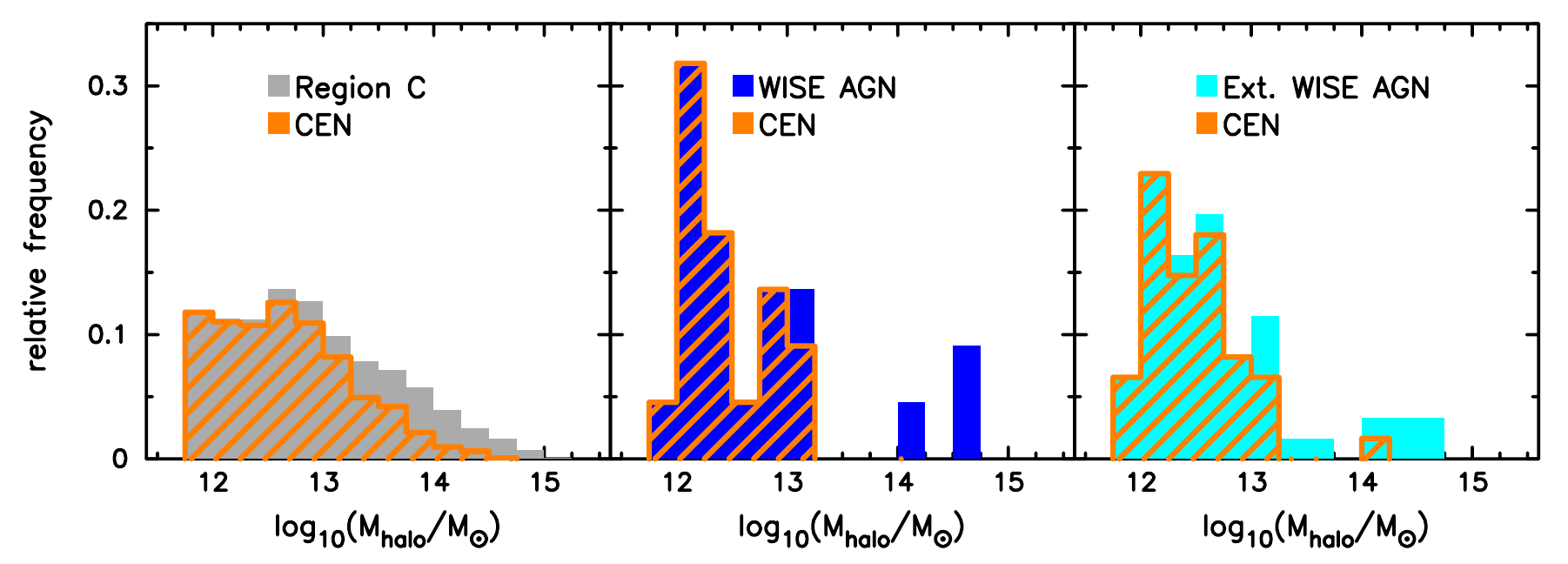}}
	\caption{Relative frequency distribution of halo mass for the following subsamples of merging and interacting galaxies: Region C (left, grey), {\it WISE} AGNs (middle, dark blue), and Extended {\it WISE} AGNs (right, light blue). For each plot, the orange-hatched regions represent the subset of that halo mass bin with centrally located galaxies.}
	\label{Props_Mhalo}
\end{figure*}

In Figure~\ref{Props_Mhalo}, we show the relative group halo mass distributions for the three subpopulations described previously. For each, we separately highlight the CEN subsample. We find that the {\it WISE} and Extended {\it WISE} AGN subpopulations are found mostly in small groups with ${\rm M}_{\rm halo} < 10 ^ {13} ~ {\rm M}_{\odot}$. In contrast, a fair fraction of Region C mergers and interactions reside in larger groups, although their numbers drop sharply for cluster-sized halos (${\rm M}_{\rm halo} > 10 ^ {14} ~ {\rm M}_{\odot}$) in which velocity dispersions are too high for mergers to occur frequently \citep{Brodwin+13}, and the mergers that do occur in such environments tend to be dry and involve the central galaxy \citep{McIntosh+08}. If we consider only the CEN subsets, the fraction found in low-mass halos increases for all three subpopulations. The K-S test results (see Table~\ref{KS_test_values}) show that the overall and CEN-only Extended {\it WISE} AGNs differ from the Region C control at 98\% significance, suggesting that halo mass plays an important role in the occurrence of dusty AGNs in mergers and interactions. As with the pair separation K-S test results in \S~4.1, we speculate that a larger sample of {\it WISE} AGNs may result in a more significant difference from the Region C sample than we report in Table~\ref{KS_test_values}.

\subsubsection{Central Fraction}

We find a similarly high CEN fraction ($\sim80\%$) for mergers and interactions in all three subpopulations, which could suggest that this ranking does not play a role in the occurrence of {\it WISE} AGNs. Yet, we find that ${90_{-9}^{+5}}$\% of {\it WISE} and Extended {\it WISE} AGNs that are CENs reside in $< 10 ^ {13} ~ {\rm M}_{\odot}$ groups compared to only ${73_{-2}^{+1}}$\% for Region C. This is consistent with theoretical calculations that predict gas can continue to accrete efficiently to the centers of such halos \citep{Keres+05,Keres+12,Nelson+13} and, thereby, feed supermassive black hole growth. Additionally, the small number of SAT systems with a dusty AGN that are found in larger halos may mark the dynamical centers of low-mass subhalos that the group finder cannot identify \citep{McIntosh+08}, but in which gas accretion would still be very efficient. In terms of environment, the most likely scenario is that merger-driven dusty AGNs occur only at the centers of small groups, which are expected to have sufficient gas supply to support the significant central star-forming event needed to produce the enhanced dust obscuration.

\subsection{Star-forming Fraction}
\label{sec:43starformation} 

Bursts of new SF are theoretically tied to galaxy merging \citep[e.g.,][]{Cox+08} and have been observationally linked to mergers and  interactions \citep{Kennicutt+87, Barton+00, Lambas+03, Alonso+04, Ellison+08, Ellison+11}. Moreover, an association between younger stellar populations, presumably from recent or ongoing SF, and more powerful AGNs is known \citep{Kauffmann+03}. Thus, in this section we test whether the occurrence of dusty AGNs in mergers and interactions is associated with SF. 

\begin{figure}
	\center{\includegraphics[scale=0.7, angle=0]{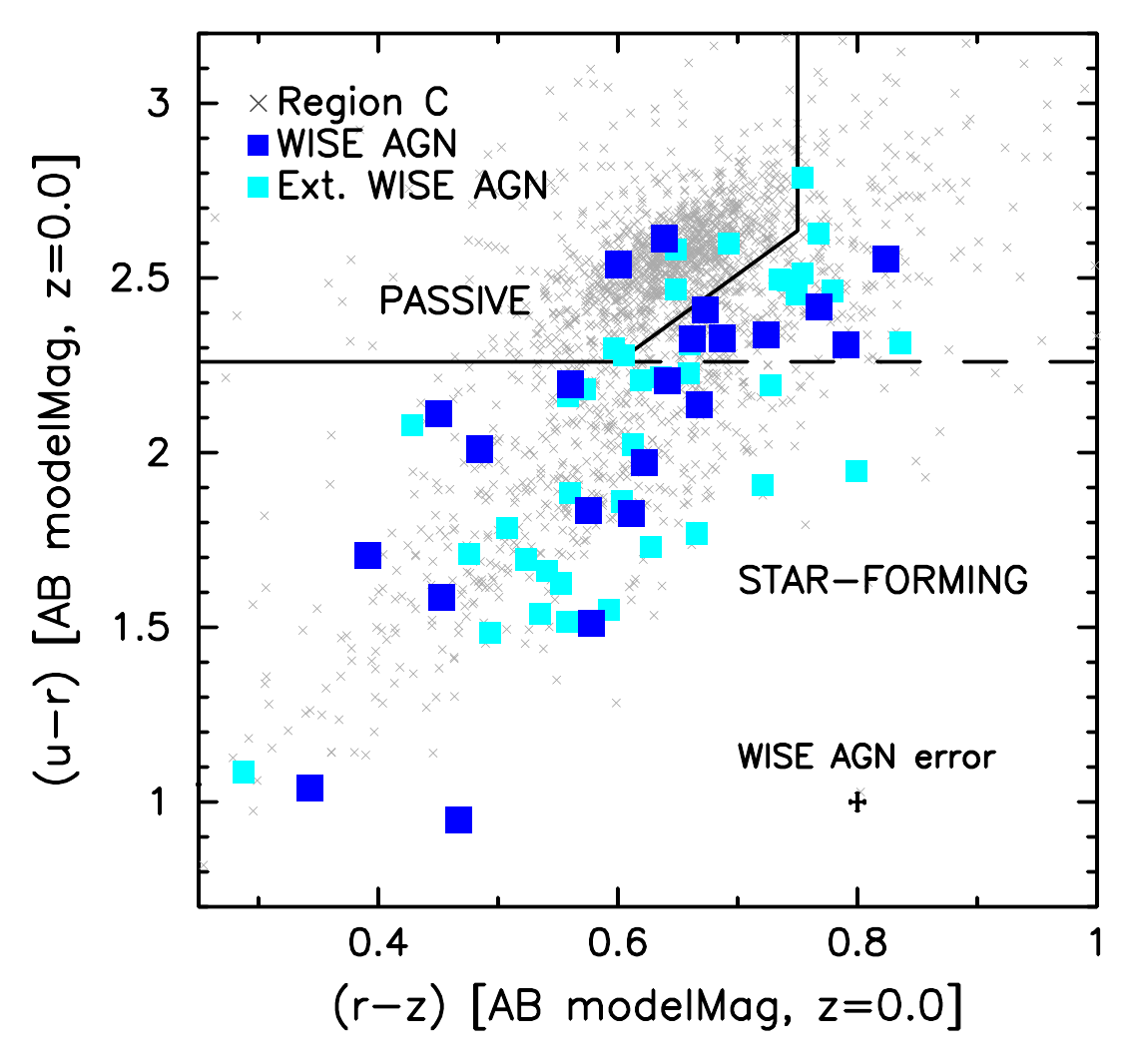}}
	\caption{SDSS ($u-r$) vs. ($r-z$) colors for interacting and merging systems. Subsamples are denoted as follows: {\it WISE} AGNs are shown as dark blue squares, Extended {\it WISE} AGNs as light blue squares, and Region C as small grey x\textquoteright{}s. The $urz$ colors are based on SDSS Model magnitudes (see text for details). The solid lines by \citet{Holden+12} define the boundary between passive and star-forming systems. We adopt the dashed line to separate optically blue star formers (below) from star-forming galaxies with dust (above). The error bars in the lower right corner are the median color errors for the {\it WISE} AGN subsample.}
	\label{Props_urz_SF}
\end{figure}

For this analysis, we use SDSS ($u-r$) versus ($r-z$) colors to distinguish passive (non-star-forming) galaxies from star-forming systems following \citet{Holden+12} and \citet{McIntosh+14} for the subset of our sample with SDSS spectroscopy (see \S~\ref{sec:22SDSSspec}). This $urz$ method is analogous to using rest-frame $UVJ$ data to separate optically red passive galaxies from dust-reddened star-formers \citep[e.g.,][]{Labbe+05,Williams+09}. Thus, this method is advantageous for identifying star-forming galaxies over fiber spectroscopic emission types (\S~\ref{sec:22SDSSspec}) that may be impacted by nuclear obscuration. We note that this method is only \textit{indicative} of star formation and does not test the amount of star formation. The ($u-r$) and ($r-z$) colors are based on SDSS model magnitudes corrected for Galactic extinction and K$+$evolution corrected to $z=0$, as described in detail in \citet{McIntosh+14}. In Figure~\ref{Props_urz_SF}, we show the $urz$ color-color plane for three merging/interacting subpopulations. For interactions, we define the pair to be star-forming if at least one of the galaxies is found in the star-forming region of the $urz$ plane; i.e., mixed pairs are considered star-forming. The vast majority of {\it WISE} (72 -- 97\%) and Extended {\it WISE} (78 -- 94\%) AGNs in merging or interacting galaxies are star-forming according to $urz$ colors. These high fractions imply a strong connection between dusty AGNs and SF in merging and interacting galaxies. Only 55 - 61\% of Region C galaxies fall in the star-forming region. We note that, because this method indicates the overall galaxy color, the contamination from nuclear emission is negligible for obscured AGNs (which would appear red in the optical) and non-AGNs. For the subsample of unobscured AGNs (Seyferts in Region C), the nuclear emission could push them into the star-forming region, making our fraction of Region C galaxies in the star-forming region an upper-limit.

\begin{figure*}
	\center{\includegraphics[scale=0.7, angle=0]{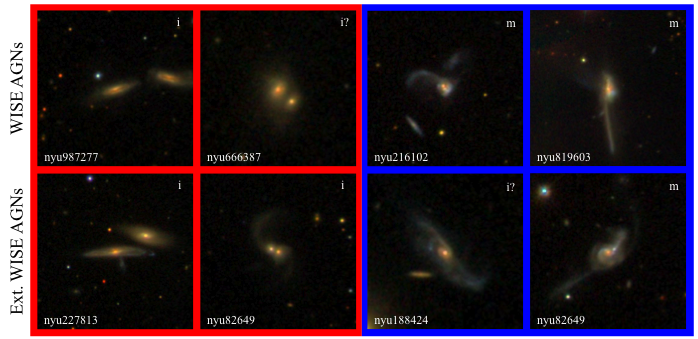}}
	\caption{Examples of {\it WISE} AGN/Ext. {\it WISE} AGN mergers and interactions plotted in Figure~\ref{Props_urz_SF}. All images are $100 \times 100$ kpc cutouts of $gri$-combined color images, centered on the {\it WISE} AGN, downloaded from the SDSS Image List Tool. Images outlined in red are centered on a galaxy with dusty $urz$ star-forming colors, while images outlined in blue are centered on galaxies with $(u-r)<2.25$ colors. Galaxy identification numbers are given in the lower left corner of each thumbnail (from the DR4 NYU-VAGC; \citealt{Blanton+05}), and the merger stage is given in the upper right corner (see \S \ref{sec:2sample}).}
	\label{Props_WAGN_images}
\end{figure*}

A significant fraction of the dusty AGN subpopulations in Figure~\ref{Props_urz_SF} have optically blue colors and, thus, unobscured ongoing or recent SF in contrast with their obscured nuclear activity \footnote{\citet{McIntosh+14} found that the passive locus extends a bit past the passive $(u-r)=2.25$ boundary and contains {\it recently-quenched} galaxies.}. To explore this further, we adopt a $(u-r)=2.25$ cut (dashed line) to separate blue from dusty star-forming colors. This selection isolates a population of dusty AGNs (with a dust-reddened nucleus) that are also very blue in the optical. We explore this seemingly contradictory relationship in Figure~\ref{Props_WAGN_images}, in which we compare examples of dusty AGNs with optically blue (blue box) and red (red box) colors. We find that dusty AGNs with blue $(u-r)$ colors clearly have unobscured regions of SF along with very dusty cores, while optically-red dusty AGNs do not. This illustrates how a galaxy can be optically blue, but still host an obscured (optically-red) central AGN.

\subsection{{[}OIII{]} Luminosity as an Indicator of AGN Power}
\label{sec:44AGNpower} 

The {[}OIII{]} ${\lambda 5007}$ emission line is the strongest emission line in a typical Type-2 (obscured) AGN, is less contaminated by SF than other optical emission lines, and its luminosity is commonly used as a measure of the strength of nuclear activity \citep{Kauffmann+03, Heckman+04}. \citet{Toba+14} found that Type-2 AGNs also identified as {\it WISE} AGNs have typically larger L{[}OIII{]} / H${\beta}$ values than Type-2 AGNs not identified as {\it WISE} AGNs. Similarly, \citet{Satyapal+14} performed a study on close pairs of galaxies and found that a majority of the most powerful optical AGNs are also classified as {\it WISE} AGNs, though it should be noted that this population of optical AGNs also included LINERs, which are controversial and weak AGNs at best \citep{Annibali+10,Kehrig+12,Yan_Blanton+12}. To test whether or not dusty AGNs are preferentially stronger than non-dusty AGNs in merging and interacting galaxies, we compare the {[}OIII{]} luminosities of Region C emission-line Seyfert galaxies to those of our {\it WISE} AGNs and Extended {\it WISE} AGNs.  

\begin{figure*}
	\center{\includegraphics[scale=0.85, angle=0]{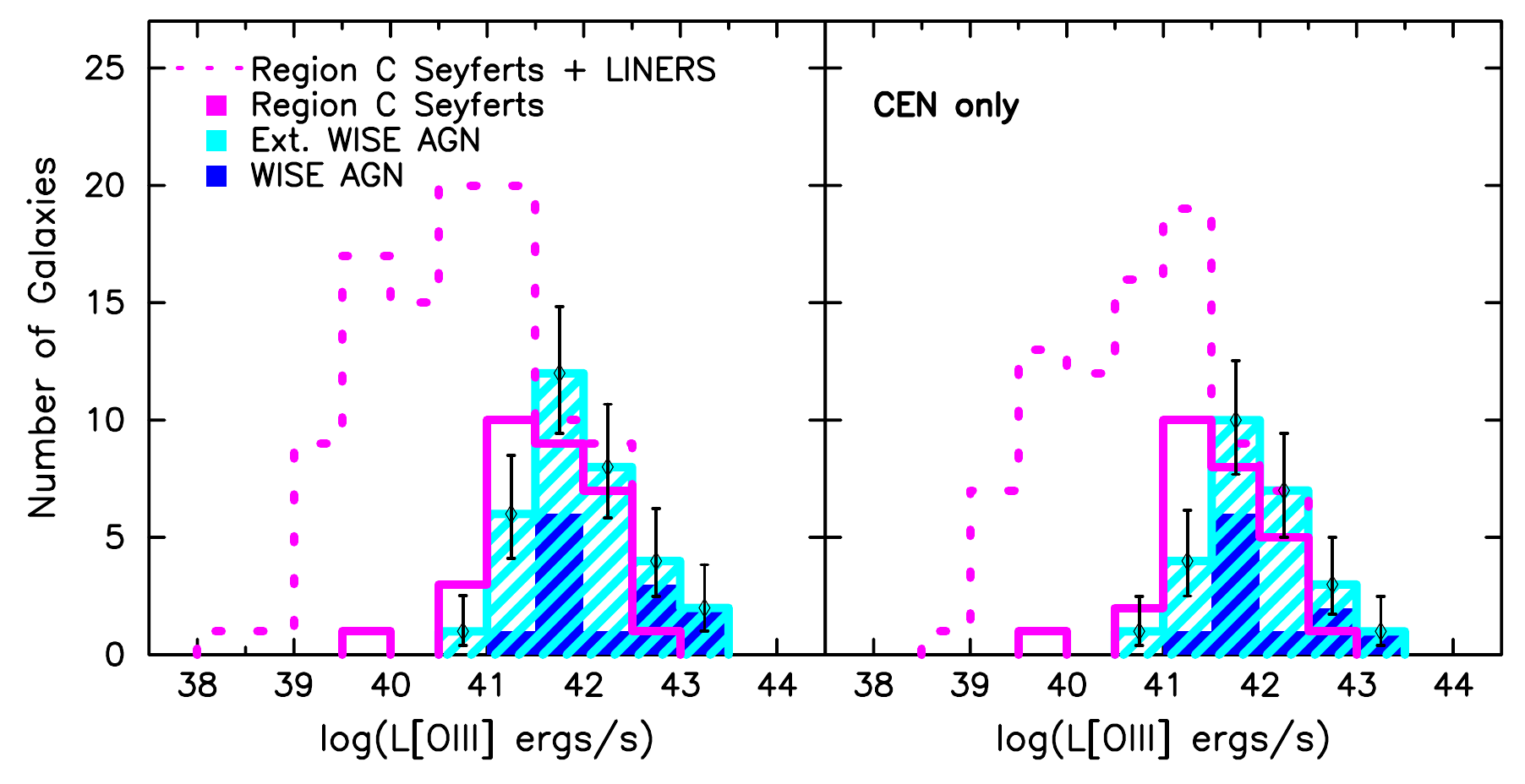}}
	\caption{Relative frequency distribution of {[}OIII{]} luminosity for the following subsamples of merging and interacting galaxies: {\it WISE} AGNs (dark blue solid), Extended {\it WISE} AGNs (light blue hatching), and Region C emission-line Seyferts (pink outline) selected by BPT analysis (see \S~\ref{sec:22SDSSspec} for details). The dotted line represents the combined Region C Seyferts and LINERs. Black error bars are the 68 percent binomial confidence intervals for the Extended {\it WISE} AGN population. Right: {[}OIII{]} luminosity distribution for subsamples described above that are CEN-ranked systems (see \S~\ref{sec:42enviro}).}
	\label{Props_LOIII}
\end{figure*}

We calculate luminosities of the {[}OIII{]} ${\lambda 5007}$ line using fluxes provided in the MPA-JHU spectroscopic catalog (see \S~\ref{sec:22SDSSspec}). We correct the {[}OIII{]} fluxes for dust within the host galaxy using $\tau_{\lambda}\propto\lambda^{-0.7}$ and assuming an intrinsic Balmer decrement ${H\alpha/H\beta}$ of 2.86 \citep{Charlot+00}. Additionally, we remove a handful of systems with {[}OIII{]} SNR $<$ 10. In the left panel of Figure~\ref{Props_LOIII}, we plot the $L[{\rm OIII}]$ distributions for three subsamples of AGNs in merging and interacting systems: {\it WISE} AGNs, Extended {\it WISE} AGNs, and Region C emission-line Seyferts. The {\it WISE} AGNs and Extended {\it WISE} AGNs appear to have slightly higher $L[{\rm OIII}]$ values on average than the non-dusty Seyferts and, therefore, higher nuclear accretion rates. We perform K-S tests to compare both dusty AGN subsamples to the Region C Seyfert subpopulation and find statistically insignificant percent differences of 82\% ({\it WISE} AGNs) and 52\% (Extended {\it WISE} AGNs), suggesting no strong connection between AGN power and merging or interacting dusty AGNs. 

If we restrict our K-S test comparisons to the CEN subsets of each subpopulation, shown in the right panel of Figure~\ref{Props_LOIII}, we find a tentative 97\% difference between {\it WISE} AGNs and Region C Seyferts, suggesting a possible connection between AGN power and central obscuration in mergers and interactions at the centers of small dark matter halos. We note that we may not find as strong of a trend between $L[{\rm OIII}]$ and {\it WISE}-detected AGNs as previous works owing to the fact that we use only emission-line Seyferts, which are unambiguous optical AGNs. The inclusion of LINERs, as done in \citet{Satyapal+14}, would shift the trend of the optical AGN to the lower power side (shown by the dotted line in Figure~\ref{Props_LOIII}), skewing the results to make {\it WISE} AGNs appear preferentially stronger. We also test the effect of the Seyfert population in Region C by performing all K-S tests in \S~\ref{sec:4nature} with the Region C Seyferts removed. We find no significant difference from the statistics presented in Table~\ref{KS_test_values}.

\section{The Dusty AGN-Merger Connection}
\label{sec:5discussion} 

We perform a simple test of whether or not highly disturbed galaxies and major pairs with indicators of tidal activity have an excess obscured-AGN frequency. The results presented in \S \ref{sec:3WISEanalysis} and \S \ref{sec:4nature} raise several interesting questions about the merger process and the AGN-merger connection. We find that merging (interacting) galaxies are 5 -- 17 (2.6 -- 4.9) times more likely to host a dusty {\it WISE} AGN than a non-merging, non-interacting galaxy. These results are similar for the four {\it WISE}-based dusty AGN criteria that we explore. This finding demonstrates a link between dusty AGNs and galaxy merging, in agreement with a similar study done by \citet{Satyapal+14}, who found that post-mergers are 10 -- 20 times more likely to host a dusty AGN than non-merging galaxies. \citet{Satyapal+14} studied close projected pairs and post-mergers with $z\leq 0.2$, using a different definition of {\it WISE} AGNs (see \S \ref{sec:31WISEcolors}). In general, a {\it dusty} AGN-merger connection is consistent with theoretical predictions of a link between gas-rich major mergers, AGNs and central SF as outlined in the Introduction. A dusty AGN-merger connection has been inferred in the literature \citep[e.g.,][]{Hopkins+10b} owing to the strong connection between ULIRGs and galaxy merging first noted by \citet{Sanders+88a}. This connection is observationally supported by the work of \citet{Kocevski+15}, who found that obscured AGNs are three times more likely to display merger or interaction signatures over unobscured AGNs. This scenario naturally explains the studies that found no AGN-merger link \citep{Cisternas+11, Scott+14, Villforth+14}, as these were at wavelengths incompatible with severe dust attenuation \citep{Alexander+01,Brandt+05, Kartaltepe+10, Treister+10}. Our results suggests that many AGNs in merging galaxies may be missed due to nuclear obscuration, in agreement with merger simulations that include the impact of dust obscuration \citep{Snyder+13} and IR observations \citep{Goulding+09, Goulding+10, Treister+10, Satyapal+14}. In this section, we discuss the frequency of nuclear obscuration in merging galaxies hosting an AGN, the incidences of different AGNs to be hosted by a merger or interaction, the likely physical reasons for why many AGNs are {\it not} hosted by merging/interacting galaxies, and the role that SF plays in the dusty AGN-merger connection. We also address the various limitations of infrared color selection of AGNs.


\subsection{How Common is Obscured AGN Activity in Mergers?}

Studies done in the optical and X-ray may miss important AGN activity due to dust attenuation in the galaxy \citep{Papovich+04, Goulding+09, Goulding+10}. As predicted in major merger simulations \citep{Barnes+91,Mihos+96} and demonstrated observationally \citep{Haan+09}, the gravitational torques in a merging pair cause gas in the pair to lose angular momentum and fall to the nucleus of the system, fueling new SF and AGN activity. The lack of a merger-AGN connection in optical and X-ray studies begs the question: how likely is the AGN to be obscured in a merging or interacting galaxy? To answer this, we compute an obscured AGN fraction:
\begin{equation}
	f_{\rm obscured} = \frac{N({\rm {\it WISE}\,\, AGN})}{N({\rm all\,\, AGN})} ,
\end{equation}
\noindent where the numerator is the number of {\it WISE} AGNs in a given population and the denominator is the sum of {\it WISE} AGNs plus non-{\it WISE}, emission-line Seyfert galaxies. We also calculate the obscured AGN fraction based on the Extended {\it WISE} AGN population. We present these fractions in Table~\ref{Obscuration_fraction_table} for merging, interacting, and possibly interacting galaxies, as well as the control sample described in \S \ref{sec:31WISEcolors}. We find that an AGN in an ongoing merger is 2 -- 6 times more likely to be obscured than an AGN of a non-merging and non-interacting host galaxy. Galaxies that are clearly interacting are nearly as likely as mergers to host a dusty AGN, while the likelihood drops for possibly interacting galaxies that may be simply chance projections of normal galaxies. When considering our less conservative Extended {\it WISE} definition of dusty AGNs, the high obscured AGN fraction remains for merging and interacting subsets compared to the control sample, but is not as large as for {\it WISE} AGNs. We emphasize that $f_{\rm obscured}$ is an estimate based solely on {\it WISE} and SDSS identifications of AGNs, which does not include other detections using shorter and longer wavelengths. Yet, this simple analysis shows that a large fraction (here $f_{\rm obscured} \sim 50\%$) of AGNs in mergers and interactions are dusty or at least partially obscured. This value is consistent with work done by \citet{Treister+10}. Coupled with the fact that merging and interacting galaxies are significantly more likely to host a dusty AGN, our results suggests that the major merging process typically produces AGNs accompanied by substantial amounts of dust. This is consistent with the idea that these mergers trigger fresh centrally-concentrated SF that supplies the dust that obscures the AGN in optical and X-ray surveys. We expand on the SF connection in \S~\ref{sec:54discussion_sf}.

\begin{table}
	\caption{The fraction of AGNs that are obscured ("obscuration fraction") for the following populations given in Column (1): merging galaxies, interacting pairs, possibly interacting pairs, and non-merging, non-interacting control galaxies. Column (2): obscured AGN percentages for combining the Seyfert population and {\it WISE} AGN population using the criteria defined by J11. Column (3): obscured AGN percentages for combining the Seyfert population and Extended {\it WISE} AGN population using the criteria defined in \S \ref{sec:312extendedWISEAGN}.}
	\label{Obscuration_fraction_table}
	\begin{tabular}{ccc}
		\hline
		Type & {\it WISE} AGN & Ext. {\it WISE} AGN\\
		(1) & (2) & (3)\\
		\hline 
		\hline
		mergers & $46_{-25}^{+26}$\% & $58_{-26}^{+23}$\%\\
		int. pair & $37_{-18}^{+22}$\% & $75_{-18}^{+12}$\%\\
		poss. int. pair & $23_{-10}^{+14}$\% & $55_{-12}^{+12}$\%\\
		control & $9.9_{-1.3}^{+1.4}$\% & $29.6_{-2.0}^{+2.0}$\%\\
		\hline
	\end{tabular}
\end{table}


\subsection{Which AGNs are Hosted by Mergers or Interactions?}

We find a connection between merging/interacting galaxies and dusty AGNs, which implies that galaxy merging produces such AGNs. Yet, not all AGNs are dusty and not all AGNs are associated with galaxy mergers and interactions, which suggests that other physical processes besides galaxy merging can produce AGNs. For example, \citet{Scott+14} studied the incidence of BPT-selected AGN in merging galaxies and found no clear connection. Here, we combine IR ({\it WISE} and Extended {\it WISE}) and optical (emission-line Seyfert) AGNs to find the fraction of different AGN subpopulations that are hosted by a merger or interaction. In Table~\ref{AGN_Fractions_table}, we tabulate the separate host merger and host interaction fractions for {\it WISE} AGNs, Extended {\it WISE} AGNs, emission-line Seyferts (including {\it WISE} AGNs), and combinations of each, which we compare to the fractions among non-AGN galaxies. In all cases, the fraction of AGNs that are hosted by a merging or interacting galaxy is quite small ($<10\%$). Yet, we find that the dusty AGNs are the most likely to be involved with galaxy merging; e.g., {\it WISE} AGNs are 4 -- 16 (2 -- 5) times more likely to be hosted by a merger (interaction) than non-AGN galaxies. We find a similar likelihood for Extended {\it WISE} AGNs. In contrast, the percentage of mergers and interactions found in the Seyfert population is similar to that found in the non-AGN population. Thus, while many AGNs must be triggered by non-merging processes, dusty AGNs are much more likely to be associated with merging activity than non-dusty AGNs, which are just as likely to be hosted by normal galaxies as by merging/interacting systems.


\begin{table*}
	\caption{Fraction of merging and interacting systems for the following AGN populations given in Column (1): {\it WISE} AGN as defined by J11, the Extended {\it WISE} AGN cut discussed in \S \ref{sec:31WISEcolors}, emission-lineSeyfert galaxies, the {\it WISE} AGN population combined with Seyfert galaxies, the Extended {\it WISE} AGN population combined with Seyfert galaxies, and all non-AGN systems. Column (2): the number (N) of galaxy systems contained in that population. Columns (3) and (4): the percentage of an AGN population that consists of merging galaxies or interacting pairs, respectively.}
	\label{AGN_Fractions_table}
	\begin{tabular}{cccc}
		\hline
		Population & N & Merging Galaxies & Interacting Pairs\\
		(1) & (2) & (3) & (4)\\
		\hline 
		\hline
		{\it WISE} AGN & 210 & $2.38_{-1.36}^{+3.07}$\% & $8.10_{-2.98}^{+4.48}$\%\\
		Extended {\it WISE} AGN & 673 & $1.04_{-0.54}^{+1.09}$\% & $8.02_{-1.82}^{+2.30}$\%\\
		Seyfert & 1909 & $0.37_{-0.19}^{+0.38}$\% & $2.83_{-0.66}^{+0.84}$\%\\
		{\it WISE} AGN \& Seyfert & 1973 & $0.56_{-0.25}^{+0.44}$\% & $3.19_{-0.69}^{+0.87}$\%\\
		Ext. {\it WISE} AGN \& Seyfert & 2171 & $0.55_{-0.31}^{+0.41}$\% & $4.05_{-0.75}^{+0.92}$\%\\
		non-AGNs & 41670 & $0.28_{-0.04}^{+0.06}$\% & $2.35_{-0.14}^{+0.15}$\%\\
		\hline
	\end{tabular}
\end{table*}

\begin{figure}
	\center{\includegraphics[scale=0.34, angle=0]{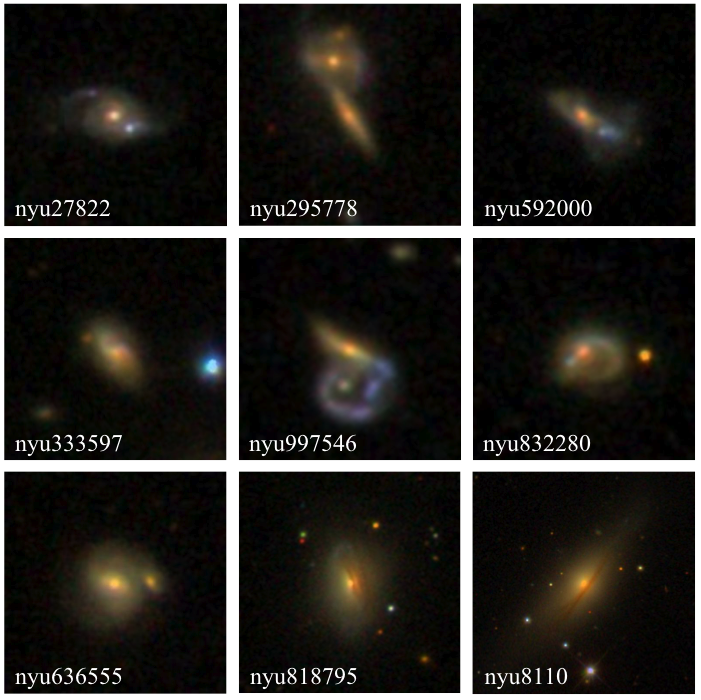}}
	\caption{Examples of Misclassified Control Galaxies in the {\it WISE} AGN Population (described in \S \ref{sec:31WISEcolors}). All images are $50 \times 50$ kpc cutouts of $gri$-combined color images, centered on the {\it WISE} AGN, downloaded from the SDSS Image List Tool. The top six galaxies all show clear signs of merger or interaction activity, including double nuclei and tidal features. The bottom three galaxies are all candidates for a minor merger (nyu636555) or the post-merger phase. The galaxy identification numbers for the galaxies are provided (from the DR4 NYU-VAGC; \citealt{Blanton+05}).}
	\label{control_WAGN_images}
\end{figure}

Owing to the strong association between dusty AGNs and merging, we visually reinspected the images of 188 {\it WISE} AGNs with normal (non-merging and non-interacting) classifications and find that 15 $\pm$ 6\% of this population are visually disturbed or show signs of recent or ongoing merger or interaction (major or minor) activity; examples shown in Figure~\ref{control_WAGN_images}. Most of these galaxies appear to be ongoing minor interactions (nyu27822, nyu592000, nyu636555) or later-stage post-merger remnant ellipticals with dust lanes (nyu818795 and nyu8110), which may be the result of minor merging \citep{Kauffmann+03, Kaviraj+09, Martini+13}. The fact that a closer inspection of {\it WISE} AGNs reveals additional examples with signs of recent tidal activity indicates that the association between dusty AGNs and merging is stronger than our results indicate, and that minor interactions may also trigger dusty AGNs.

In addition to visually inspecting those control galaxies classified as dusty AGNs, we also visually inspect the subset of 1711 control Seyfert galaxies that are not classified as {\it WISE} AGNs, and are presumably unobscured. We find that 1.2 -- 2.5 \% of these galaxies show signs of recent or ongoing minor mergers or interactions. While this result further supports the AGN-merger connection, it also illustrates a deeper connection between major mergers and {\it dusty} AGNs, as demonstrated throughout this work. This discovery also supports the hypothesis that merging (major or minor) is not the cause of the majority of AGNs. We note that deeper data could reveal a larger selection of merging or interacting galaxies in our control sample, whose signatures of interaction are too faint to detect in SDSS DR4 \citep{Bessiere+12}.

\begin{figure}
	\center{\includegraphics[scale=0.275, angle=0]{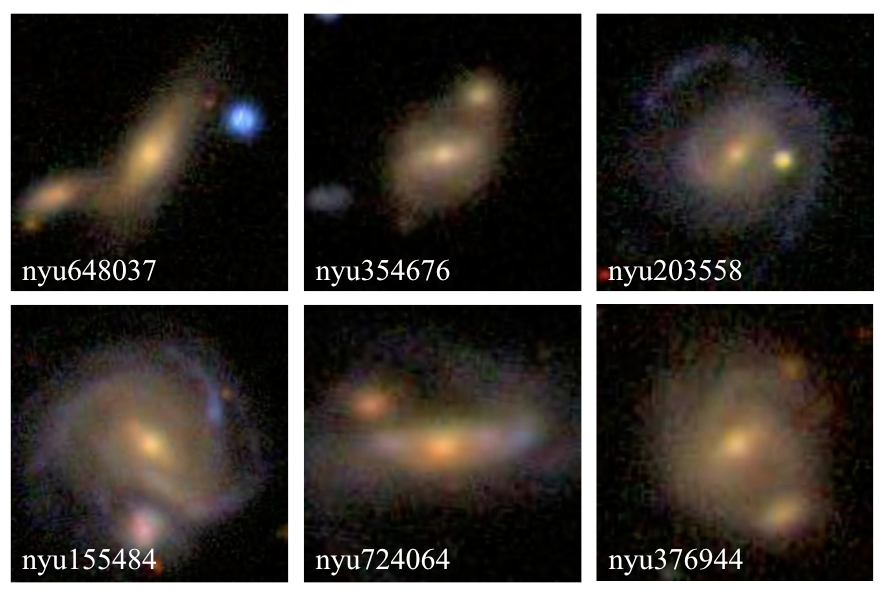}}
	\caption{Examples of Misclassified Control Galaxies in the unobscured Seyfert AGN Population. All images are $50 \times 50$ kpc cutouts of $gri$-combined color images, centered on the unobscured AGN, downloaded from the SDSS Image List Tool. All galaxies show signs of an ongoing minor merger or interaction. The galaxy identification numbers for the galaxies are provided (from the DR4 NYU-VAGC; \citealt{Blanton+05}).}
	\label{control_unobsSey_images}
\end{figure}


\subsection{Why Do Most Mergers Lack an AGN?}

While we find a clear dusty AGN-merger connection, we also find that only 5 -- 15 \% of merging galaxies are classified as optical Seyferts or {\it WISE} AGNs. Here, we explore several reasons for why the majority of ongoing mergers lack an AGN.

First, our AGN selection criteria are based solely on optical emission-lines and near-IR {\it WISE} colors, thus, we may be missing some mergers that host AGNs that are detected only at other wavelengths. While studies comparing AGN detection at different wavelengths find some fraction of previously unidentified sources in one passband compared to another, they find substantial overlap when comparing identifications in the optical and X-ray \citep[e.g.,][]{Grupe+99,Anderson+03}, the X-ray and radio \citep[e.g.,][]{Brinkmann+00}, and the optical and radio \citep[e.g.,][]{Ivezic+02}. By using {\it WISE} we are accounting for dust obscuration which is the strongest cause of missed AGNs in optical surveys \citep{Goulding+09}, although \citet{Snyder+13} demonstrated that significant dust can even obscure near-IR detections of very luminous AGNs during merging coalescence. Given these findings, we do not expect the frequency of mergers hosting AGNs to be significantly higher than what we find, but we note that a thorough accounting of merger-driven AGNs over a wide range of wavelengths is needed to constrain this definitively.

Second, not all major galaxy-galaxy interactions are predicted to produce an AGN. The simulations that predict AGNs in mergers \citep{Volonteri+03, DiMatteo+05, Springel+05, Hopkins+08, DeBuhr+11} require the presence of gas, which can be funneled into the nucleus of the system by gravitational torques \citep{Barnes+91,Mihos+96}. If the merger is dry (little to no gas), it is reasonable to assume that a dusty AGN will not occur, which is consistent with our finding that statistically zero {\it WISE} AGNs are hosted by passive mergers/interactions (\S~\ref{sec:43starformation}). Even if gas is present, simulations \citep{Cox+08,Johansson+09} and observations \citep{Haines+15} have found that not all gas-rich mergers make a nuclear starburst (and presumably a dusty AGN). \citet{Cox+08} showed that an excess amount of gas can decrease the starburst efficiency in a merger. If the starburst efficiency is decreased, and AGNs and SF are thought to be intimately connected \citep{Hickox+14}, it may be possible that too much gas can also decrease AGN efficiency. \citet{Johansson+09} demonstrated that the nuclear black hole accretion rate decreases with increasing merger progenitor mass ratio; i.e., 1:1 mergers will produce stronger AGNs than 4:1 mergers. It is impossible to discern progenitor mass ratios for visually identified `train wreck' mergers. Additionally, the orbital geometry can have a large effect on gas dynamics in a merger, with coplanar mergers having the highest gas inflow to the nucleus \citep{Mihos+96}. In short, only {\it idealized} major encounters (finely-tuned gas fraction, mass ratio, and orbital configuration of the merger progenitors) produce the brief starbursts and dusty AGNs seen in simulations of the modern merger hypothesis such as presented in Fig. 1 of \citet{Hopkins+08}. To give a quantitative constraint on the number of mergers we would expect to fit this perfect scenario, we look to our sample of 307 interacting pairs. Using $urz$ star-forming colors as an indicator of gas content and mass ratios  less than two (described in \S \ref{sec:21SDSSsample}), we find that 11 -- 19\% of interacting pairs fit the idealized case and should, in theory, produce an obscured AGN. This number is higher than our {\it WISE} AGN merger sample would suggest.

Yet another probable reason why the majority of mergers lack AGNs is differences in the merger observability timescales and AGN lifetimes. For example, depending on progenitor mass ratio and gas fractions, the strong asymmetric features typically used to identify major mergers at or near the time of coalescence (i.e., our ongoing merger definition; see \S \ref{sec:21SDSSsample}) have timescales that range from $\sim$0.2--0.5\,Gyr \citep{Lotz+10a,Lotz+10b}. In comparison, AGN phase lifetime estimates range from 0.01--1\,Gyr \citep{Martini+01,Marconi+04} to as little as a repeating $\sim 10^5$\,years `AGN flicker' \citep{Schawinski+15}. Given these ranges of timescales it is, therefore, possible that all/most major gas-rich mergers do produce AGNs if the AGN phase is typically on order 10 times shorter than the merger observability time. But, owing to our lack of knowledge about progenitor mass ratios, gas content and orbits in our sample of visually identified mergers, it is impossible to test this.


\subsection{The Role of Star Formation in the Dusty AGN-Merger Connection}
\label{sec:54discussion_sf} 

We find that a high majority (72 -- 97\%) of merging/interacting galaxies hosting a {\it WISE} AGN have $urz$ star-forming colors. This indicates a link between merging, SF, and the triggering of an obscured AGN. This is consistent with the major merger simulations by \citet{Hopkins+08} and others \citep{Barnes+91, Mihos+96}, which predict that the gas brought into the nucleus by gravitational torques in a gas-rich merger should simultaneously fuel the production of new stars and accrete onto the black hole. A natural consequence of the new SF would be centrally-concentrated dust obscuration, the amount of which depends on the rate of SF \citep{Goulding+12}. A sufficient amount of dust obscuration will produce an infrared (IR) AGN. Assuming that SF plays a critical role in obscuring AGNs hosted by merging/interacting galaxies, we test whether the incidence of dusty AGNs is higher among star-forming mergers and interactions. We find no increase in the incidence of WISE AGNs in mergers over the control sample (5 -- 18 times). However, we find that the vast majority of {\it WISE} AGNs are found in star-forming systems for both merging or interacting galaxies and the control sample, confirming that star formation in the host galaxy is intimately linked to infrared AGNs, as other studies have found \citep[e.g.,][]{Hickox+09, Snyder+13}.


Additionally, we repeat all of the K-S tests described in \S \ref{sec:4nature} to test for the unique nature of $urz$-selected star-forming mergers and interactions that host dusty AGNs. Specifically, for each property we compare the {\it WISE} (Extended {\it WISE}) subset distribution against that of the Region C control sample distribution using only star-forming merging/interacting galaxies. We find no statistical differences for any of the properties explored except for a $\sim$3-sigma difference between the Extended {\it WISE} and Region C star-forming systems in terms of pair separation, compared to the 2-sigma result discussed in \S \ref{sec:pairsep}. As such, it appears that star-forming merging or interacting galaxies that host dusty AGNs are not different from others that lack a dusty AGN. This lack of difference could support the idea that all star-forming merging and interacting galaxies have a dusty AGN phase.



\subsection{The Limitations of Infrared Color Analysis for Isolating Merging AGNs}
\label{sec:55discussion_IRlimitations} 

There are many AGN selection techniques across the wavelength spectrum. While optical and ultraviolet can select unobscured AGNs, they will often miss obscured sources \citep{Alexander+01}. The X-ray wavelengths are commonly regarded as the most complete AGN selection method available, but X-ray surveys can miss obscured AGNs \citep{Alexander+01,Brandt+05, Kartaltepe+10, Treister+10}. In our search for obscured AGNs, we choose to use an infrared color selection, as many other studies have done \citep{Jarrett+11, Stern+12, Assef+13, Yan+13, Satyapal+14}. While the main advantage of the infrared is sensitivity to obscured AGNs \citep{Cardamone+08}, it does have limitations. One downside is the possible contamination of the mid-infrared bands from SF in the host galaxy \citep{DelMoro+16, Lange+16}. Cold dust emission from SF tends to dominate the far-IR, peaking in the 100 - 160 ${\mu}$m range, but the mid-IR bands can also be affected. Another possible source of error that arises out of mid-IR AGN selection is incompleteness of low-luminosity AGNs. The mid-IR color selection techniques tend to preferentially select the brightest AGNs \citep{Cardamone+08}, and miss a significant number of low-luminosity sources. Using other selection methods could find this missing population of low-luminosity AGNs in our sample, thus strengthening our AGN-merger connection. However, the true effect of this bias cannot be accurately determined. \citet{Treister+12} found that major mergers are tied only to the most luminous AGNs, thus we should expect this bias not from the data but from the merger process. While we select use of the infrared to account for dust obscuration, \citet{Snyder+13} demonstrated that significant dust can even obscure near-IR detections of very luminous AGNs during merging coalescence. All of these limitations suggest that our merger-AGN rate is a lower limit, and would benefit from the addition of X-ray and radio data for a full picture of the AGN-merger connection.

Throughout this paper, we assume that {\it WISE} color AGN selection should select obscured AGNs. From the major merger model, which predicts high amounts of SF and therefore dust, we also assume that the dust from SF acts as the obscuring material in these special systems. We note that the AGN Unification Model \citep{Urry+95} describes AGN obscuration as a result of the viewing angle; the torus around an AGN can obscure the AGN itself. This would imply that obscured AGNs selected in this sample are not different from other AGNs due to merging, but rather to viewing angle. However, recent work by \citet{Kocevski+15} shows that the viewing angle of an AGN cannot be the only property differentiating populations of AGNs, but cites a recent merger event as a plausible source of differing AGN populations.


\section{Summary}
\label{sec:6summary} 

We combine data from {\it WISE} and the SDSS to explore the relationship between dust-obscured AGNs and galaxy mergers within the context of the current major merger model \citep{Volonteri+03, DiMatteo+05, Springel+05, Hopkins+08, DeBuhr+11}. We present a new, volume-limited ($z\leq 0.08$) catalog of visually-selected major mergers and galaxy-galaxy interactions from the SDSS, with stellar masses ${\rm M}_{\rm star} > 2 \times 10 ^ {10} ~ {\rm M}_{\odot}$. We use SDSS fiber spectroscopy diagnostics from the MPA-JHU emission-line analysis \citep{Kauffmann+03, Brinchmann+04} to map the locations of over 40,000 normal galaxies with different emission types from \citet{McIntosh+14} in the $[3.4]-[4.6]$ versus $[4.6]-[12]$ color-color plane. We test multiple dusty AGN selection methods and find that one-quarter of Seyferts have redder $[3.4]-[4.6]$ colors than $\sim99\%$ of non-Seyferts. We use this empirical criterion to define an `Extended' {\it WISE} AGN selection of dusty AGNs. We perform a simple test of whether or not highly disturbed galaxies and major pairs with indicators of tidal activity have an excess obscured-AGN frequency. We use the normal galaxies as a control sample against which we quantify the amount of dusty AGN activity in mergers and interactions. We confirm a dusty AGN-merger connection, consistent with the major merger model by \citet{Hopkins+08} and others \citep{Volonteri+03, DiMatteo+05, Springel+05, DeBuhr+11, Snyder+13}, and observationally supported by \citet{Satyapal+14}, in which gravitational torques drive gas inflows to the center of the merging system, feeding the AGN. Our key results are summarized as follows:

(i) We find an excess of obscured AGN activity in merging galaxies when compared to a control sample with the same redshift and stellar mass constraints, indicating that merging (interacting) systems are 5 -- 17 (3 -- 5) times more likely to host an obscured AGN compared with non-merging, non-interacting galaxies, in agreement with \citet{Satyapal+14}.

(ii) We find that mergers hosting a dusty AGN favor smaller pair separations and smaller dark matter halo masses than other mergers and interactions. We find that most dusty AGN mergers are located at the centers of ${\rm M}_{\rm halo} < 10 ^ {13} ~ {\rm M}_{\odot}$ groups; this relationship also supports the \citet{Hopkins+08} major merger picture, in which mergers favor halos at the small group scale. 

(iii) We find that the vast majority of mergers hosting dusty AGNs have star-forming SDSS $urz$ colors. This connection is also consistent with major merger models which predict heightened SF at the time of the merger (and the concomitant AGN).

(iv) We find that AGNs also classified as ongoing mergers are five times more likely to be obscured than AGNs in non-merging, non-interacting galaxies. Half of all AGNs hosted by a merger are dusty, suggesting that shorter wavelengths are inadequate in selecting AGNs in merging systems. 

(v) We find no enhanced frequency of optical BPT-selected AGNs in merging over non-merging galaxies at this redshift, indicating that the missed detection of dusty AGNs at optical and shorter wavelengths is likely the reason for the ongoing AGN-merger connection debate.

From this study, we find strong evidence in favor of the major merger model, in which gas-rich mergers produce central bursts of SF and fuel AGNs. The SF produces dust, which obscures the AGN and reradiates in the thermal infrared. Because of this obscuration, surveys at shorter wavelengths may see not only an incomplete picture of the merger-AGN connection, but are also biased against measuring the true merger, SF, and AGN rates.


\section*{Acknowledgments}

We are grateful for the support and advice of C. Tremonti, J. Rigby, T. Jarrett, X. Her, B. Decker, and members of the UMKC Galaxy Evolution Group. Special thanks go to S. Ellison, D. W. Darg, and S. Kaviraj for providing Galaxy Zoo catalogues for comparison.
The authors would like to thank the anonymous reviewer, whose enlightening comments helped to strengthen the paper.
M.E.W. and D.H.M. thankfully acknowledge support from NASA grant NNX13AE96G.
M.E.W. acknowledges support from the Missouri Consortium of NASA's National Space Grant College and Fellowship Program.
This publication also makes use of data products from the Wide-field Infrared Survey Explorer, which is a joint project of the University of California, Los Angeles, and the Jet Propulsion Laboratory/California Institute of Technology, funded by the National Aeronautics and Space Administration.
This publication makes use of the Sloan Digital Sky Survey (SDSS). Funding for the SDSS has been provided by the Alfred P. Sloan Foundation, the Participating Institutions, the National Science Foundation, the U.S. Department of Energy, the National Aeronautics and Space Administration, the Japanese Monbukagakusho, the Max Planck Society, and the Higher Education Funding Council for England. The SDSS Web Site is http://www.sdss.org/. The SDSS is managed by the Astrophysical Research Consortium for the Participating Institutions. The Participating Institutions are the American Museum of Natural History, Astrophysical Institute Potsdam, University of Basel, University of Cambridge, Case Western Reserve University, University of Chicago, Drexel University, Fermilab, the Institute for Advanced Study, the Japan Participation Group, Johns Hopkins University, the Joint Institute for Nuclear Astrophysics, the Kavli Institute for Particle Astrophysics and Cosmology, the Korean Scientist Group, the Chinese Academy of Sciences (LAMOST), Los Alamos National Laboratory, the Max-Planck-Institute for Astronomy (MPIA), the Max-Planck-Institute for Astrophysics (MPA), New Mexico State University, Ohio State University, University of Pittsburgh, University of Portsmouth, Princeton University, the United States Naval Observatory, and the University of Washington.
This publication also made use of NASA's Astrophysics Data System Bibliographic Services and TOPCAT \citep[Tools for OPerations on Catalogues And Tables,][]{Taylor+05}.

\bibliographystyle{mnras}
\bibliography{Weston_AGNpaper_bib}

\begin{thebibliography}{}
\makeatletter
\relax
\def\mn@urlcharsother{\let\do\@makeother \do\$\do\&\do\#\do\^\do\_\do\%\do\~}
\def\mn@doi{\begingroup\mn@urlcharsother \@ifnextchar [ {\mn@doi@}
  {\mn@doi@[]}}
\def\mn@doi@[#1]#2{\def\@tempa{#1}\ifx\@tempa\@empty \href
  {http://dx.doi.org/#2} {doi:#2}\else \href {http://dx.doi.org/#2} {#1}\fi
  \endgroup}
\def\mn@eprint#1#2{\mn@eprint@#1:#2::\@nil}
\def\mn@eprint@arXiv#1{\href {http://arxiv.org/abs/#1} {{\tt arXiv:#1}}}
\def\mn@eprint@dblp#1{\href {http://dblp.uni-trier.de/rec/bibtex/#1.xml}
  {dblp:#1}}
\def\mn@eprint@#1:#2:#3:#4\@nil{\def\@tempa {#1}\def\@tempb {#2}\def\@tempc
  {#3}\ifx \@tempc \@empty \let \@tempc \@tempb \let \@tempb \@tempa \fi \ifx
  \@tempb \@empty \def\@tempb {arXiv}\fi \@ifundefined
  {mn@eprint@\@tempb}{\@tempb:\@tempc}{\expandafter \expandafter \csname
  mn@eprint@\@tempb\endcsname \expandafter{\@tempc}}}

\bibitem[\protect\citeauthoryear{{Adelman-McCarthy} \& et
  al.}{{Adelman-McCarthy} \& et~al.}{2006}]{Adelman-McCarthy+06}
{Adelman-McCarthy} J.~K.,  et al. 2006, \mn@doi [ApJS] {10.1086/497917}, \href
  {http://adsabs.harvard.edu/abs/2006ApJS..162...38A} {162, 38}

\bibitem[\protect\citeauthoryear{{Alexander}, {Brandt}, {Hornschemeier},
  {Garmire}, {Schneider}, {Bauer}  \& {Griffiths}}{{Alexander}
  et~al.}{2001}]{Alexander+01}
{Alexander} D.~M.,  {Brandt} W.~N.,  {Hornschemeier} A.~E.,  {Garmire} G.~P.,
  {Schneider} D.~P.,  {Bauer} F.~E.,   {Griffiths} R.~E.,  2001, \mn@doi [\aj]
  {10.1086/323540}, \href {http://adsabs.harvard.edu/abs/2001AJ....122.2156A}
  {122, 2156}

\bibitem[\protect\citeauthoryear{{Alonso}, {Tissera}, {Coldwell}  \&
  {Lambas}}{{Alonso} et~al.}{2004}]{Alonso+04}
{Alonso} M.~S.,  {Tissera} P.~B.,  {Coldwell} G.,   {Lambas} D.~G.,  2004,
  \mn@doi [MNRAS] {10.1111/j.1365-2966.2004.08002.x}, \href
  {http://adsabs.harvard.edu/abs/2004MNRAS.352.1081A} {352, 1081}

\bibitem[\protect\citeauthoryear{{Anderson} et~al.,}{{Anderson}
  et~al.}{2003}]{Anderson+03}
{Anderson} S.~F.,  et~al., 2003, \mn@doi [\aj] {10.1086/378999}, \href
  {http://adsabs.harvard.edu/abs/2003AJ....126.2209A} {126, 2209}

\bibitem[\protect\citeauthoryear{{Annibali}, {Bressan}, {Rampazzo},
  {Zeilinger}, {Vega}  \& {Panuzzo}}{{Annibali} et~al.}{2010}]{Annibali+10}
{Annibali} F.,  {Bressan} A.,  {Rampazzo} R.,  {Zeilinger} W.~W.,  {Vega} O.,
  {Panuzzo} P.,  2010, \mn@doi [\aap] {10.1051/0004-6361/200913774}, \href
  {http://adsabs.harvard.edu/abs/2010A%26A...519A..40A} {519, A40}

\bibitem[\protect\citeauthoryear{{Ashby} et~al.,}{{Ashby}
  et~al.}{2009}]{Ashby+09}
{Ashby} M.~L.~N.,  et~al., 2009, \mn@doi [ApJ] {10.1088/0004-637X/701/1/428},
  \href {http://adsabs.harvard.edu/abs/2009ApJ...701..428A} {701, 428}

\bibitem[\protect\citeauthoryear{{Assef} et~al.,}{{Assef}
  et~al.}{2013}]{Assef+13}
{Assef} R.~J.,  et~al., 2013, \mn@doi [ApJ] {10.1088/0004-637X/772/1/26}, \href
  {http://adsabs.harvard.edu/abs/2013ApJ...772...26A} {772, 26}

\bibitem[\protect\citeauthoryear{{Baldwin}, {Phillips}  \&
  {Terlevich}}{{Baldwin} et~al.}{1981}]{Baldwin+81}
{Baldwin} J.~A.,  {Phillips} M.~M.,   {Terlevich} R.,  1981, \mn@doi [PASP]
  {10.1086/130766}, \href {http://adsabs.harvard.edu/abs/1981PASP...93....5B}
  {93, 5}

\bibitem[\protect\citeauthoryear{{Barnes}}{{Barnes}}{2002}]{Barnes+02}
{Barnes} J.~E.,  2002, \mn@doi [MNRAS] {10.1046/j.1365-8711.2002.05335.x},
  \href {http://adsabs.harvard.edu/abs/2002MNRAS.333..481B} {333, 481}

\bibitem[\protect\citeauthoryear{{Barnes} \& {Hernquist}}{{Barnes} \&
  {Hernquist}}{1991}]{Barnes+91}
{Barnes} J.~E.,  {Hernquist} L.~E.,  1991, \mn@doi [\apjl] {10.1086/185978},
  \href {http://adsabs.harvard.edu/abs/1991ApJ...370L..65B} {370, L65}

\bibitem[\protect\citeauthoryear{{Barton}, {de Carvalho}  \& {Geller}}{{Barton}
  et~al.}{1998}]{Barton+98}
{Barton} E.~J.,  {de Carvalho} R.~R.,   {Geller} M.~J.,  1998, \mn@doi [\aj]
  {10.1086/300531}, \href {http://adsabs.harvard.edu/abs/1998AJ....116.1573B}
  {116, 1573}

\bibitem[\protect\citeauthoryear{{Barton}, {Geller}  \& {Kenyon}}{{Barton}
  et~al.}{2000}]{Barton+00}
{Barton} E.~J.,  {Geller} M.~J.,   {Kenyon} S.~J.,  2000, \mn@doi [ApJ]
  {10.1086/308392}, \href {http://adsabs.harvard.edu/abs/2000ApJ...530..660B}
  {530, 660}

\bibitem[\protect\citeauthoryear{{Batuski}, {Burns}, {Newberry}, {Hill},
  {Deeg}, {Laubscher}  \& {Elston}}{{Batuski} et~al.}{1991}]{Batuski+91}
{Batuski} D.~J.,  {Burns} J.~O.,  {Newberry} M.~V.,  {Hill} J.~M.,  {Deeg}
  H.-J.,  {Laubscher} B.~E.,   {Elston} R.~J.,  1991, \mn@doi [\aj]
  {10.1086/115822}, \href {http://adsabs.harvard.edu/abs/1991AJ....101.1983B}
  {101, 1983}

\bibitem[\protect\citeauthoryear{{Baugh}, {Cole}  \& {Frenk}}{{Baugh}
  et~al.}{1996}]{Baugh+96}
{Baugh} C.~M.,  {Cole} S.,   {Frenk} C.~S.,  1996, MNRAS, \href
  {http://adsabs.harvard.edu/abs/1996MNRAS.283.1361B} {283, 1361}

\bibitem[\protect\citeauthoryear{{Beers}, {Gebhardt}, {Forman}, {Huchra}  \&
  {Jones}}{{Beers} et~al.}{1991}]{Beers+91}
{Beers} T.~C.,  {Gebhardt} K.,  {Forman} W.,  {Huchra} J.~P.,   {Jones} C.,
  1991, \mn@doi [\aj] {10.1086/115982}, \href
  {http://adsabs.harvard.edu/abs/1991AJ....102.1581B} {102, 1581}

\bibitem[\protect\citeauthoryear{{Bell}, {McIntosh}, {Katz}  \&
  {Weinberg}}{{Bell} et~al.}{2003}]{Bell+03}
{Bell} E.~F.,  {McIntosh} D.~H.,  {Katz} N.,   {Weinberg} M.~D.,  2003, \mn@doi
  [ApJS] {10.1086/378847}, \href
  {http://adsabs.harvard.edu/abs/2003ApJS..149..289B} {149, 289}

\bibitem[\protect\citeauthoryear{{Bessiere}, {Tadhunter}, {Ramos Almeida}  \&
  {Villar Mart{\'{\i}}n}}{{Bessiere} et~al.}{2012}]{Bessiere+12}
{Bessiere} P.~S.,  {Tadhunter} C.~N.,  {Ramos Almeida} C.,   {Villar
  Mart{\'{\i}}n} M.,  2012, \mn@doi [\mnras]
  {10.1111/j.1365-2966.2012.21701.x}, \href
  {http://adsabs.harvard.edu/abs/2012MNRAS.426..276B} {426, 276}

\bibitem[\protect\citeauthoryear{{Best}, {Kauffmann}, {Heckman}, {Brinchmann},
  {Charlot}, {Ivezi{\'c}}  \& {White}}{{Best} et~al.}{2005}]{Best+05}
{Best} P.~N.,  {Kauffmann} G.,  {Heckman} T.~M.,  {Brinchmann} J.,  {Charlot}
  S.,  {Ivezi{\'c}} {\v Z}.,   {White} S.~D.~M.,  2005, \mn@doi [MNRAS]
  {10.1111/j.1365-2966.2005.09192.x}, \href
  {http://adsabs.harvard.edu/abs/2005MNRAS.362...25B} {362, 25}

\bibitem[\protect\citeauthoryear{{Blanton}, {Lin}, {Lupton}, {Maley}, {Young},
  {Zehavi}  \& {Loveday}}{{Blanton} et~al.}{2003}]{Blanton+03b}
{Blanton} M.~R.,  {Lin} H.,  {Lupton} R.~H.,  {Maley} F.~M.,  {Young} N.,
  {Zehavi} I.,   {Loveday} J.,  2003, \mn@doi [AJ] {10.1086/344761}, \href
  {http://adsabs.harvard.edu/abs/2003AJ....125.2276B} {125, 2276}

\bibitem[\protect\citeauthoryear{{Blanton} et~al.,}{{Blanton}
  et~al.}{2005}]{Blanton+05}
{Blanton} M.~R.,  et~al., 2005, \mn@doi [AJ] {10.1086/429803}, \href
  {http://adsabs.harvard.edu/abs/2005AJ....129.2562B} {129, 2562}

\bibitem[\protect\citeauthoryear{{B{\"o}hm} et~al.,}{{B{\"o}hm}
  et~al.}{2013}]{Bohm+13}
{B{\"o}hm} A.,  et~al., 2013, \mn@doi [A\&A] {10.1051/0004-6361/201015444},
  \href {http://www.aanda.org/articles/aa/pdf/2013/01/aa15444-10.pdf} {549,
  A46}

\bibitem[\protect\citeauthoryear{{Brandt} \& {Hasinger}}{{Brandt} \&
  {Hasinger}}{2005}]{Brandt+05}
{Brandt} W.~N.,  {Hasinger} G.,  2005, \mn@doi [\araa]
  {10.1146/annurev.astro.43.051804.102213}, \href
  {http://adsabs.harvard.edu/abs/2005ARA%26A..43..827B} {43, 827}

\bibitem[\protect\citeauthoryear{{Brinchmann}, {Charlot}, {White}, {Tremonti},
  {Kauffmann}, {Heckman}  \& {Brinkmann}}{{Brinchmann}
  et~al.}{2004}]{Brinchmann+04}
{Brinchmann} J.,  {Charlot} S.,  {White} S.~D.~M.,  {Tremonti} C.,  {Kauffmann}
  G.,  {Heckman} T.,   {Brinkmann} J.,  2004, \mn@doi [MNRAS]
  {10.1111/j.1365-2966.2004.07881.x}, \href
  {http://adsabs.harvard.edu/abs/2004MNRAS.351.1151B} {351, 1151}

\bibitem[\protect\citeauthoryear{{Brinkmann}, {Laurent-Muehleisen}, {Voges},
  {Siebert}, {Becker}, {Brotherton}, {White}  \& {Gregg}}{{Brinkmann}
  et~al.}{2000}]{Brinkmann+00}
{Brinkmann} W.,  {Laurent-Muehleisen} S.~A.,  {Voges} W.,  {Siebert} J.,
  {Becker} R.~H.,  {Brotherton} M.~S.,  {White} R.~L.,   {Gregg} M.~D.,  2000,
  \aap, \href {http://adsabs.harvard.edu/abs/2000A%26A...356..445B} {356, 445}

\bibitem[\protect\citeauthoryear{{Brodwin} et~al.,}{{Brodwin}
  et~al.}{2013}]{Brodwin+13}
{Brodwin} M.,  et~al., 2013, \mn@doi [ApJ] {10.1088/0004-637X/779/2/138}, \href
  {http://adsabs.harvard.edu/abs/2013ApJ...779..138B} {779, 138}

\bibitem[\protect\citeauthoryear{{Capelo}, {Volonteri}, {Dotti}, {Bellovary},
  {Mayer}  \& {Governato}}{{Capelo} et~al.}{2015}]{Capelo+15}
{Capelo} P.~R.,  {Volonteri} M.,  {Dotti} M.,  {Bellovary} J.~M.,  {Mayer} L.,
   {Governato} F.,  2015, \mn@doi [MNRAS] {10.1093/mnras/stu2500}, \href
  {http://adsabs.harvard.edu/abs/2015MNRAS.447.2123C} {447, 2123}

\bibitem[\protect\citeauthoryear{{Cappi}, {Benoist}, {da Costa}  \&
  {Maurogordato}}{{Cappi} et~al.}{2003}]{Cappi+03}
{Cappi} A.,  {Benoist} C.,  {da Costa} L.~N.,   {Maurogordato} S.,  2003,
  \mn@doi [\aap] {10.1051/0004-6361:20031016}, \href
  {http://adsabs.harvard.edu/abs/2003A%26A...408..905C} {408, 905}

\bibitem[\protect\citeauthoryear{{Cardamone} et~al.,}{{Cardamone}
  et~al.}{2008}]{Cardamone+08}
{Cardamone} C.~N.,  et~al., 2008, \mn@doi [\apj] {10.1086/587800}, \href
  {http://adsabs.harvard.edu/abs/2008ApJ...680..130C} {680, 130}

\bibitem[\protect\citeauthoryear{{Cava} et~al.,}{{Cava} et~al.}{2009}]{Cava+09}
{Cava} A.,  et~al., 2009, \mn@doi [\aap] {10.1051/0004-6361:200810997}, \href
  {http://adsabs.harvard.edu/abs/2009A%26A...495..707C} {495, 707}

\bibitem[\protect\citeauthoryear{{Charlot} \& {Fall}}{{Charlot} \&
  {Fall}}{2000}]{Charlot+00}
{Charlot} S.,  {Fall} S.~M.,  2000, \mn@doi [ApJ] {10.1086/309250}, \href
  {http://adsabs.harvard.edu/abs/2000ApJ...539..718C} {539, 718}

\bibitem[\protect\citeauthoryear{{Cisternas} et~al.,}{{Cisternas}
  et~al.}{2011}]{Cisternas+11}
{Cisternas} M.,  et~al., 2011, \mn@doi [ApJ] {10.1088/0004-637X/726/2/57},
  \href {http://adsabs.harvard.edu/abs/2011ApJ...726...57C} {726, 57}

\bibitem[\protect\citeauthoryear{{Clemens} et~al.,}{{Clemens}
  et~al.}{2013}]{Clemens+13}
{Clemens} M.~S.,  et~al., 2013, \mn@doi [MNRAS] {10.1093/mnras/stt760}, \href
  {http://adsabs.harvard.edu/abs/2013MNRAS.433..695C} {433, 695}

\bibitem[\protect\citeauthoryear{{Cole}, {Lacey}, {Baugh}  \& {Frenk}}{{Cole}
  et~al.}{2000}]{Cole+00}
{Cole} S.,  {Lacey} C.~G.,  {Baugh} C.~M.,   {Frenk} C.~S.,  2000, \mn@doi
  [MNRAS] {10.1046/j.1365-8711.2000.03879.x}, \href
  {http://adsabs.harvard.edu/abs/2000MNRAS.319..168C} {319, 168}

\bibitem[\protect\citeauthoryear{{Colless} \& {et al.}}{{Colless} \& {et
  al.}}{2001}]{Colless+01}
{Colless} M.,  {et al.} 2001, \mn@doi [\mnras]
  {10.1046/j.1365-8711.2001.04902.x}, \href
  {http://adsabs.harvard.edu/cgi-bin/nph-bib_query?bibcode=2001MNRAS.328.1039C&db_key=AST}
  {328, 1039}

\bibitem[\protect\citeauthoryear{{Conselice}}{{Conselice}}{2003}]{Conselice+03a}
{Conselice} C.~J.,  2003, \mn@doi [\apjs] {10.1086/375001}, \href
  {http://adsabs.harvard.edu/abs/2003ApJS..147....1C} {147, 1}

\bibitem[\protect\citeauthoryear{{Cotini}, {Ripamonti}, {Caccianiga}, {Colpi},
  {Della Ceca}, {Mapelli}, {Severgnini}  \& {Segreto}}{{Cotini}
  et~al.}{2013}]{Cotini+13}
{Cotini} S.,  {Ripamonti} E.,  {Caccianiga} A.,  {Colpi} M.,  {Della Ceca} R.,
  {Mapelli} M.,  {Severgnini} P.,   {Segreto} A.,  2013, \mn@doi [MNRAS]
  {10.1093/mnras/stt358}, \href
  {http://adsabs.harvard.edu/abs/2013MNRAS.431.2661C} {431, 2661}

\bibitem[\protect\citeauthoryear{{Cox}, {Jonsson}, {Somerville}, {Primack}  \&
  {Dekel}}{{Cox} et~al.}{2008}]{Cox+08}
{Cox} T.~J.,  {Jonsson} P.,  {Somerville} R.~S.,  {Primack} J.~R.,   {Dekel}
  A.,  2008, \mn@doi [MNRAS] {10.1111/j.1365-2966.2007.12730.x}, \href
  {http://adsabs.harvard.edu/abs/2008MNRAS.384..386C} {384, 386}

\bibitem[\protect\citeauthoryear{{Darg} et~al.,}{{Darg}
  et~al.}{2010}]{Darg+10a}
{Darg} D.~W.,  et~al., 2010, \mn@doi [\mnras]
  {10.1111/j.1365-2966.2009.15686.x}, \href
  {http://adsabs.harvard.edu/abs/2010MNRAS.401.1043D} {401, 1043}

\bibitem[\protect\citeauthoryear{{Davoust} \& {Considere}}{{Davoust} \&
  {Considere}}{1995}]{Davoust+95}
{Davoust} E.,  {Considere} S.,  1995, \aaps, \href
  {http://adsabs.harvard.edu/abs/1995A%26AS..110...19D} {110, 19}

\bibitem[\protect\citeauthoryear{{Davoust} \& {Contini}}{{Davoust} \&
  {Contini}}{2004}]{Davoust+04}
{Davoust} E.,  {Contini} T.,  2004, \mn@doi [\aap]
  {10.1051/0004-6361:20031726}, \href
  {http://adsabs.harvard.edu/abs/2004A%26A...416..515D} {416, 515}

\bibitem[\protect\citeauthoryear{{Debuhr}, {Quataert}  \& {Ma}}{{Debuhr}
  et~al.}{2011}]{DeBuhr+11}
{Debuhr} J.,  {Quataert} E.,   {Ma} C.-P.,  2011, \mn@doi [MNRAS]
  {10.1111/j.1365-2966.2010.17992.x}, \href
  {http://adsabs.harvard.edu/abs/2011MNRAS.412.1341D} {412, 1341}

\bibitem[\protect\citeauthoryear{{Dekel} \& {Birnboim}}{{Dekel} \&
  {Birnboim}}{2006}]{Dekel+06}
{Dekel} A.,  {Birnboim} Y.,  2006, \mn@doi [MNRAS]
  {10.1111/j.1365-2966.2006.10145.x}, \href
  {http://adsabs.harvard.edu/abs/2006MNRAS.368....2D} {368, 2}

\bibitem[\protect\citeauthoryear{{Del Moro} et~al.,}{{Del Moro}
  et~al.}{2016}]{DelMoro+16}
{Del Moro} A.,  et~al., 2016, \mn@doi [\mnras] {10.1093/mnras/stv2748}, \href
  {http://adsabs.harvard.edu/abs/2016MNRAS.456.2105D} {456, 2105}

\bibitem[\protect\citeauthoryear{{Di Matteo}, {Springel}  \& {Hernquist}}{{Di
  Matteo} et~al.}{2005}]{DiMatteo+05}
{Di Matteo} T.,  {Springel} V.,   {Hernquist} L.,  2005, \mn@doi [Nat]
  {10.1038/nature03335}, \href
  {http://adsabs.harvard.edu/abs/2005Natur.433..604D} {433, 604}

\bibitem[\protect\citeauthoryear{{Di Matteo}, {Colberg}, {Springel},
  {Hernquist}  \& {Sijacki}}{{Di Matteo} et~al.}{2008}]{DiMatteo+08}
{Di Matteo} T.,  {Colberg} J.,  {Springel} V.,  {Hernquist} L.,   {Sijacki} D.,
   2008, \mn@doi [\apj] {10.1086/524921}, \href
  {http://adsabs.harvard.edu/abs/2008ApJ...676...33D} {676, 33}

\bibitem[\protect\citeauthoryear{{Domingue}, {Xu}, {Jarrett}  \&
  {Cheng}}{{Domingue} et~al.}{2009}]{Domingue+09}
{Domingue} D.~L.,  {Xu} C.~K.,  {Jarrett} T.~H.,   {Cheng} Y.,  2009, \mn@doi
  [\apj] {10.1088/0004-637X/695/2/1559}, \href
  {http://adsabs.harvard.edu/abs/2009ApJ...695.1559D} {695, 1559}

\bibitem[\protect\citeauthoryear{{Donoso} et~al.,}{{Donoso}
  et~al.}{2012}]{Donoso+12}
{Donoso} E.,  et~al., 2012, \mn@doi [ApJ] {10.1088/0004-637X/748/2/80}, \href
  {http://adsabs.harvard.edu/abs/2012ApJ...748...80D} {748, 80}

\bibitem[\protect\citeauthoryear{{Donoso}, {Yan}, {Stern}  \& {Assef}}{{Donoso}
  et~al.}{2014}]{Donoso+14}
{Donoso} E.,  {Yan} L.,  {Stern} D.,   {Assef} R.~J.,  2014, \mn@doi [ApJ]
  {10.1088/0004-637X/789/1/44}, \href
  {http://adsabs.harvard.edu/abs/2014ApJ...789...44D} {789, 44}

\bibitem[\protect\citeauthoryear{{Ellison}, {Patton}, {Simard}  \&
  {McConnachie}}{{Ellison} et~al.}{2008}]{Ellison+08}
{Ellison} S.~L.,  {Patton} D.~R.,  {Simard} L.,   {McConnachie} A.~W.,  2008,
  \mn@doi [AJ] {10.1088/0004-6256/135/5/1877}, \href
  {http://adsabs.harvard.edu/abs/2008AJ....135.1877E} {135, 1877}

\bibitem[\protect\citeauthoryear{{Ellison}, {Patton}, {Mendel}  \&
  {Scudder}}{{Ellison} et~al.}{2011}]{Ellison+11}
{Ellison} S.~L.,  {Patton} D.~R.,  {Mendel} J.~T.,   {Scudder} J.~M.,  2011,
  \mn@doi [MNRAS] {10.1111/j.1365-2966.2011.19624.x}, \href
  {http://adsabs.harvard.edu/abs/2011MNRAS.418.2043E} {418, 2043}

\bibitem[\protect\citeauthoryear{{Ellison}, {Mendel}, {Patton}  \&
  {Scudder}}{{Ellison} et~al.}{2013}]{Ellison+13b}
{Ellison} S.~L.,  {Mendel} J.~T.,  {Patton} D.~R.,   {Scudder} J.~M.,  2013,
  \mn@doi [MNRAS] {10.1093/mnras/stt1562}, \href
  {http://adsabs.harvard.edu/abs/2013MNRAS.435.3627E} {435, 3627}

\bibitem[\protect\citeauthoryear{{Fairall} et~al.,}{{Fairall}
  et~al.}{1992}]{Fairall+92}
{Fairall} A.~P.,  et~al., 1992, \mn@doi [\aj] {10.1086/116037}, \href
  {http://adsabs.harvard.edu/abs/1992AJ....103...11F} {103, 11}

\bibitem[\protect\citeauthoryear{{Falco} et~al.,}{{Falco}
  et~al.}{1999}]{Falco+99}
{Falco} E.~E.,  et~al., 1999, \mn@doi [\pasp] {10.1086/316343}, \href
  {http://adsabs.harvard.edu/abs/1999PASP..111..438F} {111, 438}

\bibitem[\protect\citeauthoryear{{Fan}, {Fang}, {Chen}, {Li}, {Lv}, {Knudsen}
  \& {Kong}}{{Fan} et~al.}{2014}]{Fan+14}
{Fan} L.,  {Fang} G.,  {Chen} Y.,  {Li} J.,  {Lv} X.,  {Knudsen} K.~K.,
  {Kong} X.,  2014, \mn@doi [ApJ] {10.1088/2041-8205/784/1/L9}, \href
  {http://adsabs.harvard.edu/abs/2014ApJ...784L...9F} {784, L9}

\bibitem[\protect\citeauthoryear{{Feldmann}, {Mayer}  \& {Carollo}}{{Feldmann}
  et~al.}{2008}]{Feldmann+08}
{Feldmann} R.,  {Mayer} L.,   {Carollo} C.~M.,  2008, \mn@doi [\apj]
  {10.1086/590235}, \href {http://adsabs.harvard.edu/abs/2008ApJ...684.1062F}
  {684, 1062}

\bibitem[\protect\citeauthoryear{{Fisher}, {Huchra}, {Strauss}, {Davis},
  {Yahil}  \& {Schlegel}}{{Fisher} et~al.}{1995}]{Fisher+95}
{Fisher} K.~B.,  {Huchra} J.~P.,  {Strauss} M.~A.,  {Davis} M.,  {Yahil} A.,
  {Schlegel} D.,  1995, \mn@doi [\apjs] {10.1086/192208}, \href
  {http://adsabs.harvard.edu/abs/1995ApJS..100...69F} {100, 69}

\bibitem[\protect\citeauthoryear{{Francis}, {Nelson}  \& {Cutri}}{{Francis}
  et~al.}{2004}]{Francis+04}
{Francis} P.~J.,  {Nelson} B.~O.,   {Cutri} R.~M.,  2004, \mn@doi [\aj]
  {10.1086/380939}, \href {http://adsabs.harvard.edu/abs/2004AJ....127..646F}
  {127, 646}

\bibitem[\protect\citeauthoryear{{Goulding} \& {Alexander}}{{Goulding} \&
  {Alexander}}{2009}]{Goulding+09}
{Goulding} A.~D.,  {Alexander} D.~M.,  2009, \mn@doi [MNRAS]
  {10.1111/j.1365-2966.2009.15194.x}, \href
  {http://adsabs.harvard.edu/abs/2009MNRAS.398.1165G} {398, 1165}

\bibitem[\protect\citeauthoryear{{Goulding}, {Alexander}, {Lehmer}  \&
  {Mullaney}}{{Goulding} et~al.}{2010}]{Goulding+10}
{Goulding} A.~D.,  {Alexander} D.~M.,  {Lehmer} B.~D.,   {Mullaney} J.~R.,
  2010, \mn@doi [\mnras] {10.1111/j.1365-2966.2010.16700.x}, \href
  {http://adsabs.harvard.edu/abs/2010MNRAS.406..597G} {406, 597}

\bibitem[\protect\citeauthoryear{{Goulding}, {Alexander}, {Mullaney},
  {Gelbord}, {Hickox}, {Ward}  \& {Watson}}{{Goulding}
  et~al.}{2011}]{Goulding+11}
{Goulding} A.~D.,  {Alexander} D.~M.,  {Mullaney} J.~R.,  {Gelbord} J.~M.,
  {Hickox} R.~C.,  {Ward} M.,   {Watson} M.~G.,  2011, \mn@doi [MNRAS]
  {10.1111/j.1365-2966.2010.17755.x}, \href
  {http://adsabs.harvard.edu/abs/2011MNRAS.411.1231G} {411, 1231}

\bibitem[\protect\citeauthoryear{{Goulding}, {Alexander}, {Bauer}, {Forman},
  {Hickox}, {Jones}, {Mullaney}  \& {Trichas}}{{Goulding}
  et~al.}{2012}]{Goulding+12}
{Goulding} A.~D.,  {Alexander} D.~M.,  {Bauer} F.~E.,  {Forman} W.~R.,
  {Hickox} R.~C.,  {Jones} C.,  {Mullaney} J.~R.,   {Trichas} M.,  2012,
  \mn@doi [ApJ] {10.1088/0004-637X/755/1/5}, \href
  {http://adsabs.harvard.edu/abs/2012ApJ...755....5G} {755, 5}

\bibitem[\protect\citeauthoryear{{Granato}, {De Zotti}, {Silva}, {Bressan}  \&
  {Danese}}{{Granato} et~al.}{2004}]{Granato+14}
{Granato} G.~L.,  {De Zotti} G.,  {Silva} L.,  {Bressan} A.,   {Danese} L.,
  2004, \mn@doi [ApJ] {10.1086/379875}, \href
  {http://adsabs.harvard.edu/abs/2004ApJ...600..580G} {600, 580}

\bibitem[\protect\citeauthoryear{{Grogin}, {Geller}  \& {Huchra}}{{Grogin}
  et~al.}{1998}]{Grogin+98}
{Grogin} N.~A.,  {Geller} M.~J.,   {Huchra} J.~P.,  1998, \mn@doi [\apjs]
  {10.1086/313164}, \href {http://adsabs.harvard.edu/abs/1998ApJS..119..277G}
  {119, 277}

\bibitem[\protect\citeauthoryear{{Gronwall}, {Salzer}, {Sarajedini}, {Jangren},
  {Chomiuk}, {Moody}, {Frattare}  \& {Boroson}}{{Gronwall}
  et~al.}{2004}]{Gronwall+04}
{Gronwall} C.,  {Salzer} J.~J.,  {Sarajedini} V.~L.,  {Jangren} A.,  {Chomiuk}
  L.,  {Moody} J.~W.,  {Frattare} L.~M.,   {Boroson} T.~A.,  2004, \mn@doi
  [\aj] {10.1086/382717}, \href
  {http://adsabs.harvard.edu/abs/2004AJ....127.1943G} {127, 1943}

\bibitem[\protect\citeauthoryear{{Grupe}, {Beuermann}, {Mannheim}  \&
  {Thomas}}{{Grupe} et~al.}{1999}]{Grupe+99}
{Grupe} D.,  {Beuermann} K.,  {Mannheim} K.,   {Thomas} H.-C.,  1999, \aap,
  \href {http://adsabs.harvard.edu/abs/1999A%26A...350..805G} {350, 805}

\bibitem[\protect\citeauthoryear{{G{\"u}rkan}, {Hardcastle}  \&
  {Jarvis}}{{G{\"u}rkan} et~al.}{2014}]{Gurkan+14}
{G{\"u}rkan} G.,  {Hardcastle} M.~J.,   {Jarvis} M.~J.,  2014, \mn@doi [MNRAS]
  {10.1093/mnras/stt2264}, \href
  {http://adsabs.harvard.edu/abs/2014MNRAS.438.1149G} {438, 1149}

\bibitem[\protect\citeauthoryear{{Haan}, {Schinnerer}, {Emsellem},
  {Garc{\'{\i}}a-Burillo}, {Combes}, {Mundell}  \& {Rix}}{{Haan}
  et~al.}{2009}]{Haan+09}
{Haan} S.,  {Schinnerer} E.,  {Emsellem} E.,  {Garc{\'{\i}}a-Burillo} S.,
  {Combes} F.,  {Mundell} C.~G.,   {Rix} H.-W.,  2009, \mn@doi [ApJ]
  {10.1088/0004-637X/692/2/1623}, \href
  {http://adsabs.harvard.edu/abs/2009ApJ...692.1623H} {692, 1623}

\bibitem[\protect\citeauthoryear{{Haines}, {McIntosh}, {S{\'a}nchez},
  {Tremonti}  \& {Rudnick}}{{Haines} et~al.}{2015}]{Haines+15}
{Haines} T.,  {McIntosh} D.~H.,  {S{\'a}nchez} S.~F.,  {Tremonti} C.,
  {Rudnick} G.,  2015, preprint, \href
  {http://adsabs.harvard.edu/abs/2015arXiv150501493H} {} (\mn@eprint {arXiv}
  {1505.01493})

\bibitem[\protect\citeauthoryear{{Heckman}, {Kauffmann}, {Brinchmann},
  {Charlot}, {Tremonti}  \& {White}}{{Heckman} et~al.}{2004}]{Heckman+04}
{Heckman} T.~M.,  {Kauffmann} G.,  {Brinchmann} J.,  {Charlot} S.,  {Tremonti}
  C.,   {White} S.~D.~M.,  2004, \mn@doi [ApJ] {10.1086/422872}, \href
  {http://adsabs.harvard.edu/abs/2004ApJ...613..109H} {613, 109}

\bibitem[\protect\citeauthoryear{{Hickox} et~al.,}{{Hickox}
  et~al.}{2009}]{Hickox+09}
{Hickox} R.~C.,  et~al., 2009, \mn@doi [\apj] {10.1088/0004-637X/696/1/891},
  \href {http://adsabs.harvard.edu/abs/2009ApJ...696..891H} {696, 891}

\bibitem[\protect\citeauthoryear{{Hickox}, {Mullaney}, {Alexander}, {Chen},
  {Civano}, {Goulding}  \& {Hainline}}{{Hickox} et~al.}{2014}]{Hickox+14}
{Hickox} R.~C.,  {Mullaney} J.~R.,  {Alexander} D.~M.,  {Chen} C.-T.~J.,
  {Civano} F.~M.,  {Goulding} A.~D.,   {Hainline} K.~N.,  2014, \mn@doi [ApJ]
  {10.1088/0004-637X/782/1/9}, \href
  {http://adsabs.harvard.edu/abs/2014ApJ...782....9H} {782, 9}

\bibitem[\protect\citeauthoryear{{Hickson}, {Mendes de Oliveira}, {Huchra}  \&
  {Palumbo}}{{Hickson} et~al.}{1992}]{Hickson+92}
{Hickson} P.,  {Mendes de Oliveira} C.,  {Huchra} J.~P.,   {Palumbo} G.~G.,
  1992, \mn@doi [\apj] {10.1086/171932}, \href
  {http://adsabs.harvard.edu/abs/1992ApJ...399..353H} {399, 353}

\bibitem[\protect\citeauthoryear{{Hill} \& {Oegerle}}{{Hill} \&
  {Oegerle}}{1998}]{Hill+98}
{Hill} J.~M.,  {Oegerle} W.~R.,  1998, \mn@doi [\aj] {10.1086/300575}, \href
  {http://adsabs.harvard.edu/abs/1998AJ....116.1529H} {116, 1529}

\bibitem[\protect\citeauthoryear{{Hogg} et~al.,}{{Hogg} et~al.}{2004}]{Hogg+04}
{Hogg} D.~W.,  et~al., 2004, \apjl, \href
  {http://adsabs.harvard.edu/cgi-bin/nph-bib_query?bibcode=2004ApJ...601L..29H&amp;db_key=AST}
  {601, L29}

\bibitem[\protect\citeauthoryear{{Holden}, {van der Wel}, {Rix}  \&
  {Franx}}{{Holden} et~al.}{2012}]{Holden+12}
{Holden} B.~P.,  {van der Wel} A.,  {Rix} H.-W.,   {Franx} M.,  2012, \mn@doi
  [ApJ] {10.1088/0004-637X/749/2/96}, \href
  {http://adsabs.harvard.edu/abs/2012ApJ...749...96H} {749, 96}

\bibitem[\protect\citeauthoryear{{Hopkins} \& {Elvis}}{{Hopkins} \&
  {Elvis}}{2010}]{Hopkins+10a}
{Hopkins} P.~F.,  {Elvis} M.,  2010, \mn@doi [MNRAS]
  {10.1111/j.1365-2966.2009.15643.x}, \href
  {http://adsabs.harvard.edu/abs/2010MNRAS.401....7H} {401, 7}

\bibitem[\protect\citeauthoryear{{Hopkins}, {Hernquist}, {Cox}, {Di Matteo},
  {Robertson}  \& {Springel}}{{Hopkins} et~al.}{2006}]{Hopkins+06}
{Hopkins} P.~F.,  {Hernquist} L.,  {Cox} T.~J.,  {Di Matteo} T.,  {Robertson}
  B.,   {Springel} V.,  2006, \mn@doi [ApJS] {10.1086/499298}, \href
  {http://adsabs.harvard.edu/abs/2006ApJS..163....1H} {163, 1}

\bibitem[\protect\citeauthoryear{{Hopkins}, {Hernquist}, {Cox}  \& {Kere{\v
  s}}}{{Hopkins} et~al.}{2008}]{Hopkins+08}
{Hopkins} P.~F.,  {Hernquist} L.,  {Cox} T.~J.,   {Kere{\v s}} D.,  2008,
  \mn@doi [ApJS] {10.1086/524362}, \href
  {http://adsabs.harvard.edu/abs/2008ApJS..175..356H} {175, 356}

\bibitem[\protect\citeauthoryear{{Hopkins}, {Cox}, {Younger}  \&
  {Hernquist}}{{Hopkins} et~al.}{2009}]{Hopkins+09}
{Hopkins} P.~F.,  {Cox} T.~J.,  {Younger} J.~D.,   {Hernquist} L.,  2009,
  \mn@doi [ApJ] {10.1088/0004-637X/691/2/1168}, \href
  {http://adsabs.harvard.edu/abs/2009ApJ...691.1168H} {691, 1168}

\bibitem[\protect\citeauthoryear{{Hopkins} et~al.,}{{Hopkins}
  et~al.}{2010}]{Hopkins+10b}
{Hopkins} P.~F.,  et~al., 2010, \mn@doi [\apj] {10.1088/0004-637X/715/1/202},
  \href {http://adsabs.harvard.edu/abs/2010ApJ...715..202H} {715, 202}

\bibitem[\protect\citeauthoryear{{Huchra}, {Vogeley}  \& {Geller}}{{Huchra}
  et~al.}{1999}]{Huchra+99}
{Huchra} J.~P.,  {Vogeley} M.~S.,   {Geller} M.~J.,  1999, \mn@doi [\apjs]
  {10.1086/313194}, \href {http://adsabs.harvard.edu/abs/1999ApJS..121..287H}
  {121, 287}

\bibitem[\protect\citeauthoryear{{Ivezi{\'c}} et~al.,}{{Ivezi{\'c}}
  et~al.}{2002}]{Ivezic+02}
{Ivezi{\'c}} {\v Z}.,  et~al., 2002, \mn@doi [\aj] {10.1086/344069}, \href
  {http://adsabs.harvard.edu/abs/2002AJ....124.2364I} {124, 2364}

\bibitem[\protect\citeauthoryear{{Jarrett} et~al.,}{{Jarrett}
  et~al.}{2011}]{Jarrett+11}
{Jarrett} T.~H.,  et~al., 2011, \mn@doi [ApJ] {10.1088/0004-637X/735/2/112},
  \href {http://adsabs.harvard.edu/abs/2011ApJ...735..112J} {735, 112}

\bibitem[\protect\citeauthoryear{{Johansson}, {Naab}  \& {Burkert}}{{Johansson}
  et~al.}{2009}]{Johansson+09}
{Johansson} P.~H.,  {Naab} T.,   {Burkert} A.,  2009, \mn@doi [\apj]
  {10.1088/0004-637X/690/1/802}, \href
  {http://adsabs.harvard.edu/abs/2009ApJ...690..802J} {690, 802}

\bibitem[\protect\citeauthoryear{{Jones} et~al.,}{{Jones}
  et~al.}{2009}]{Jones+09}
{Jones} D.~H.,  et~al., 2009, \mn@doi [\mnras]
  {10.1111/j.1365-2966.2009.15338.x}, \href
  {http://adsabs.harvard.edu/abs/2009MNRAS.399..683J} {399, 683}

\bibitem[\protect\citeauthoryear{{Karachentsev}, {Lebedev}  \&
  {Shcherbanovski}}{{Karachentsev} et~al.}{1985}]{Karachentsev+85}
{Karachentsev} I.,  {Lebedev} V.,   {Shcherbanovski} A.,  1985, Bulletin
  d'Information du Centre de Donnees Stellaires, \href
  {http://adsabs.harvard.edu/abs/1985BICDS..29...87K} {29, 87}

\bibitem[\protect\citeauthoryear{{Karouzos}, {Britzen}, {Eckart}, {Witzel}  \&
  {Zensus}}{{Karouzos} et~al.}{2010}]{Karouzos+10}
{Karouzos} M.,  {Britzen} S.,  {Eckart} A.,  {Witzel} A.,   {Zensus} A.,  2010,
  \mn@doi [A\&A] {10.1051/0004-6361/200913550}, \href
  {http://www.aanda.org/articles/aa/pdf/2010/11/aa13550-09.pdf} {519, A62}

\bibitem[\protect\citeauthoryear{{Kartaltepe} et~al.,}{{Kartaltepe}
  et~al.}{2010}]{Kartaltepe+10}
{Kartaltepe} J.~S.,  et~al., 2010, \mn@doi [\apj]
  {10.1088/0004-637X/709/2/572}, \href
  {http://adsabs.harvard.edu/abs/2010ApJ...709..572K} {709, 572}

\bibitem[\protect\citeauthoryear{{Kauffmann} \& {Haehnelt}}{{Kauffmann} \&
  {Haehnelt}}{2000}]{Kauffmann+00}
{Kauffmann} G.,  {Haehnelt} M.,  2000, \mn@doi [MNRAS]
  {10.1046/j.1365-8711.2000.03077.x}, \href
  {http://adsabs.harvard.edu/abs/2000MNRAS.311..576K} {311, 576}

\bibitem[\protect\citeauthoryear{{Kauffmann}, {White}  \&
  {Guiderdoni}}{{Kauffmann} et~al.}{1993}]{Kauffmann+93}
{Kauffmann} G.,  {White} S.~D.~M.,   {Guiderdoni} B.,  1993, MNRAS, \href
  {http://adsabs.harvard.edu/abs/1993MNRAS.264..201K} {264, 201}

\bibitem[\protect\citeauthoryear{{Kauffmann} et~al.,}{{Kauffmann}
  et~al.}{2003}]{Kauffmann+03}
{Kauffmann} G.,  et~al., 2003, \mn@doi [MNRAS]
  {10.1111/j.1365-2966.2003.07154.x}, \href
  {http://adsabs.harvard.edu/abs/2003MNRAS.346.1055K} {346, 1055}

\bibitem[\protect\citeauthoryear{{Kaviraj}, {Peirani}, {Khochfar}, {Silk}  \&
  {Kay}}{{Kaviraj} et~al.}{2009}]{Kaviraj+09}
{Kaviraj} S.,  {Peirani} S.,  {Khochfar} S.,  {Silk} J.,   {Kay} S.,  2009,
  \mn@doi [\mnras] {10.1111/j.1365-2966.2009.14403.x}, \href
  {http://adsabs.harvard.edu/abs/2009MNRAS.394.1713K} {394, 1713}

\bibitem[\protect\citeauthoryear{{Kaviraj}, {Schawinski}, {Silk}  \&
  {Shabala}}{{Kaviraj} et~al.}{2011}]{Kaviraj+11}
{Kaviraj} S.,  {Schawinski} K.,  {Silk} J.,   {Shabala} S.~S.,  2011, \mn@doi
  [MNRAS] {10.1111/j.1365-2966.2011.19002.x}, \href
  {http://adsabs.harvard.edu/abs/2011MNRAS.415.3798K} {415, 3798}

\bibitem[\protect\citeauthoryear{{Keel}}{{Keel}}{1996}]{Keel+96}
{Keel} W.~C.,  1996, \mn@doi [\apjs] {10.1086/192326}, \href
  {http://adsabs.harvard.edu/abs/1996ApJS..106...27K} {106, 27}

\bibitem[\protect\citeauthoryear{{Kehrig} et~al.,}{{Kehrig}
  et~al.}{2012}]{Kehrig+12}
{Kehrig} C.,  et~al., 2012, \mn@doi [\aap] {10.1051/0004-6361/201118357}, \href
  {http://adsabs.harvard.edu/abs/2012A%26A...540A..11K} {540, A11}

\bibitem[\protect\citeauthoryear{{Kennicutt}}{{Kennicutt}}{1998}]{Kennicutt+98}
{Kennicutt} Jr. R.~C.,  1998, \mn@doi [RA\&A] {10.1146/annurev.astro.36.1.189},
  \href {http://www.annualreviews.org/doi/pdf/10.1146/annurev.astro.36.1.189}
  {36, 189}

\bibitem[\protect\citeauthoryear{{Kennicutt}, {Roettiger}, {Keel}, {van der
  Hulst}  \& {Hummel}}{{Kennicutt} et~al.}{1987}]{Kennicutt+87}
{Kennicutt} Jr. R.~C.,  {Roettiger} K.~A.,  {Keel} W.~C.,  {van der Hulst}
  J.~M.,   {Hummel} E.,  1987, \mn@doi [AJ] {10.1086/114384}, \href
  {http://adsabs.harvard.edu/abs/1987AJ.....93.1011K} {93, 1011}

\bibitem[\protect\citeauthoryear{{Kere{\v s}}, {Katz}, {Weinberg}  \&
  {Dav{\'e}}}{{Kere{\v s}} et~al.}{2005}]{Keres+05}
{Kere{\v s}} D.,  {Katz} N.,  {Weinberg} D.~H.,   {Dav{\'e}} R.,  2005, \mn@doi
  [MNRAS] {10.1111/j.1365-2966.2005.09451.x}, \href
  {http://adsabs.harvard.edu/abs/2005MNRAS.363....2K} {363, 2}

\bibitem[\protect\citeauthoryear{{Kere{\v s}}, {Vogelsberger}, {Sijacki},
  {Springel}  \& {Hernquist}}{{Kere{\v s}} et~al.}{2012}]{Keres+12}
{Kere{\v s}} D.,  {Vogelsberger} M.,  {Sijacki} D.,  {Springel} V.,
  {Hernquist} L.,  2012, \mn@doi [\mnras] {10.1111/j.1365-2966.2012.21548.x},
  \href {http://adsabs.harvard.edu/abs/2012MNRAS.425.2027K} {425, 2027}

\bibitem[\protect\citeauthoryear{{Kewley}, {Dopita}, {Sutherland}, {Heisler}
  \& {Trevena}}{{Kewley} et~al.}{2001}]{Kewley+01}
{Kewley} L.~J.,  {Dopita} M.~A.,  {Sutherland} R.~S.,  {Heisler} C.~A.,
  {Trevena} J.,  2001, \mn@doi [ApJ] {10.1086/321545}, \href
  {http://adsabs.harvard.edu/abs/2001ApJ...556..121K} {556, 121}

\bibitem[\protect\citeauthoryear{{Kirshner}, {Oemler}, {Schechter}  \&
  {Shectman}}{{Kirshner} et~al.}{1983}]{Kirshner+83}
{Kirshner} R.~P.,  {Oemler} Jr. A.,  {Schechter} P.~L.,   {Shectman} S.~A.,
  1983, \mn@doi [\aj] {10.1086/113419}, \href
  {http://adsabs.harvard.edu/abs/1983AJ.....88.1285K} {88, 1285}

\bibitem[\protect\citeauthoryear{{Kocevski} et~al.,}{{Kocevski}
  et~al.}{2012}]{Kocevski+12}
{Kocevski} D.~D.,  et~al., 2012, \mn@doi [\apj] {10.1088/0004-637X/744/2/148},
  \href {http://adsabs.harvard.edu/abs/2012ApJ...744..148K} {744, 148}

\bibitem[\protect\citeauthoryear{{Kocevski} et~al.,}{{Kocevski}
  et~al.}{2015}]{Kocevski+15}
{Kocevski} D.~D.,  et~al., 2015, \mn@doi [\apj] {10.1088/0004-637X/814/2/104},
  \href {http://adsabs.harvard.edu/abs/2015ApJ...814..104K} {814, 104}

\bibitem[\protect\citeauthoryear{{Kopylova} \& {Kopylov}}{{Kopylova} \&
  {Kopylov}}{2001}]{Kopylova+01}
{Kopylova} F.~G.,  {Kopylov} A.~I.,  2001, \mn@doi [Astronomy Letters]
  {10.1134/1.1374671}, \href
  {http://adsabs.harvard.edu/abs/2001AstL...27..345K} {27, 345}

\bibitem[\protect\citeauthoryear{{Koranyi} \& {Geller}}{{Koranyi} \&
  {Geller}}{2002}]{Koranyi+02}
{Koranyi} D.~M.,  {Geller} M.~J.,  2002, \mn@doi [\aj] {10.1086/338096}, \href
  {http://adsabs.harvard.edu/abs/2002AJ....123..100K} {123, 100}

\bibitem[\protect\citeauthoryear{{Koss}, {Mushotzky}, {Veilleux}  \&
  {Winter}}{{Koss} et~al.}{2010}]{Koss+10}
{Koss} M.,  {Mushotzky} R.,  {Veilleux} S.,   {Winter} L.,  2010, \mn@doi
  [\apjl] {10.1088/2041-8205/716/2/L125}, \href
  {http://adsabs.harvard.edu/abs/2010ApJ...716L.125K} {716, L125}

\bibitem[\protect\citeauthoryear{{Labb{\'e}} et~al.,}{{Labb{\'e}}
  et~al.}{2005}]{Labbe+05}
{Labb{\'e}} I.,  et~al., 2005, \mn@doi [\apjl] {10.1086/430700}, \href
  {http://adsabs.harvard.edu/abs/2005ApJ...624L..81L} {624, L81}

\bibitem[\protect\citeauthoryear{{Lake} \& {Dressler}}{{Lake} \&
  {Dressler}}{1986}]{Lake+86}
{Lake} G.,  {Dressler} A.,  1986, \mn@doi [ApJ] {10.1086/164713}, \href
  {http://adsabs.harvard.edu/abs/1986ApJ...310..605L} {310, 605}

\bibitem[\protect\citeauthoryear{{Lambas}, {Tissera}, {Alonso}  \&
  {Coldwell}}{{Lambas} et~al.}{2003}]{Lambas+03}
{Lambas} D.~G.,  {Tissera} P.~B.,  {Alonso} M.~S.,   {Coldwell} G.,  2003,
  \mn@doi [MNRAS] {10.1111/j.1365-2966.2003.07179.x}, \href
  {http://adsabs.harvard.edu/abs/2003MNRAS.346.1189L} {346, 1189}

\bibitem[\protect\citeauthoryear{{Lange}, {van Dokkum}, {Momcheva}, {Nelson},
  {Leja}, {Brammer}, {Whitaker}  \& {Franx}}{{Lange} et~al.}{2016}]{Lange+16}
{Lange} J.~U.,  {van Dokkum} P.~G.,  {Momcheva} I.~G.,  {Nelson} E.~J.,  {Leja}
  J.,  {Brammer} G.,  {Whitaker} K.~E.,   {Franx} M.,  2016, \mn@doi [\apjl]
  {10.3847/2041-8205/819/1/L4}, \href
  {http://adsabs.harvard.edu/abs/2016ApJ...819L...4L} {819, L4}

\bibitem[\protect\citeauthoryear{{Lotz}, {Primack}  \& {Madau}}{{Lotz}
  et~al.}{2004}]{Lotz+04}
{Lotz} J.~M.,  {Primack} J.,   {Madau} P.,  2004, \mn@doi [\aj]
  {10.1086/421849}, \href {http://adsabs.harvard.edu/abs/2004AJ....128..163L}
  {128, 163}

\bibitem[\protect\citeauthoryear{{Lotz}, {Jonsson}, {Cox}  \& {Primack}}{{Lotz}
  et~al.}{2010a}]{Lotz+10b}
{Lotz} J.~M.,  {Jonsson} P.,  {Cox} T.~J.,   {Primack} J.~R.,  2010a, \mn@doi
  [\mnras] {10.1111/j.1365-2966.2010.16268.x}, \href
  {http://adsabs.harvard.edu/abs/2010MNRAS.404..575L} {404, 575}

\bibitem[\protect\citeauthoryear{{Lotz}, {Jonsson}, {Cox}  \& {Primack}}{{Lotz}
  et~al.}{2010b}]{Lotz+10a}
{Lotz} J.~M.,  {Jonsson} P.,  {Cox} T.~J.,   {Primack} J.~R.,  2010b, \mn@doi
  [MNRAS] {10.1111/j.1365-2966.2010.16269.x}, \href
  {http://adsabs.harvard.edu/abs/2010MNRAS.404..590L} {404, 590}

\bibitem[\protect\citeauthoryear{{Lu} \& {Freudling}}{{Lu} \&
  {Freudling}}{1995}]{Lu+95}
{Lu} N.~Y.,  {Freudling} W.,  1995, \mn@doi [\apj] {10.1086/176077}, \href
  {http://adsabs.harvard.edu/abs/1995ApJ...449..527L} {449, 527}

\bibitem[\protect\citeauthoryear{{Magorrian} et~al.,}{{Magorrian}
  et~al.}{1998}]{Magorrian+98}
{Magorrian} J.,  et~al., 1998, \mn@doi [\aj] {10.1086/300353}, \href
  {http://adsabs.harvard.edu/abs/1998AJ....115.2285M} {115, 2285}

\bibitem[\protect\citeauthoryear{{Mahdavi} \& {Geller}}{{Mahdavi} \&
  {Geller}}{2004}]{Mahdavi+04}
{Mahdavi} A.,  {Geller} M.~J.,  2004, \mn@doi [\apj] {10.1086/383458}, \href
  {http://adsabs.harvard.edu/abs/2004ApJ...607..202M} {607, 202}

\bibitem[\protect\citeauthoryear{{Marconi}, {Risaliti}, {Gilli}, {Hunt},
  {Maiolino}  \& {Salvati}}{{Marconi} et~al.}{2004}]{Marconi+04}
{Marconi} A.,  {Risaliti} G.,  {Gilli} R.,  {Hunt} L.~K.,  {Maiolino} R.,
  {Salvati} M.,  2004, \mn@doi [\mnras] {10.1111/j.1365-2966.2004.07765.x},
  \href {http://adsabs.harvard.edu/abs/2004MNRAS.351..169M} {351, 169}

\bibitem[\protect\citeauthoryear{{Mart{\'{\i}}nez-Sansigre} \&
  {Taylor}}{{Mart{\'{\i}}nez-Sansigre} \&
  {Taylor}}{2009}]{Martinez-Sansigre+09}
{Mart{\'{\i}}nez-Sansigre} A.,  {Taylor} A.~M.,  2009, \mn@doi [\apj]
  {10.1088/0004-637X/692/2/964}, \href
  {http://adsabs.harvard.edu/abs/2009ApJ...692..964M} {692, 964}

\bibitem[\protect\citeauthoryear{{Martini} \& {Weinberg}}{{Martini} \&
  {Weinberg}}{2001}]{Martini+01}
{Martini} P.,  {Weinberg} D.~H.,  2001, \mn@doi [\apj] {10.1086/318331}, \href
  {http://adsabs.harvard.edu/abs/2001ApJ...547...12M} {547, 12}

\bibitem[\protect\citeauthoryear{{Martini}, {Dicken}  \&
  {Storchi-Bergmann}}{{Martini} et~al.}{2013}]{Martini+13}
{Martini} P.,  {Dicken} D.,   {Storchi-Bergmann} T.,  2013, \mn@doi [\apj]
  {10.1088/0004-637X/766/2/121}, \href
  {http://adsabs.harvard.edu/abs/2013ApJ...766..121M} {766, 121}

\bibitem[\protect\citeauthoryear{{Marzke}, {Huchra}  \& {Geller}}{{Marzke}
  et~al.}{1996}]{Marzke+96}
{Marzke} R.~O.,  {Huchra} J.~P.,   {Geller} M.~J.,  1996, \mn@doi [\aj]
  {10.1086/118142}, \href {http://adsabs.harvard.edu/abs/1996AJ....112.1803M}
  {112, 1803}

\bibitem[\protect\citeauthoryear{{Masjedi} et~al.,}{{Masjedi}
  et~al.}{2006}]{Masjedi+06}
{Masjedi} M.,  et~al., 2006, \mn@doi [\apj] {10.1086/503536}, \href
  {http://adsabs.harvard.edu/abs/2006ApJ...644...54M} {644, 54}

\bibitem[\protect\citeauthoryear{{Mateos} et~al.,}{{Mateos}
  et~al.}{2012}]{Mateos+12}
{Mateos} S.,  et~al., 2012, \mn@doi [MNRAS] {10.1111/j.1365-2966.2012.21843.x},
  \href {http://adsabs.harvard.edu/abs/2012MNRAS.426.3271M} {426, 3271}

\bibitem[\protect\citeauthoryear{{McIntosh}, {Guo}, {Hertzberg}, {Katz}, {Mo},
  {van den Bosch}  \& {Yang}}{{McIntosh} et~al.}{2008}]{McIntosh+08}
{McIntosh} D.~H.,  {Guo} Y.,  {Hertzberg} J.,  {Katz} N.,  {Mo} H.~J.,  {van
  den Bosch} F.~C.,   {Yang} X.,  2008, \mn@doi [MNRAS]
  {10.1111/j.1365-2966.2008.13531.x}, \href
  {http://adsabs.harvard.edu/abs/2008MNRAS.388.1537M} {388, 1537}

\bibitem[\protect\citeauthoryear{{McIntosh} et~al.,}{{McIntosh}
  et~al.}{2014}]{McIntosh+14}
{McIntosh} D.~H.,  et~al., 2014, \mn@doi [MNRAS] {10.1093/mnras/stu808}, \href
  {http://adsabs.harvard.edu/abs/2014MNRAS.442..533M} {442, 533}

\bibitem[\protect\citeauthoryear{{Mihos} \& {Hernquist}}{{Mihos} \&
  {Hernquist}}{1996}]{Mihos+96}
{Mihos} J.~C.,  {Hernquist} L.,  1996, \mn@doi [\apj] {10.1086/177353}, \href
  {http://adsabs.harvard.edu/abs/1996ApJ...464..641M} {464, 641}

\bibitem[\protect\citeauthoryear{{Miller} \& {Owen}}{{Miller} \&
  {Owen}}{2003}]{Miller+03}
{Miller} N.~A.,  {Owen} F.~N.,  2003, \mn@doi [\aj] {10.1086/374767}, \href
  {http://adsabs.harvard.edu/abs/2003AJ....125.2427M} {125, 2427}

\bibitem[\protect\citeauthoryear{{Miller}, {Krughoff}, {Batuski}  \&
  {Hill}}{{Miller} et~al.}{2002}]{Miller+02}
{Miller} C.~J.,  {Krughoff} K.~S.,  {Batuski} D.~J.,   {Hill} J.~M.,  2002,
  \mn@doi [\aj] {10.1086/342536}, \href
  {http://adsabs.harvard.edu/abs/2002AJ....124.1918M} {124, 1918}

\bibitem[\protect\citeauthoryear{{Nazaryan}, {Petrosian}, {Hakobyan}, {McLean}
  \& {Kunth}}{{Nazaryan} et~al.}{2014}]{Nazaryan+14}
{Nazaryan} T.~A.,  {Petrosian} A.~R.,  {Hakobyan} A.~A.,  {McLean} B.~J.,
  {Kunth} D.,  2014, in IAU Symposium. pp 327--330 (\mn@eprint {arXiv}
  {1312.4733}), \mn@doi{10.1017/S1743921314004189}

\bibitem[\protect\citeauthoryear{{Nelson}, {Vogelsberger}, {Genel}, {Sijacki},
  {Kere{\v s}}, {Springel}  \& {Hernquist}}{{Nelson} et~al.}{2013}]{Nelson+13}
{Nelson} D.,  {Vogelsberger} M.,  {Genel} S.,  {Sijacki} D.,  {Kere{\v s}} D.,
  {Springel} V.,   {Hernquist} L.,  2013, \mn@doi [\mnras]
  {10.1093/mnras/sts595}, \href
  {http://adsabs.harvard.edu/abs/2013MNRAS.429.3353N} {429, 3353}

\bibitem[\protect\citeauthoryear{{Owen}, {White}  \& {Thronson}}{{Owen}
  et~al.}{1988}]{Owen+88}
{Owen} F.~N.,  {White} R.~A.,   {Thronson} Jr. H.~A.,  1988, \mn@doi [\aj]
  {10.1086/114605}, \href {http://adsabs.harvard.edu/abs/1988AJ.....95....1O}
  {95, 1}

\bibitem[\protect\citeauthoryear{{Papovich} et~al.,}{{Papovich}
  et~al.}{2004}]{Papovich+04}
{Papovich} C.,  et~al., 2004, \mn@doi [\apjs] {10.1086/422880}, \href
  {http://adsabs.harvard.edu/abs/2004ApJS..154...70P} {154, 70}

\bibitem[\protect\citeauthoryear{{Papovich} et~al.,}{{Papovich}
  et~al.}{2006}]{Papovich+06}
{Papovich} C.,  et~al., 2006, \mn@doi [\aj] {10.1086/504598}, \href
  {http://adsabs.harvard.edu/abs/2006AJ....132..231P} {132, 231}

\bibitem[\protect\citeauthoryear{{Patton}, {Torrey}, {Ellison}, {Mendel}  \&
  {Scudder}}{{Patton} et~al.}{2013}]{Patton+13}
{Patton} D.~R.,  {Torrey} P.,  {Ellison} S.~L.,  {Mendel} J.~T.,   {Scudder}
  J.~M.,  2013, \mn@doi [MNRAS] {10.1093/mnrasl/slt058}, \href
  {http://adsabs.harvard.edu/abs/2013MNRAS.433L..59P} {433, L59}

\bibitem[\protect\citeauthoryear{{Peterson}, {Ellis}, {Efstathiou}, {Shanks},
  {Bean}, {Fong}  \& {Zen-Long}}{{Peterson} et~al.}{1986}]{Peterson+86}
{Peterson} B.~A.,  {Ellis} R.~S.,  {Efstathiou} G.,  {Shanks} T.,  {Bean}
  A.~J.,  {Fong} R.,   {Zen-Long} Z.,  1986, \mn@doi [\mnras]
  {10.1093/mnras/221.2.233}, \href
  {http://adsabs.harvard.edu/abs/1986MNRAS.221..233P} {221, 233}

\bibitem[\protect\citeauthoryear{{Pier} \& {Krolik}}{{Pier} \&
  {Krolik}}{1992}]{Pier+92}
{Pier} E.~A.,  {Krolik} J.~H.,  1992, \mn@doi [\apj] {10.1086/172042}, \href
  {http://adsabs.harvard.edu/abs/1992ApJ...401...99P} {401, 99}

\bibitem[\protect\citeauthoryear{{Press}, {Vetterling}, {Teukolsky}  \&
  {Flannery}}{{Press} et~al.}{1992}]{Press+92}
{Press} W.~H.,  {Vetterling} W.~T.,  {Teukolsky} S.~A.,   {Flannery} B.~P.,
  1992, {Numerical Recipes in Fortran 77}.
Press Syndicate of the University of Cambridge

\bibitem[\protect\citeauthoryear{{Rines}, {Geller}, {Diaferio}, {Mahdavi},
  {Mohr}  \& {Wegner}}{{Rines} et~al.}{2002}]{Rines+02}
{Rines} K.,  {Geller} M.~J.,  {Diaferio} A.,  {Mahdavi} A.,  {Mohr} J.~J.,
  {Wegner} G.,  2002, \mn@doi [\aj] {10.1086/342344}, \href
  {http://adsabs.harvard.edu/abs/2002AJ....124.1266R} {124, 1266}

\bibitem[\protect\citeauthoryear{{Rines}, {Geller}, {Diaferio}, {Kurtz}  \&
  {Jarrett}}{{Rines} et~al.}{2004}]{Rines+04}
{Rines} K.,  {Geller} M.~J.,  {Diaferio} A.,  {Kurtz} M.~J.,   {Jarrett} T.~H.,
   2004, \mn@doi [\aj] {10.1086/423218}, \href
  {http://adsabs.harvard.edu/abs/2004AJ....128.1078R} {128, 1078}

\bibitem[\protect\citeauthoryear{{Rosario} et~al.,}{{Rosario}
  et~al.}{2015}]{Rosario+15}
{Rosario} D.~J.,  et~al., 2015, \mn@doi [\aap] {10.1051/0004-6361/201423782},
  \href {http://arxiv.org/abs/1409.5122} {573, A85}

\bibitem[\protect\citeauthoryear{{Rothberg} \& {Joseph}}{{Rothberg} \&
  {Joseph}}{2006}]{Rothberg+06}
{Rothberg} B.,  {Joseph} R.~D.,  2006, \mn@doi [AJ] {10.1086/498452}, \href
  {http://adsabs.harvard.edu/abs/2006AJ....131..185R} {131, 185}

\bibitem[\protect\citeauthoryear{{Sabater}, {Best}  \&
  {Argudo-Fern{\'a}ndez}}{{Sabater} et~al.}{2013}]{Sabater+13}
{Sabater} J.,  {Best} P.~N.,   {Argudo-Fern{\'a}ndez} M.,  2013, \mn@doi
  [MNRAS] {10.1093/mnras/sts675}, \href
  {http://adsabs.harvard.edu/abs/2013MNRAS.430..638S} {430, 638}

\bibitem[\protect\citeauthoryear{{Sanders}, {Soifer}, {Elias}, {Madore},
  {Matthews}, {Neugebauer}  \& {Scoville}}{{Sanders}
  et~al.}{1988}]{Sanders+88a}
{Sanders} D.~B.,  {Soifer} B.~T.,  {Elias} J.~H.,  {Madore} B.~F.,  {Matthews}
  K.,  {Neugebauer} G.,   {Scoville} N.~Z.,  1988, \mn@doi [\apj]
  {10.1086/165983}, \href {http://adsabs.harvard.edu/abs/1988ApJ...325...74S}
  {325, 74}

\bibitem[\protect\citeauthoryear{{Sanders}, {Phinney}, {Neugebauer}, {Soifer}
  \& {Matthews}}{{Sanders} et~al.}{1989}]{Sanders+89}
{Sanders} D.~B.,  {Phinney} E.~S.,  {Neugebauer} G.,  {Soifer} B.~T.,
  {Matthews} K.,  1989, \mn@doi [\apj] {10.1086/168094}, \href
  {http://adsabs.harvard.edu/abs/1989ApJ...347...29S} {347, 29}

\bibitem[\protect\citeauthoryear{{Satyapal}, {Ellison}, {McAlpine}, {Hickox},
  {Patton}  \& {Mendel}}{{Satyapal} et~al.}{2014}]{Satyapal+14}
{Satyapal} S.,  {Ellison} S.~L.,  {McAlpine} W.,  {Hickox} R.~C.,  {Patton}
  D.~R.,   {Mendel} J.~T.,  2014, \mn@doi [MNRAS] {10.1093/mnras/stu650}, \href
  {http://adsabs.harvard.edu/abs/2014MNRAS.441.1297S} {441, 1297}

\bibitem[\protect\citeauthoryear{{Saunders} et~al.,}{{Saunders}
  et~al.}{2000}]{Saunders+00}
{Saunders} W.,  et~al., 2000, \mn@doi [\mnras]
  {10.1046/j.1365-8711.2000.03528.x}, \href
  {http://adsabs.harvard.edu/abs/2000MNRAS.317...55S} {317, 55}

\bibitem[\protect\citeauthoryear{{Schawinski}, {Thomas}, {Sarzi}, {Maraston},
  {Kaviraj}, {Joo}, {Yi}  \& {Silk}}{{Schawinski} et~al.}{2007}]{Schawinski+07}
{Schawinski} K.,  {Thomas} D.,  {Sarzi} M.,  {Maraston} C.,  {Kaviraj} S.,
  {Joo} S.-J.,  {Yi} S.~K.,   {Silk} J.,  2007, \mn@doi [MNRAS]
  {10.1111/j.1365-2966.2007.12487.x}, \href
  {http://adsabs.harvard.edu/abs/2007MNRAS.382.1415S} {382, 1415}

\bibitem[\protect\citeauthoryear{{Schawinski}, {Virani}, {Simmons}, {Urry},
  {Treister}, {Kaviraj}  \& {Kushkuley}}{{Schawinski}
  et~al.}{2009}]{Schawinski+09}
{Schawinski} K.,  {Virani} S.,  {Simmons} B.,  {Urry} C.~M.,  {Treister} E.,
  {Kaviraj} S.,   {Kushkuley} B.,  2009, \mn@doi [ApJ]
  {10.1088/0004-637X/692/1/L19}, \href
  {http://adsabs.harvard.edu/abs/2009ApJ...692L..19S} {692, L19}

\bibitem[\protect\citeauthoryear{{Schawinski}, {Koss}, {Berney}  \&
  {Sartori}}{{Schawinski} et~al.}{2015}]{Schawinski+15}
{Schawinski} K.,  {Koss} M.,  {Berney} S.,   {Sartori} L.~F.,  2015, \mn@doi
  [\mnras] {10.1093/mnras/stv1136}, \href
  {http://adsabs.harvard.edu/abs/2015MNRAS.451.2517S} {451, 2517}

\bibitem[\protect\citeauthoryear{{Scott} \& {Kaviraj}}{{Scott} \&
  {Kaviraj}}{2014}]{Scott+14}
{Scott} C.,  {Kaviraj} S.,  2014, \mn@doi [MNRAS] {10.1093/mnras/stt2014},
  \href {http://adsabs.harvard.edu/abs/2014MNRAS.437.2137S} {437, 2137}

\bibitem[\protect\citeauthoryear{{Shao}, {Kauffmann}, {Li}, {Wang}  \&
  {Heckman}}{{Shao} et~al.}{2013}]{Shao+13}
{Shao} L.,  {Kauffmann} G.,  {Li} C.,  {Wang} J.,   {Heckman} T.~M.,  2013,
  \mn@doi [MNRAS] {10.1093/mnras/stt1832}, \href
  {http://adsabs.harvard.edu/abs/2013MNRAS.436.3451S} {436, 3451}

\bibitem[\protect\citeauthoryear{{Sharples}, {Ellis}  \& {Gray}}{{Sharples}
  et~al.}{1988}]{Sharples+88}
{Sharples} R.~M.,  {Ellis} R.~S.,   {Gray} P.~M.,  1988, \mn@doi [\mnras]
  {10.1093/mnras/231.3.479}, \href
  {http://adsabs.harvard.edu/abs/1988MNRAS.231..479S} {231, 479}

\bibitem[\protect\citeauthoryear{{Shectman}, {Landy}, {Oemler}, {Tucker},
  {Lin}, {Kirshner}  \& {Schechter}}{{Shectman} et~al.}{1996}]{Shectman+96}
{Shectman} S.~A.,  {Landy} S.~D.,  {Oemler} A.,  {Tucker} D.~L.,  {Lin} H.,
  {Kirshner} R.~P.,   {Schechter} P.~L.,  1996, \mn@doi [\apj]
  {10.1086/177858}, \href {http://adsabs.harvard.edu/abs/1996ApJ...470..172S}
  {470, 172}

\bibitem[\protect\citeauthoryear{{Shier} \& {Fischer}}{{Shier} \&
  {Fischer}}{1998}]{Shier+98}
{Shier} L.~M.,  {Fischer} J.,  1998, \mn@doi [ApJ] {10.1086/305434}, \href
  {http://adsabs.harvard.edu/abs/1998ApJ...497..163S} {497, 163}

\bibitem[\protect\citeauthoryear{{Slinglend}, {Batuski}, {Miller}, {Haase},
  {Michaud}  \& {Hill}}{{Slinglend} et~al.}{1998}]{Slinglend+98}
{Slinglend} K.,  {Batuski} D.,  {Miller} C.,  {Haase} S.,  {Michaud} K.,
  {Hill} J.~M.,  1998, \mn@doi [\apjs] {10.1086/313079}, \href
  {http://adsabs.harvard.edu/abs/1998ApJS..115....1S} {115, 1}

\bibitem[\protect\citeauthoryear{{Small}, {Sargent}  \& {Hamilton}}{{Small}
  et~al.}{1997}]{Small+97}
{Small} T.~A.,  {Sargent} W.~L.~W.,   {Hamilton} D.,  1997, \mn@doi [\apjs]
  {10.1086/313009}, \href {http://adsabs.harvard.edu/abs/1997ApJS..111....1S}
  {111, 1}

\bibitem[\protect\citeauthoryear{{Snyder}, {Hayward}, {Sajina}, {Jonsson},
  {Cox}, {Hernquist}, {Hopkins}  \& {Yan}}{{Snyder} et~al.}{2013}]{Snyder+13}
{Snyder} G.~F.,  {Hayward} C.~C.,  {Sajina} A.,  {Jonsson} P.,  {Cox} T.~J.,
  {Hernquist} L.,  {Hopkins} P.~F.,   {Yan} L.,  2013, \mn@doi [\apj]
  {10.1088/0004-637X/768/2/168}, \href
  {http://adsabs.harvard.edu/abs/2013ApJ...768..168S} {768, 168}

\bibitem[\protect\citeauthoryear{{Springel} \& {Hernquist}}{{Springel} \&
  {Hernquist}}{2005}]{Springel-Hernquist+05}
{Springel} V.,  {Hernquist} L.,  2005, \mn@doi [ApJ] {10.1086/429486}, \href
  {http://adsabs.harvard.edu/abs/2005ApJ...622L...9S} {622, L9}

\bibitem[\protect\citeauthoryear{{Springel}, {Di Matteo}  \&
  {Hernquist}}{{Springel} et~al.}{2005}]{Springel+05}
{Springel} V.,  {Di Matteo} T.,   {Hernquist} L.,  2005, \mn@doi [MNRAS]
  {10.1111/j.1365-2966.2005.09238.x}, \href
  {http://adsabs.harvard.edu/abs/2005MNRAS.361..776S} {361, 776}

\bibitem[\protect\citeauthoryear{{Springob}, {Haynes}, {Giovanelli}  \&
  {Kent}}{{Springob} et~al.}{2005}]{Springob+05}
{Springob} C.~M.,  {Haynes} M.~P.,  {Giovanelli} R.,   {Kent} B.~R.,  2005,
  \mn@doi [\apjs] {10.1086/431550}, \href
  {http://adsabs.harvard.edu/abs/2005ApJS..160..149S} {160, 149}

\bibitem[\protect\citeauthoryear{{Stern} et~al.,}{{Stern}
  et~al.}{2005}]{Stern+05}
{Stern} D.,  et~al., 2005, \mn@doi [ApJ] {10.1086/432523}, \href
  {http://adsabs.harvard.edu/abs/2005ApJ...631..163S} {631, 163}

\bibitem[\protect\citeauthoryear{{Stern} et~al.,}{{Stern}
  et~al.}{2012}]{Stern+12}
{Stern} D.,  et~al., 2012, \mn@doi [ApJ] {10.1088/0004-637X/753/1/30}, \href
  {http://adsabs.harvard.edu/abs/2012ApJ...753...30S} {753, 30}

\bibitem[\protect\citeauthoryear{{Storchi-Bergmann}, {Gonz{\'a}lez Delgado},
  {Schmitt}, {Cid Fernandes}  \& {Heckman}}{{Storchi-Bergmann}
  et~al.}{2001}]{Storchi-Bergmann+01}
{Storchi-Bergmann} T.,  {Gonz{\'a}lez Delgado} R.~M.,  {Schmitt} H.~R.,  {Cid
  Fernandes} R.,   {Heckman} T.,  2001, \mn@doi [ApJ] {10.1086/322290}, \href
  {http://adsabs.harvard.edu/abs/2001ApJ...559..147S} {559, 147}

\bibitem[\protect\citeauthoryear{{Strauss} et~al.,}{{Strauss}
  et~al.}{2002}]{Strauss+02}
{Strauss} M.~A.,  et~al., 2002, \mn@doi [AJ] {10.1086/342343}, \href
  {http://adsabs.harvard.edu/abs/2002AJ....124.1810S} {124, 1810}

\bibitem[\protect\citeauthoryear{{Taylor}}{{Taylor}}{2005}]{Taylor+05}
{Taylor} M.~B.,  2005, in {Shopbell} P.,  {Britton} M.,   {Ebert} R.,  eds,
  Astronomical Society of the Pacific Conference Series Vol. 347, Astronomical
  Data Analysis Software and Systems XIV. p.~29

\bibitem[\protect\citeauthoryear{{Theureau}, {Bottinelli}, {Coudreau-Durand},
  {Gouguenheim}, {Hallet}, {Loulergue}, {Paturel}  \& {Teerikorpi}}{{Theureau}
  et~al.}{1998}]{Theureau+98}
{Theureau} G.,  {Bottinelli} L.,  {Coudreau-Durand} N.,  {Gouguenheim} L.,
  {Hallet} N.,  {Loulergue} M.,  {Paturel} G.,   {Teerikorpi} P.,  1998,
  \mn@doi [\aaps] {10.1051/aas:1998416}, \href
  {http://adsabs.harvard.edu/abs/1998A%26AS..130..333T} {130, 333}

\bibitem[\protect\citeauthoryear{{Toba} et~al.,}{{Toba} et~al.}{2014}]{Toba+14}
{Toba} Y.,  et~al., 2014, \mn@doi [ApJ] {10.1088/0004-637X/788/1/45}, \href
  {http://adsabs.harvard.edu/abs/2014ApJ...788...45T} {788, 45}

\bibitem[\protect\citeauthoryear{{Treister}, {Urry}, {Schawinski}, {Cardamone}
  \& {Sanders}}{{Treister} et~al.}{2010}]{Treister+10}
{Treister} E.,  {Urry} C.~M.,  {Schawinski} K.,  {Cardamone} C.~N.,   {Sanders}
  D.~B.,  2010, \mn@doi [\apjl] {10.1088/2041-8205/722/2/L238}, \href
  {http://adsabs.harvard.edu/abs/2010ApJ...722L.238T} {722, L238}

\bibitem[\protect\citeauthoryear{{Treister}, {Schawinski}, {Urry}  \&
  {Simmons}}{{Treister} et~al.}{2012}]{Treister+12}
{Treister} E.,  {Schawinski} K.,  {Urry} C.~M.,   {Simmons} B.~D.,  2012,
  \mn@doi [ApJ] {10.1088/2041-8205/758/2/L39}, \href
  {http://adsabs.harvard.edu/abs/2012ApJ...758L..39T} {758, L39}

\bibitem[\protect\citeauthoryear{{Urry} \& {Padovani}}{{Urry} \&
  {Padovani}}{1995}]{Urry+95}
{Urry} C.~M.,  {Padovani} P.,  1995, \mn@doi [\pasp] {10.1086/133630}, \href
  {http://adsabs.harvard.edu/abs/1995PASP..107..803U} {107, 803}

\bibitem[\protect\citeauthoryear{{Veilleux}, {Kim}  \& {Sanders}}{{Veilleux}
  et~al.}{2002}]{Veilleux+02}
{Veilleux} S.,  {Kim} D.-C.,   {Sanders} D.~B.,  2002, \mn@doi [\apjs]
  {10.1086/343844}, \href {http://adsabs.harvard.edu/abs/2002ApJS..143..315V}
  {143, 315}

\bibitem[\protect\citeauthoryear{{Villforth} et~al.,}{{Villforth}
  et~al.}{2014}]{Villforth+14}
{Villforth} C.,  et~al., 2014, \mn@doi [MNRAS] {10.1093/mnras/stu173}, \href
  {http://adsabs.harvard.edu/abs/2014MNRAS.439.3342V} {439, 3342}

\bibitem[\protect\citeauthoryear{{Volonteri}, {Haardt}  \& {Madau}}{{Volonteri}
  et~al.}{2003}]{Volonteri+03}
{Volonteri} M.,  {Haardt} F.,   {Madau} P.,  2003, \mn@doi [ApJ]
  {10.1086/344675}, \href {http://adsabs.harvard.edu/abs/2003ApJ...582..559V}
  {582, 559}

\bibitem[\protect\citeauthoryear{{Wegner}, {Colless}, {Saglia}, {McMahan},
  {Davies}, {Burstein}  \& {Baggley}}{{Wegner} et~al.}{1999}]{Wegner+99}
{Wegner} G.,  {Colless} M.,  {Saglia} R.~P.,  {McMahan} R.~K.,  {Davies} R.~L.,
   {Burstein} D.,   {Baggley} G.,  1999, \mn@doi [\mnras]
  {10.1046/j.1365-8711.1999.02339.x}, \href
  {http://adsabs.harvard.edu/abs/1999MNRAS.305..259W} {305, 259}

\bibitem[\protect\citeauthoryear{{Wegner} et~al.,}{{Wegner}
  et~al.}{2001}]{Wegner+01}
{Wegner} G.,  et~al., 2001, \mn@doi [\aj] {10.1086/323915}, \href
  {http://adsabs.harvard.edu/abs/2001AJ....122.2893W} {122, 2893}

\bibitem[\protect\citeauthoryear{{White} \& {Rees}}{{White} \&
  {Rees}}{1978}]{White+78}
{White} S.~D.~M.,  {Rees} M.~J.,  1978, MNRAS, \href
  {http://adsabs.harvard.edu/abs/1978MNRAS.183..341W} {183, 341}

\bibitem[\protect\citeauthoryear{{Williams}, {Quadri}, {Franx}, {van Dokkum}
  \& {Labb{\'e}}}{{Williams} et~al.}{2009}]{Williams+09}
{Williams} R.~J.,  {Quadri} R.~F.,  {Franx} M.,  {van Dokkum} P.,   {Labb{\'e}}
  I.,  2009, \mn@doi [\apj] {10.1088/0004-637X/691/2/1879}, \href
  {http://adsabs.harvard.edu/abs/2009ApJ...691.1879W} {691, 1879}

\bibitem[\protect\citeauthoryear{{Willick}, {Bowyer}  \& {Brodie}}{{Willick}
  et~al.}{1990}]{Willick+90}
{Willick} J.~A.,  {Bowyer} S.,   {Brodie} J.~P.,  1990, \mn@doi [\apj]
  {10.1086/168772}, \href {http://adsabs.harvard.edu/abs/1990ApJ...355..393W}
  {355, 393}

\bibitem[\protect\citeauthoryear{{Woods}, {Geller}  \& {Barton}}{{Woods}
  et~al.}{2006}]{Woods+06}
{Woods} D.~F.,  {Geller} M.~J.,   {Barton} E.~J.,  2006, \mn@doi [\aj]
  {10.1086/504834}, \href {http://adsabs.harvard.edu/abs/2006AJ....132..197W}
  {132, 197}

\bibitem[\protect\citeauthoryear{{Wright} et~al.,}{{Wright}
  et~al.}{2010}]{Wright+10}
{Wright} E.~L.,  et~al., 2010, \mn@doi [AJ] {10.1088/0004-6256/140/6/1868},
  \href {http://adsabs.harvard.edu/abs/2010AJ....140.1868W} {140, 1868}

\bibitem[\protect\citeauthoryear{{Wu}, {Zou}, {Xia}  \& {Deng}}{{Wu}
  et~al.}{1998}]{Wu+98}
{Wu} H.,  {Zou} Z.~L.,  {Xia} X.~Y.,   {Deng} Z.~G.,  1998, \mn@doi [\aaps]
  {10.1051/aas:1998374}, \href
  {http://adsabs.harvard.edu/abs/1998A%26AS..127..521W} {127, 521}

\bibitem[\protect\citeauthoryear{{Xu}, {Wei}  \& {Hu}}{{Xu}
  et~al.}{2001}]{Xu+01}
{Xu} D.-W.,  {Wei} J.-Y.,   {Hu} J.-Y.,  2001, \cjaa, \href
  {http://adsabs.harvard.edu/abs/2001ChJAA...1...46X} {1, 46}

\bibitem[\protect\citeauthoryear{{Yagi}, {Goto}  \& {Hattori}}{{Yagi}
  et~al.}{2006}]{Yagi+06}
{Yagi} M.,  {Goto} T.,   {Hattori} T.,  2006, \mn@doi [\apj] {10.1086/500795},
  \href {http://adsabs.harvard.edu/abs/2006ApJ...642..152Y} {642, 152}

\bibitem[\protect\citeauthoryear{{Yan} \& {Blanton}}{{Yan} \&
  {Blanton}}{2012}]{Yan_Blanton+12}
{Yan} R.,  {Blanton} M.~R.,  2012, \mn@doi [\apj] {10.1088/0004-637X/747/1/61},
  \href {http://adsabs.harvard.edu/abs/2012ApJ...747...61Y} {747, 61}

\bibitem[\protect\citeauthoryear{{Yan}, {Newman}, {Faber}, {Konidaris}, {Koo}
  \& {Davis}}{{Yan} et~al.}{2006}]{Yan+06}
{Yan} R.,  {Newman} J.~A.,  {Faber} S.~M.,  {Konidaris} N.,  {Koo} D.,
  {Davis} M.,  2006, \mn@doi [ApJ] {10.1086/505629}, \href
  {http://adsabs.harvard.edu/abs/2006ApJ...648..281Y} {648, 281}

\bibitem[\protect\citeauthoryear{{Yan} et~al.,}{{Yan} et~al.}{2013}]{Yan+13}
{Yan} L.,  et~al., 2013, \mn@doi [AJ] {10.1088/0004-6256/145/3/55}, \href
  {http://adsabs.harvard.edu/abs/2013AJ....145...55Y} {145, 55}

\bibitem[\protect\citeauthoryear{{Yang}, {Mo}, {van den Bosch}, {Pasquali},
  {Li}  \& {Barden}}{{Yang} et~al.}{2007}]{Yang+07}
{Yang} X.,  {Mo} H.~J.,  {van den Bosch} F.~C.,  {Pasquali} A.,  {Li} C.,
  {Barden} M.,  2007, \mn@doi [ApJ] {10.1086/522027}, \href
  {http://adsabs.harvard.edu/abs/2007ApJ...671..153Y} {671, 153}

\bibitem[\protect\citeauthoryear{{Zabludoff} \& {Mulchaey}}{{Zabludoff} \&
  {Mulchaey}}{1998}]{Zabludoff+98}
{Zabludoff} A.~I.,  {Mulchaey} J.~S.,  1998, \mn@doi [\apj] {10.1086/305355},
  \href {http://adsabs.harvard.edu/abs/1998ApJ...496...39Z} {496, 39}

\bibitem[\protect\citeauthoryear{{da Costa} et~al.,}{{da Costa}
  et~al.}{1998}]{daCosta+98}
{da Costa} L.~N.,  et~al., 1998, \mn@doi [\aj] {10.1086/300410}, \href
  {http://adsabs.harvard.edu/abs/1998AJ....116....1D} {116, 1}

\bibitem[\protect\citeauthoryear{{de Vaucouleurs}, {de Vaucouleurs}, {Corwin},
  {Buta}, {Paturel}  \& {Fouque}}{{de Vaucouleurs}
  et~al.}{1991}]{deVaucouleurs+91}
{de Vaucouleurs} G.,  {de Vaucouleurs} A.,  {Corwin} Jr. H.~G.,  {Buta} R.~J.,
  {Paturel} G.,   {Fouque} P.,  1991, \skytel, \href
  {http://adsabs.harvard.edu/abs/1991S%26T....82Q.621D} {82, 621}

\makeatother
\end{thebibliography}

\label{lastpage}
\end{document}